\documentclass[11pt, a4paper]{article}
\usepackage{a4wide, url}

\usepackage{graphicx}
\usepackage{natbib}
\usepackage{enumitem}
\usepackage{hyperref}
\usepackage{setspace}
\usepackage{multirow, booktabs}
\usepackage{amsmath, amsthm, amssymb, amsfonts}
\usepackage{bm, color}  
\usepackage{epstopdf}
\usepackage{subfig}
\usepackage{comment}
\usepackage{arydshln}
\usepackage{mathtools}
\usepackage[utf8]{inputenc}
\usepackage{bbm}

\usepackage{algorithm}
\usepackage{algpseudocode}
\usepackage{tikz}
\usetikzlibrary{matrix,decorations.pathreplacing, calc, positioning, fit}

\allowdisplaybreaks

\DeclareMathAlphabet\mathbfcal{OMS}{cmsy}{b}{n}

\newcommand{\PreserveBackslash}[1]{\let\temp=\\#1\let\\=\temp}

\newcommand{\nn}{\nonumber}
\newcommand{\mbf}{\mathbf}
\newcommand{\mc}{\mathcal}
\newcommand{\bmx}{\begin{bmatrix}}
\newcommand{\emx}{\end{bmatrix}}

\renewcommand{\l}{\left}
\renewcommand{\r}{\right}

\def\wh{\widehat}
\def\wt{\widetilde}

\DeclareMathOperator*{\argmin}{arg\,min}
\newcommand{\E}[0]{\mathsf{E}}
\newcommand{\Var}[0]{\mathsf{Var}}

\newcommand{\iid}{\text{iid}}

\newcommand{\rain}{\text{rain}}
\newcommand{\air}{\text{air}}
\newcommand{\hod}{\text{hod}}
\newcommand{\dow}{\text{dow}}

\newcommand{\dnet}{\mbf D_{\text{net}}}

\newcommand{\lnet}{\mbf L_{\text{net}}}
\newcommand{\wtlnet}{\wt{\mbf L}_{\text{net}}}
\newcommand{\scard}{\vert \mc S \vert}
\newcommand{\dcard}{\vert \mathcal{D}^{\circ} \vert}
\newcommand{\hcard}{\vert \mathcal{H}^{\circ} \vert}

\algnewcommand{\Inputs}[1]{%
  \State \textbf{Inputs:}
  \Statex \hspace*{\algorithmicindent}\parbox[t]{.8\linewidth}{\raggedright #1}
}
\algnewcommand{\Initialize}[1]{%
  \State \textbf{Initialize:}
  \Statex \hspace*{\algorithmicindent}\parbox[t]{.8\linewidth}{\raggedright #1}
}

\algnewcommand{\Outputs}[1]{%
  \State \textbf{Ouputs:}
  \Statex \hspace*{\algorithmicindent}\parbox[t]{.8\linewidth}{\raggedright #1}
}

\algnewcommand\algorithmicgenerate{\textbf{GENERATE}}
\algnewcommand\GENERATE[1]{\State\algorithmicgenerate\ #1}
\algnewcommand\algorithmicset{\textbf{set}}
\algnewcommand\SET[1]{\State\algorithmicset\ #1}
\algrenewcommand\algorithmicfor{\textbf{for}} 
\algrenewcommand\algorithmicend{\textbf{end}} 
\algrenewcommand\algorithmicwhile{\textbf{while}} %

\theoremstyle{definition}

\theoremstyle{definition}

\theoremstyle{definition}

\theoremstyle{definition}

\theoremstyle{definition}

\theoremstyle{remark}
\newtheorem{rem}{Remark}[section]
\theoremstyle{definition}

\theoremstyle{definition}

\title{Capturing usage patterns in bike sharing system via \\ multilayer network fused Lasso}
\author{Yunjin Choi$^1$ \and Haeran Cho$^2$ \and Hyelim Son$^3$}

\begin{document}

\maketitle

\begin{abstract}
Data collected from a bike-sharing system exhibit complex temporal and spatial features. We analyze shared-bike usage data collected in three large cities at the level of individual stations, accounting for station-specific behavior and covariate effects. For this, we adopt a penalized regression approach with a multilayer network fused Lasso penalty. These fusion penalties are imposed on networks which embed spatio-temporal linkages, and capture the homogeneity in bike usage that is attributed to intricate spatio-temporal features without arbitrarily partitioning the data. On the real-life datasets, we demonstrate that the proposed approach yields competitive predictive performance and provides a new interpretation of the data.
\end{abstract}

\footnotetext[1]{Department of Statistics, University of Seoul, South Korea. Email: \url{ycstat@uos.ac.kr}.} 

\footnotetext[2]{School of Mathematics, University of Bristol, UK.
Email: \url{haeran.cho@bristol.ac.uk}.}

\footnotetext[3]{School of Economics, University of Seoul, South Korea. 
Email: \url{hlson@uos.ac.kr}.} 

\noindent%

\section{Introduction}

Bike-sharing systems (BSS) have become increasingly popular in urban areas and have successfully  complemented public transportation systems in dense metropolitan cities.
In addition to its utility to bike users, the installation of BSSs has been found
to reduce the usage of automobiles \citep{fishman2014bike} and thus traffic congestion and possibly green house emissions \citep{hamilton2018bicycle}.
To fully realize these benefits, efficient allocation of docking stations and bike docks is essential,
which in turn requires understanding the user behaviour based on the abundant data collected on the BSS and other urban and environmental factors that are known to influence bike usage.

In line with the increasing popularity of BSSs,
there exists a vast literature on the analysis of bike usage patterns; for and overview, see \cite{shaheen2010bikesharing} and \cite{fishman2016bikeshare}.
Below we provide a brief summary of the literature on quantitative or statistical analysis of BSS usage data which is categorized into two, following \cite{etienne2014model}.
The first branch addresses the problem of clustering stations based on usage patterns 
\citep{froehlich2009sensing, vogel2011understanding, etienne2014model}.
Regarding the BSS as a network, community detection algorithms have also been adopted for this purpose \citep{austwick2013structure, borgnat2013dynamical, zhou2015understanding}.
\cite{gervini2019exploring} 
cluster the stations based on functional canonical correlations
of log-intensity functions.

The second line of research concerns the problem of 
predicting the station occupancy or the state of the system at a given time.
\cite{faghih2016incorporating} model incoming and outgoing traffic at multiple stations as a panel with variables accounting for spatial and temporal autoregressive structures.
\cite{liu2016rebalancing} model inter-station bike transitioning
for improving the effectiveness of rebalancing operations
by predicting the station drop-off demand.
\cite{torti2021modelling} adopt functional linear regression
to model the directed flow between pairs of administrative divisions that aggregate multiple stations.

In all above, it is well-documented that BSS data show temporal and spatial patterns. 
To address these, some previous works pre-process the datasets e.g.\ by aggregating stations into administrative regions \citep{torti2021modelling}, partition the data using subject-specific knowledge \citep{faghih2016incorporating}, or analyze the data collected on weekdays and at weekends separately \citep{liu2016rebalancing}.
In complex urban environments, however, it may be difficult to find a single clustering of the data that comprehensively accounts for the usage patterns since there exist multiple approaches to produce geographical or temporal divisions according to socioeconomic characteristics, land zones, traffic infrastructure or population composition.
Besides, collecting in-depth information about the multifaceted nature of a large metropolitan city is costly or even impossible.

In this paper, we analyze the hourly bike rental data collected from BSSs in three metropolitan cities (Seoul, New York and San Francisco) by adopting a penalized regression modeling approach. 
Without arbitrary partitioning of the data, we model each BSS dataset at the granularity of individual stations by including station-specific parameters as well as trends and variables related to precipitation and air quality (where available).
Such a model enjoys considerable flexibility and captures time-dependent usage patterns at individual stations but it potentially suffers from the risk of overfitting as the number of parameters 
increases linearly with the number of stations. 

To address this issue and accommodate the characteristics of the BSS, we propose a multilayer network fused Lasso penalty which extends the fused Lasso penalty \citep{tibshirani2005}. 
In the absence of a natural ordering among the stations, the proposed penalty imposes the penalization using a multilayer network and promotes fusion of the parameters linked by edges in the network.
In doing so, we view the BSS as a multilayer network where the stations serve as its nodes, layers correspond to different hours of a day and within-layer and cross-layer edges are given by spatial and temporal proximity of the stations.

The model fitted from the penalized regression method adaptively captures spatial and temporal homogeneity in bike usage, without (arbitrarily) partitioning the data which potentially leads to information loss. 
In our data analysis, the proposed multilayer fused Lasso exhibits superior predictive performance compared to alternative penalization methods as well as competitors separately modeling the data from each station.
Also, we propose a new network-based model complexity measure which reveals that while similarities exist, the stations exhibit fair amount of heterogeneity.
This conclusion supports that partitioning the stations into a handful of clusters may be inappropriate for such large-scale urban transportation systems.

The remainder of this paper is organised as follows.
Section~\ref{sec:method} introduces a penalized Poisson regression methodology and provides a multilayer network-based interpretation of the proposed penalty.
Section~\ref{sec:analysis} demonstrates the effectiveness of our proposed approach on the bike usage datasets collected from the three cities. 
Section~\ref{sec:conc} concludes the paper, and Appendix describes the algorithm for the penalized maximum likelihood estimation as well as providing further descriptions of the datasets and additional numerical results.

\section{Model and estimation}
\label{sec:method}

From the review of the literature and exploratory data analysis presented in Appendix~\ref{sec:data}, we list some of the commonly observed characteristics of usage patterns in large BSS:
\begin{enumerate}[noitemsep, wide, labelwidth=0pt, itemindent=!, label = (C\arabic*)]
\item \label{c:trend} There exists an overall increasing trend.
\item \label{c:temp} Bike rentals show station-specific temporal patterns that smoothly vary over time.
\item \label{c:net} These patterns are shared across stations that are geographically close. 
\item \label{c:cov} Bike rentals are influenced by the weather condition and the air quality.
\end{enumerate}
To accommodate these characteristics, Section~\ref{sec:model} proposes a Poisson regression model for hourly rental frequencies collected from the entire BSS.
Section~\ref{sec:penalty} presents the accompanying estimation strategy
and introduces the multilayer fused Lasso penalty designed to capture the characteristics~\ref{c:temp}--\ref{c:net}.
Section~\ref{sec:network} provides a network interpretation of the proposed penalization technique which aids in understanding and visualizing the penalty.

\subsection{Poisson regression model}
\label{sec:model}

Let $Y_i$ denote the $i$th observation representing the hourly rental frequency
at station $S(i) \in \mc S = \{1, \ldots, S\}$
and hour $H(i) \in \mc H = \{0, \ldots, 23\}$,
on day $D(i) \in \mc D = \{\text{Mo, $\ldots$, Su}\}$, with the time index of the $i$th observation denoted by $t(i) \in \mc T = \{0, \ldots, T - 1\}$ and the time span of the data by $T$.
In total, the dataset contains $n = S \times T \times 24$ observations.
We denote by the vector of covariates relevant for the $i$th observation in $\mbf x_i = (S(i), t(i), D(i), H(i), \mbf z_i^\top)^\top$, where $\mbf z_i$ represents additional covariate information such as
the precipitation and air quality. 


Each station in the BSS has a fixed number of docks but this does not determine the capacity of a station as for some systems, it is possible to leave bikes even if the docks are fully occupied by chaining them to existing bikes.
Based on this property of the BSS, and in view of the large number of stations (e.g.\ $S = 1505$ for Seoul) that brings in a large number of station-specific parameters in the model~\eqref{eq:model} below, we propose to adopt a Poisson distribution for modeling the conditional distribution of $Y_i$, the hourly count of the bikes rented out at station $S(i)$, given $\mbf x_i$, i.e.\
$Y_i \vert \mbf x_i \sim_{\iid} \text{Poisson}(\mu_i)$ with $\mu_i \equiv \mu(\mbf x_i) = \E(Y_i \vert \mbf x_i)$.

\begin{rem}
\label{rem:poisson}
The Poisson distribution is frequently adopted in bike usage data analysis, see e.g.\ \cite{etienne2014model}, \cite{gervini2019exploring} and \cite{torti2021modelling}.
Although the data may exhibit over-dispersion (see Appendix~\ref{sec:overdisp}), we find that the penalized regression approach proposed in Section~\ref{sec:penalty} below shows good predictive performance across multiple datasets, see Section~\ref{sec:analysis}.
From this perspective, we regard the proposed method as penalized quasi-maximum likelihood estimation.
Alternatively, generalized \citep{consul1989generalized} or Conway–Maxwell \citep{shmueli2005} Poisson distributions are also available for dealing with possible over- or under-dispersion.
\end{rem}

One option to relate $\mu_i$ to $\mbf x_i$ is via the following log-linear model:
\begin{align}
\log\l(\frac{\mu_i}{C_{S(i)}}\r) = \theta_{S(i)} + \alpha\, t(i) + \langle \bm\beta, \mbf z_i \rangle +
\theta^{\hod}_{H(i)} + \theta^{\dow}_{D(i)}.
\label{eq:model:no}
\end{align}
Here, $\theta^{\hod}_h$ and $\theta^{\dow}_d$, respectively, contain ``hour of a day'' and ``day of a week'' effects common to all stations.
The offset term $C_s$ denotes the capacity of station~$s$, i.e.\ the number of docks,
so that~\eqref{eq:model:no} can be interpreted as modeling
the expected rental frequency per hour per station capacity.
This model, referred to as the {\it no-interaction} model,
does not permit the temporal effects to be station-specific
and thus may be too simple to address~\ref{c:temp}.

Allowing for interactions between the station and temporal effects,
we consider the following {\it full-interaction} model:
\begin{align}
\log\l(\frac{\mu_i}{C_{S(i)}}\r) =& \ \theta_{S(i)} + \alpha\, t(i) 
+ \langle \bm\beta, \mbf z_{i} \rangle 
+ \theta^{\hod}_{H(i)} + \theta^{\dow}_{D(i)}
+ \theta^{\hod}_{S(i), H(i)} + \theta^{\dow}_{S(i), D(i)}.
\label{eq:model}
\end{align}
The station-hour and station-day interaction terms $\theta^{\hod}_{s, h}$ and $\theta^{\dow}_{s, d}$ permit each station to exhibit individual temporal patterns.
For model identifiability, we set the baseline parameters to zero: $\theta^{\hod}_0 = \theta^{\dow}_{\text{Mo}} = \theta^{\hod}_{s, 0} = \theta^{\dow}_{s, \text{Mo}} = \theta^{\hod}_{1, h} = \theta^{\dow}_{1, d} = 0$.

Model~(\ref{eq:model}) accounts for~\ref{c:trend}, \ref{c:cov}, and to a certain extent, \ref{c:temp},
by including the parameters $\alpha$ capturing the overall trend, $\bm\beta$ capturing the covariate effects
and $\theta^{\hod}_{s, h}$ and $\theta^{\dow}_{s, d}$ addressing station-specific temporal patterns.
In doing so, we take a different approach from those taken in previous studies in which,
after (arbitrarily) partitioning the dataset according to temporal or spatial variables, or both,
individual partitions are separately modelled (e.g.\ \cite{liu2016rebalancing}; \cite{torti2021modelling}). 
Instead, by including the interaction terms, we use the full dataset and avoid any information loss.

While the full-interaction model enjoys considerably more flexibility than the model in~\eqref{eq:model:no},
it suffers the risk of overfitting the data with a large number of parameters; 
in the case of the BSS in Seoul, it amounts to $p = 45155$. 
In addition, the characteristic identified in~\ref{c:net},
that the stations geographically close to one another tend to exhibit similar usage patterns,
is not adequately accounted for by the model fitted without any constraint.
Such an approach does not benefit from the temporal ordering inherent in the parameters $\theta^{\hod}_{s, h}$, and thus does not fully account for~\ref{c:temp}.
In the next section, we propose a penalized maximum likelihood estimation (MLE) methodology for model in~(\ref{eq:model}) with a multilayer network fused Lasso which explicitly sets out to address these issues.

\subsection{Penalized MLE via multilayer network fused Lasso}
\label{sec:penalty}

We bridge the two models~\eqref{eq:model:no} and~\eqref{eq:model} at extreme ends, by adopting a penalized MLE strategy with 
a fused Lasso penalty designed to capture spatial and temporal homogeneity in bike usage patterns observed in the data, that is, \ref{c:temp} and~\ref{c:net}.
First proposed by \cite{tibshirani2005}, the fused Lasso augments the Lasso \citep{tibshirani1996regression} with a penalty that takes advantage of a meaningful ordering of the variables when such is available.

Under~\eqref{eq:model}, we partition the parameters into
$\Theta = \{\theta_s, \, s \in \mc S\}$, 
$\Theta_H = \{\theta^{\hod}_{h}, \, \theta^{\hod}_{s, h}, \, h \in \mc H \setminus \{0\}, \, s \in \mc S \setminus \{1\} \}$,
$\Theta_D = \{\theta^{\dow}_{d}, \, \theta^{\dow}_{s, d}, \, d \in \mc D \setminus \{\text{Mo}\}, \,  s \in \mc S \setminus \{1\} \}$ and
$\Delta = \{\alpha, \bm\beta 
\}$.
We adopt the fusion penalty to pool the information (i)~from adjacent stations for the estimation of $\Theta$, $\Theta_H$ and $\Theta_D$, and (ii)~over the course of a day for the estimation of $\Theta_H$.
We also impose a standard Lasso penalty on all station-specific parameters to encourage sparsity. 
Subsequently, we minimize the following penalized negative log-likelihood
\begin{small}
\begin{align}
- \ell(\Theta, \Theta_H, \Theta_D, \Delta)
+ \lambda \sum_{s \in \mc S} \l( \sum_{h \in \mc H} \vert \theta^{\hod}_{s, h} \vert
+ \sum_{d \in \mc D} \vert \theta^{\dow}_{s, d} \vert \r)
+ \lambda_N p_N(\Theta, \Theta_H, \Theta_D) 
+ \lambda_H p_H(\Theta_H)
\label{eq:org_obj_admm}
\end{align}
\end{small}
with respect to $\Theta$, $\Theta_H$, $\Theta_D$ and $\Delta$, where $\lambda, \lambda_N, \lambda_H > 0$ control the degree of the penalization.
Here, $\ell(\Theta, \Theta_H, \Theta_D, \Delta) = -\sum_{i = 1}^n \mu_i + \sum_{i = 1}^n y_i \log(\mu_i) + \text{constant}$, denotes the log-likelihood,
and $\mu_i = \mu_i(\Theta, \Theta_H, \Theta_D, \Delta)$  
is a function of the parameter vectors under~\eqref{eq:model}.
We choose not to impose the Lasso penalization on $\theta_s$, $\theta^{\hod}_h$ and $\theta^{\dow}_d$, to avoid cancelling out station-specific usage patterns, as well as those representing overall hourly and daily patterns shared across all stations.

To capture the similarities between geographically adjacent stations,
we define a set of neighboring stations for each station indexed by $s$,
as $\mc N_r(s) = \{s^\prime \in \mc S \setminus \{s\}: \, d(s, s^\prime) < r\}$ and its cardinality by $\vert \mc N_r(s) \vert$.
Here, $d(s, s^\prime)$ denotes the distance between the two stations $s$ and $s^\prime$ measured as the crow flies, and $r$ denotes a pre-specified distance.
Note that $\mc N_r(s), \, s \in \mc S$, define overlapping groups of the stations.
Then, we define 
\begin{align}
& p_N(\Theta, \Theta_H, \Theta_D) = \sum_{s \in \mc S} \sqrt{ \vert \mc N_r(s) \vert \, \sum_{s^\prime \in \mc N_r(s)} \l[ \sum_{h \in \mc H} (\phi^{\hod}_{s, h} - \phi^{\hod}_{s', h})^2 +
\sum_{d \in \mc D} (\phi^{\dow}_{s, d} - \phi^{\dow}_{s^\prime, d})^2 \r] }
\label{def:admm_net_penalty} 
\end{align}
with $\phi^{\hod}_{s, h} = \theta_s + \theta^{\hod}_{h} + \theta^{\hod}_{s, h}$ and $\phi^{\dow}_{s, d} = \theta_s + \theta^{\dow}_{d} + \theta^{\dow}_{s, d}$,
which encode the station-specific hourly and daily patterns under~(\ref{eq:model}).
The penalization on $p_N$ forces the pairs of parameters 
$(\theta_s, \theta_{s'})$, $(\theta^{\hod}_{s, h}, \theta^{\hod}_{s', h})$ and $(\theta^{\dow}_{s, d}, \theta^{\dow}_{s', d})$
to fuse for neighboring stations~$s$ and $s'$, which encourages them to exhibit similar hourly and daily patterns and thus addresses the behavior noted in~\ref{c:net}.
In fact, the fusion of parameters is promoted between any pair of stations
which are connected on single- and multi-layer networks defined with $\mc S$ as the node set, and the edge set given by the neighborhood relationship, see Section~\ref{sec:network} for the network interpretation of the proposed penalization.
In other words, by adopting $p_N$, we pool information across the BSS to estimate the interaction parameters.
Every parameter inside the square root is considered to belong to the same group,
and the weighting applied with the size of $\mc N_r(s)$ 
follows the convention of the group Lasso \citep{yuan2006model} literature,
imposing more penalization on the parameters associated with well-connected stations.
Then, the proposed method applies the Lasso penalty to the $\ell_1$-norm of $\theta^{\hod}_{s, h}$ and $\theta^{\dow}_{s, d}$ only, which gives it the interpretation of bridging between no-interaction and full-interaction models.

There is a natural temporal ordering inherent in $\Theta_H$ that gives rise to
the fusion penalty in its canonical form imposed on $\phi^{\hod}_{s, h}$:
\begin{align}
p_H(\Theta_H) =  \sum_{s \in \mc S} 
\sum_{h = 0}^{23} \l\vert \phi^{\hod}_{s, h} - \phi^{\hod}_{s, h + 1} \r\vert 
=  \sum_{s \in \mc S} 
\sum_{h = 0}^{23} \l\vert (\theta^{\hod}_{h} +\theta^{\hod}_{s, h})- (\theta^{\hod}_{h+1} + \theta^{\hod}_{s, h + 1}) \r\vert 
\label{def:admm_hour_penalty} 
\end{align}
with $\theta^{\hod}_{s, 24} = \theta^{\hod}_{s, 0}$.
Imposing a penalty on $p_H$ encourages the consecutive (station-specific) hourly effects $(\theta^{\hod}_h, \theta^{\hod}_{h + 1})$ and $(
\theta^{\hod}_{s, h}, \theta^{\hod}_{s, h+1})$,
to become close to one another and suppress abrupt changes in usage prediction.

Jointly, $p_N$ and $p_H$ comprise the proposed multilayer network fusion penalty.
The impact of the fusion penalty is determined by the sizes of $\lambda_N$ and $\lambda_H$,
which we select via cross validation as described in Section~\ref{sec:tuning}.
The Alternating Direction Method of Multipliers (ADMM) algorithm \citep{boyd2011distributed} is employed to solve the convex optimization problem in~\eqref{eq:org_obj_admm}.
Efficient implementation of the algorithm requires careful re-parametrization of the model~\eqref{eq:model} which makes use of the data structure, see Appendix~\ref{sec:admm} for details including the discussion of the computational complexity.

\begin{rem}
\label{rem:fusion}
Since its introduction, the fusion penalty has successfully been applied to modeling temporal \citep{tibshirani2014adaptive, harchaoui2010multiple} and spatial \citep{sun2016fused, li2019spatial, sass2021flexible} patterns. 
We propose to simultaneously capture both spatial and temporal patterns underlying the BSS using the fusion penalties $p_N$ and $p_H$ 
which, to our best knowledge, is a first such attempt.
Section~\ref{sec:network} provides its novel interpretation by means of a multilayer network, which shows that our approach may be extended to accommodate general homogeneity characteristics 
by imposing the penalization through an appropriately constructed multilayer network.
\end{rem}

\subsection{Multilayer network interpretation of the fusion penalty}
\label{sec:network}

\begin{figure}[h!t!]
\centering
\resizebox*{8cm}{!}{\includegraphics{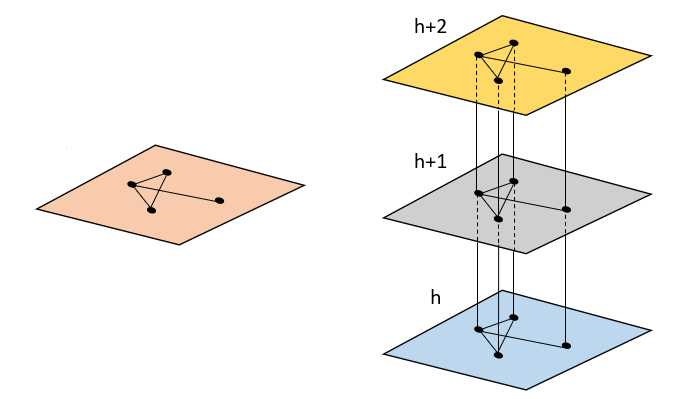}}
\caption{Illustration of a single-layer network $\mathfrak{N}_{\text{single}}$ (left) and a multilayer network $\mathfrak{N}_{\text{multi}}$ (right). Each dot represents a node (station) and a solid line represents an edge connecting the nodes within each layer and across adjacent layers.}
\label{fig:single_multi_network}
\end{figure}

We introduce the following networks that underpin the penalty functions $p_N$ and $p_H$: A single-layer network $\mathfrak{N}_{\text{single}}(r) = (\mc S, \mc E_{\text{single}}(r))$ with $\mc E_{\text{single}}(r) = \cup_{s \in \mc S} \{ (s, s'), \, s' \in \mc N_r(s)\}$, and a multilayer one 
$\mathfrak{N}_{\text{multi}}(r) = (\mc S \times \mc H, \, \mc E_{\text{multi}}(r))$ with $\vert \mc H \vert = 24$ hourly layers, where
\begin{multline*}
\mc E_{\text{multi}}(r) =
\cup_{h \in \mc H} \cup_{s \in \mc S} \l\{ \l( (s, h), (s', h) \r), \, s' \in \mc N_r(s) \r\} 
\bigcup \cup_{s \in \mc S} \l\{ \l( (s, h), (s, h + 1) \r), \, h \in \mc H \r\}. 
\end{multline*}
Below we suppress their dependency on $r$ for simplicity when it does not cause any confusion.
For an illustrative example of $\mathfrak{N}_{\text{single}}$ and $\mathfrak{N}_{\text{multi}}$, see Figure~\ref{fig:single_multi_network}.

The network $\mathfrak{N}_{\text{single}}$ is a single-layer, undirected network 
that is solely determined by the sets of neighbors $\mc N_r(s)$ of the stations.
In this network, a pair of ``day of a week'' parameters $(\phi^{\dow}_{s, d}, \phi^{\dow}_{s', d})$ for each given day $d$, are encouraged to take values close to one another by the penalty function $p_N$, provided that the stations belong to the same connected component of $\mathfrak{N}_{\text{single}}$, i.e.\ a path exists connecting the pair of stations in $\mathfrak{N}_{\text{single}}$.
On the other hand, $\mathfrak{N}_{\text{multi}}$ is a multilayer, undirected network with $\vert \mc H \vert = 24$ as the number of layers;
we follow the notational convention of \cite{kivela2014multilayer}, in which the set $\mc H$ serves as a set of elementary layers for the hourly aspect,
and each edge connects a pair of node-layer tuples $(s, h)$ and $(s', h')$ for some $s, s' \in \mc S$ and $h, h' \in \mc H$.
Each pair of the ``hour of a day'' parameters 
$\phi^{\hod}_{s, h}$ and $\phi^{\hod}_{s', h'}$
is encouraged to fuse with one another by the penalty functions $p_N$ and $p_H$,
if the corresponding pair of nodes are connected in $\mathfrak{N}_{\text{multi}}$.

\begin{figure}[h]
\centering
\resizebox*{12cm}{!}{\includegraphics{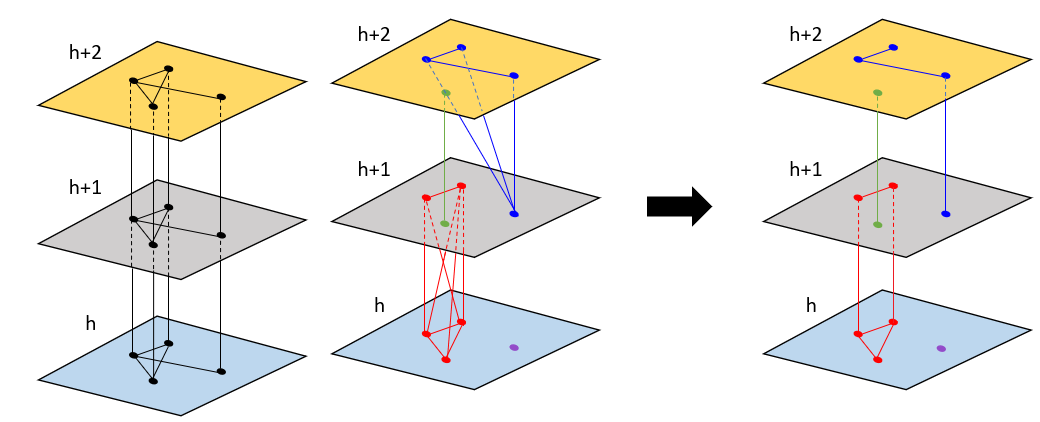}}
\caption{Illustration of multilayer networks. 
Left: Each layer of $\mathfrak{N}_{\text{multi}}$ embeds the linkages between the stations determined by their geographical distances at a given hour $h$, and the station at layer $h$ is linked to itself at layers $h - 1$ and $h + 1$, 
which underpins how $\phi^{\hod}_{s, h}$ are encouraged to be fused by $p_N$ and $p_H$.
Middle: $\wh{\mathfrak{N}}_H$ is determined by the values of the estimates of $\phi^{\hod}_{s, h}$ with an edge indicating that the connected estimates share the identical values.
Right: A multilayer network formed with its edges obtained as an intersection of the edge sets of $\mathfrak{N}_{\text{multi}}$ and $\wh{\mathfrak{N}}_H$,
which contains four connected components.} 
\label{fig:network}
\end{figure}

Next, we define networks whose edges 
are determined by the coefficient estimates from the penalized MLE.
Let 
$\wh\phi^{\hod}_{s, h}$ and $\wh\phi^{\dow}_{s, d}$ denote the estimates of the parameters $\phi^{\hod}_{s, h}$ and $\phi^{\dow}_{s, d}$, respectively;
their values depend on the tuning parameters 
$(r, \lambda, \lambda_N, \lambda_H)$ which we omit for simplicity.
Then, these networks are
\begin{align}
\wh{\mathfrak{N}}_{D, d} = \l(\mc S, \, \mc E_{D, d}\r) \text{ \ with \ } 
\mc E_{D, d} =& \l\{(s, s'), \, s \ne s': \, \wh\phi^{\dow}_{s, d} = \wh\phi^{\dow}_{s', d} \r\} \text{ \ for each \ } d \in \mc D, 
\nn \\
\wh{\mathfrak{N}}_H = \l(\mc S \times H, \mc E_H\r)  \text{ \ with \ }
\mc E_H =&  \cup_{h \in \mc H} \cup_{s \in \mc S} 
\l\{((s, h), (s', h')), \, (s,h) \ne (s',h'): \, \wh\phi^{\hod}_{s, h} = \wh\phi^{\hod}_{s', h'} \r\}. \nn
\end{align}
As with $\mathfrak{N}_{\text{single}}$, the networks $\wh{\mathfrak{N}}_{D, d}$ are single-layer networks
and an edge joins two nodes $s$ and $s'$ when their node features (i.e.\ parameter estimates of $\phi^{\dow}_{s,d}$ and $\phi^{\dow}_{s',d}$ for a given $d \in \mc D$) are identical, possibly due to the fusion penalty but not necessarily so.
The network $\wh{\mathfrak{N}}_H$, as with $\mathfrak{N}_{\text{multi}}$, 
is a multilayer network with the hourly layer given by $\mc H$, 
and an edge is formed between a pair of nodes $(s,h)$ and $(s',h')$ when the estimates of $\phi^{\hod}_{s, h}$ and $\phi^{\hod}_{s', h'}$ agree at $(s,h) \ne (s',h')$. 
For an illustrative example of $\wh{\mathfrak{N}}_H$, see Figure~\ref{fig:network}.

For two networks (either single- or multilayer) 
$\mathfrak{N}_i = (\mc V, \mc E_i), \, i = 1, 2$, sharing the same node set $\mc V$, 
denote by $\mathfrak{N}_1 \cap \mathfrak{N}_2 = (\mc V, \mc E_1 \cap \mc E_2)$ 
the network formed by taking the intersection of their edge sets.
Our proposed penalized regression method takes as an input the observable networks $\mathfrak{N}_{\text{single}}$ and $\mathfrak{N}_{\text{multi}}$, 
and outputs the networks capturing the homogeneity between the stations nodes, namely, $\mathfrak{N}_{\text{single}} \cap \wh{\mathfrak{N}}_{D, d}$ (on a given day of a week $d$) and $\mathfrak{N}_{\text{multi}} \cap \wh{\mathfrak{N}}_H$ (along the hourly layer).
Depending on the choice of penalty parameters, the output networks are not necessarily sparse; in fact, this is the case in our data analysis reported in Section~\ref{sec:analysis}.
This distinguishes our approach from the existing literature on clustering or partitioning the dataset using spatial or temporal variables prior to analysis.
\begin{rem}
\label{rem:undirected}
Although the layers in the multilayer network $\mathfrak{N}_{\text{multi}}$ represent hours of a day that admit a natural ordering, we choose to treat the inter-layer edges therein as undirected ones.
This choice stems from that $\mathfrak{N}_{\text{multi}}$ represents the penalization $p_H$ imposed on $\phi^{\hod}_{s, h}$ (see~\eqref{def:admm_hour_penalty}), which is not compatible with the notion of directed edges.
Also, it aligns with our aim in introducing the penalization for capturing the smooth transition of usage over the course of a day, as noted in~\ref{c:temp}, in view of enhancing the predictive performance of the model where temporal effects enter as covariates.
On the other hand, if our goal is to train a time series model for forecasting of the future, a directed network-based approach would be more appropriate. 
\end{rem}


\section{Data analysis}
\label{sec:analysis}

\subsection{Overview}

We analyze three bike usage datasets collected from BSSs in Seoul\footnote{The dataset is available at \url{https://data.seoul.go.kr/}.},
New York\footnote{The dataset is available at \url{https://s3.amazonaws.com/tripdata/index.html}.} and San Francisco\footnote{The dataset is available from \cite{bikeshare14}.}, which contain hourly rental records from April and May (approx.\ $60$ days), see Appendix~\ref{sec:data} for full details.
The dataset collected from the BSS in Seoul are from $S = 1505$ stations and contains $n = 2058840 (= 1505 \times 57 \times 24)$ data points, the New York dataset is from $S = 795$ stations with $n = 1144800 (= 795 \times 60 \times 24)$, and the San Francisco dataset is from $S = 71$ stations with $n = 102240 (= 71 \times 60 \times 24)$.
All three datasets come with an accompanying binary variable encoding the precipitation status ($z^{\rain}_i \in \{0, 1\}$), while only the Seoul dataset contains additional information on the air quality status ($\mbf z^{\air}_i \in \{0, 1\}^4$, representing ``very bad'', ``bad'', ``average'' or ``good''). 

For each dataset, holding out the final week as the test set for evaluating the predictive performance (see Section~\ref{sec:forecasting}), we train the proposed model (and other competitors) on the remaining training set.

\subsection{Tuning parameter selection}
\label{sec:tuning}


\paragraph{Selection of $\lambda$, $\lambda_N$ and $\lambda_H$.}
We propose to select the penalty parameters $\lambda$, $\lambda_N$ and $\lambda_H$ via cross validation (CV).
In the penalized regression literature, CV is typically performed by randomly partitioning the data into five or ten folds.
However, in the case of the BSS datasets, this approach overlooks an inherent temporal structure therein.
Therefore, we adopt a $6$-fold CV in which each fold includes a balanced number of all seven days of the week. 
As a CV measure, we adopt the mean squared Pearson residuals (MSPR): 
\begin{align}
\label{eq:cv}
\text{CV}(r, \lambda, \lambda_N, \lambda_H) = 
\frac{1}{6} \sum_{j = 1}^7 \frac{1}{n_j} \sum_{i = 1}^{n_j} 
\frac{(Y^{(j)}_i - \wh\mu^{(j)}_i(r, \lambda, \lambda_N, \lambda_H))^2}{\wh\mu^{(j)}_i(r, \lambda, \lambda_N, \lambda_H)}
\end{align}
where for the $j$th fold,
$n_j$ denotes the total number of observations,
$Y_i^{(j)}$ the $i$th observation and
$\wh\mu_i^{(j)}(r, \lambda, \lambda_N, \lambda_H)$ is the corresponding estimate of the mean from the model fitted to the remaining data with the given tuning parameters. 
Provided that model assumptions are met, MSPR is approximately $\chi^2$-distributed, and has been used for measuring goodness-of-fit in Poisson regression \citep{mccullagh1989}.
We evaluate $\text{CV}(r, \lambda, \lambda_N, \lambda_H)$ on the grids of equispaced values (in log-scale) for $(\lambda, \lambda_N, \lambda_H)$
and select the combination that returns the smallest CV.

\begin{table}[h!t!b!]
\centering
\caption{Summary statistics of the numbers of neighbor stations $\vert \mc N_r(s) \vert, \, s \in \mc S$, with varying~$r$ for BSSs in Seoul, New York (NYC) and San Francisco (SF).
We also report the number of connected components and the percentage of unconnected stations in $\mathfrak{N}_{\text{single}}(r)$.}
\label{tab:radius}
\resizebox{0.9\columnwidth}{!}{
\begin{tabular}{rr ccccc}
\toprule
$r$ & City & Mean & Median & SD & \# Connected components & \% of Unconnected stations \\
\cmidrule(lr){1-2} \cmidrule(lr){3-5} \cmidrule(lr){6-6} \cmidrule(lr){7-7}
375m & Seoul & 1.66 & 1 & 1.37 & 609 & 19.87\\
& NYC & 3.57 & 4 & 1.64 & 34 & 2.39\\
& SF & 1.07 & 1 & 0.88 & 35 & 30.99\\
\cmidrule(lr){1-2} \cmidrule(lr){3-5} \cmidrule(lr){6-6} \cmidrule(lr){7-7}
750m & Seoul &6.36 & 6 & 3.50 & 34 & 0.73\\
& NYC & 14.54 & 14 & 4.86 & 5 & 0.00 \\
& SF & 5.13 & 5 & 3.59 & 12 & 7.04\\
\cmidrule(lr){1-2} \cmidrule(lr){3-5} \cmidrule(lr){6-6}  \cmidrule(lr){7-7}
1500m & Seoul & 23.03 & 22 & 9.05 & 2 & 0.00\\
& NYC & 50.14 & 48 & 16.47 & 2 & 0.00\\
& SF & 12.79 & 12 & 8.50 & 8 & 1.41 \\
\cmidrule(lr){1-2} \cmidrule(lr){3-5} \cmidrule(lr){6-6}  \cmidrule(lr){7-7}
3000m & Seoul & 77.61 & 78 & 25.71 & 1 & 0.00\\
& NYC & 160.71 & 164 & 43.40 & 1 & 0.00\\
& SF & 21.01 & 15 & 12.48 & 5 & 0.00\\
\bottomrule
\end{tabular}}
\end{table}

\paragraph{Choice of $r$.} Table~\ref{tab:radius} illustrates that due to differences in geography and their size, BSSs in different cities exhibit varying degrees of connectivity. 
In particular, stations in New York are more heavily connected with larger numbers of neighbors compared to those in Seoul or San Francisco. 
As shown later, on the three datasets in consideration, the predictive performance of our penalized regression approach is not sensitive to the choice of $r$ thanks to the adaptive selection of the penalty parameters. 
Specifically, while the proposed CV procedure prefers a similar value of $\lambda_H$ regardless of $r$, it prefers smaller values of $\lambda_N$ with growing $r$ (see Appendix~\ref{sec:cv_table}) when controlling the level of penalization promoting homogeneity across the neighbors.
At the same time, the computational efforts increase with $r$ as the numbers of neighbors increase which is attributed to the fusion penalty $p_N$. 
Based on these considerations, we recommend to use a (system-specifically selected) small value for~$r$ which ensures that most stations have one or more neighboring stations, to enjoy the benefit of penalization while facilitating the computation.
For example, $r = 750$m for Seoul and NYC, and $r = 1500$m for SF appear sufficient in keep the percentage of unconnected stations below $1\%$.


\subsection{Analysis of bike usage in Seoul}
\label{sec:res:seoul}


Focusing on the bike usage data collected in Seoul, we first analyse the patterns of bike usage captured by the proposed fusion penalty and discuss the latter's performance. 
Results from the two other cities can be found in Appendix~\ref{sec:empiric}.

\subsubsection{Effects of penalization}
\label{sec:real:penalty}

To assess the effect of penalization, we compare our multilayer network fused Lasso-based penalized regression approach (referred to as ``fused Lasso'') with the method that adopts the Lasso penalty only, which is referred to as ``Lasso-only''.
We also consider models~\eqref{eq:model:no} and~\eqref{eq:model} fitted without any penalization, referred to as ``no-interaction'' and ``full-interaction'', respectively.

\begin{figure}[h!t!]
\centering
\includegraphics[width = 0.9\textwidth]{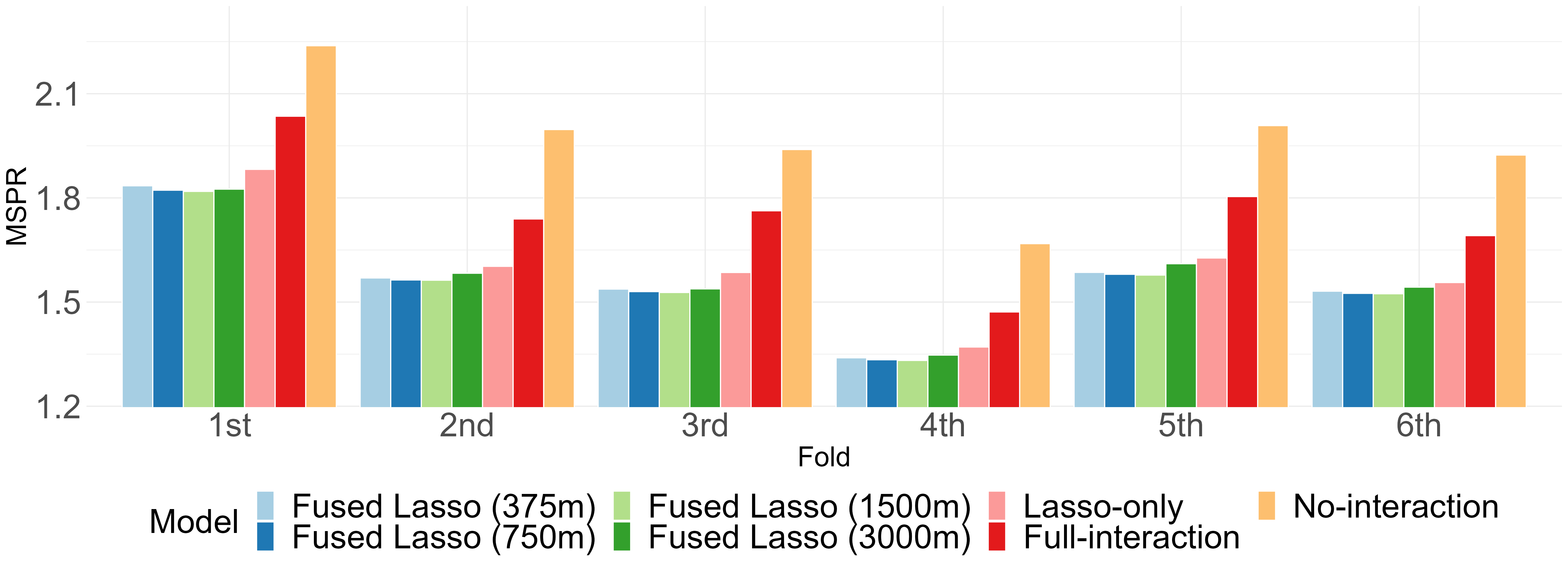}
\caption{Seoul: MSPRs from each fold used in the $6$-fold CV.
}
\label{fig:mspr}
\end{figure}

\vspace{-10pt}
\paragraph{Performance evaluation.}
We adopt the CV measures as an indicator of the out-of-sample predictive performance (see Section~\ref{sec:forecasting} for full prediction exercise results). 
Figure~\ref{fig:mspr} reports the fold-wise MSPRs involved in the $6$-fold CV in~\eqref{eq:cv}, all evaluated at the penalty parameters selected to minimize the overall CV error for the respective methods.

When comparing no- and full-interaction methods,
the flexibility afforded by allowing for station-specific temporal effects proves useful in enhancing the predictive performance as the latter model consistently attains a considerably smaller MSPR.
We observe further improvement when appropriate penalization is applied to the interaction parameters.
In particular, adopting the proposed fused Lasso penalty outperforms other penalties uniformly across the $6$ folds regardless of the choice of $r$. 
This demonstrates that capturing spatial and temporal homogeneity pays off by accounting for the stylized features of the BSS data, that neighboring stations exhibit similar usage patterns, see~\ref{c:net}.

\begin{table}[h!t!]
\caption{\small
Seoul: Model complexity (MC) out of all parameters ($\Delta \cup \Theta \cup \Theta_H  \cup \Theta_D$), as well as that out of the interaction parameters representing hour-of-a-day ($\phi^{\hod}_{s,h}$) and day-of-a-week ($\phi^{\dow}_{s,d}$) effects. Additionally, the proportion of non-zero coefficient estimates is reported.}
\label{tab:mc}
\centering
{\small 
\begin{tabular}{l cccc}
\toprule
& \multicolumn{3}{c}{Parameter set} & \multirow{2}{*}{\shortstack[c]{Proportion of \\ non-zeros}} \\
\cmidrule(lr){3-4} 
Method & All & $\Theta_H \setminus \{\theta^{\hod}_h, \, h \in \mc H\}$ & $\Theta_D \setminus \{\theta^{\dow}_d, \, d \in \mc D\}$ & \\
\cmidrule(lr){1-1} \cmidrule(lr){2-2} \cmidrule(lr){3-3} \cmidrule(lr){4-4} \cmidrule(lr){5-5}
Fused Lasso (375m) & 0.952 & 0.940 & 0.994 & 0.568\\
Fused Lasso (750m) & 0.935 & 0.923 & 0.979 & 0.605 \\
Fused Lasso (1500m) & 0.893 & 0.882 & 0.933 & 0.635 \\
Fused Lasso (3000m) & 0.736 & 0.725 & 0.774 & 0.756 \\
\cmidrule(lr){1-1} \cmidrule(lr){2-2} \cmidrule(lr){3-3} \cmidrule(lr){4-4} \cmidrule(lr){5-5}
Lasso-only & 
-- & -- & -- & 0.583 \\
\bottomrule
\end{tabular}}
\end{table}

\vspace{-10pt}
\paragraph{Sparsity and model complexity.}
We examine the reduction in complexity brought by fused Lasso penalty
in~\eqref{eq:org_obj_admm}.
The sparsity induced by the Lasso penalty is easily measured by the proportion of non-zero coefficient estimates (see the last column of Table~\ref{tab:mc}). We observe that the fused Lasso method returns a fitted model which sets approximately $30\%$ of the coefficients to be zero when $r = 750$, and this tendency is stronger with a smaller value of $r$.
To evaluate the effect of the fusion penalties, define a model complexity measure
\begin{align}
& \text{MC}(r, \lambda, \lambda_N, \lambda_H) 
=  \frac{1}{p}
\l( 34 + \mc C(\mathfrak{N}_{\text{multi}}(r) \cap \wh{\mathfrak{N}}_H) + \sum_{d \in \mc D  \setminus \{\text{Mo} \}} \mc C(\mathfrak{N}_{\text{single}}(r) \cap \wh{\mathfrak{N}}_{D, d}) \r),
\label{eq:mc}
\end{align}
utilizing the multilayer networks introduced in Section~\ref{sec:network};
we suppress the dependence on $(r, \lambda, \lambda_N, \lambda_H)$ for simplicity.
Here, $\mc C(\mathfrak{N})$ denotes the number of connected components in a network $\mathfrak{N}$, and $p$ the number of total parameters under~\eqref{eq:model} that increases linearly with the number of stations $S$.
Simply put, MC represents the proportion of coefficient estimates that are not fused by the penalties $p_N$ and~$p_H$. 
Ranging between~$0$ and~$1$, when MC is closer to~$0$, it implies that most station coefficients are heavily fused with those of the neighboring stations. 
The number of unique parameter estimates of $\phi^{\hod}_{s,h}$, which are not fused by penalization,
is given by $\mc C(\mathfrak{N}_{\text{multi}}(r) \cap \wh{\mathfrak{N}}_H)$.
As an illustration, Figure~\ref{fig:network} shows that the network $\mathfrak{N}_{\text{multi}} \cap \wh{\mathfrak{N}}_H$ contains four connected components; consequently,
$\mc C(\mathfrak{N}_{\text{multi}} \cap \wh{\mathfrak{N}}_H) = 4$. Similarly, we find the number of unique parameter estimates for $\phi^{\dow}_{s,d}$. 
\footnote{ We exclude the intersection network $\mathfrak{N}_{\text{single}}(r) \cap \wh{\mathfrak{N}}_{D, \text{Mo}}$ in~\eqref{eq:mc} since, due to model identifiability constraints, we have $\phi^{\dow}_{s,\text{Mo}} = \phi^{\hod}_{s, 0} = \theta_{s}$.
That is, the fusion among the station-specific intercept parameters~$\theta_s$ has already been accounted for by $\mathfrak{N}_{\text{multi}}(r) \cap \wh{\mathfrak{N}}_H$ at layer $h = 0$.}
Appendix~\ref{sec:mc} gives an algorithm for efficient computation of MC.

Table~\ref{tab:mc} displays that the resultant intersection networks 
$\mathfrak{N}_{\text{multi}}(r) \cap \wh{\mathfrak{N}}_H$ and
$\mathfrak{N}_{\text{single}}(r) \cap \wh{\mathfrak{N}}_{D, d}$
are highly heterogeneous, as evidenced by the MC being closer to one, i.e.\ many stations exhibit individual behavior. 
This is despite the fact that, e.g.\ $\mathfrak{N}_{\text{single}}(r)$ has $2$ (resp.\ $1$) connected components with $r = 1500$ (resp.\ $r = 3000$), see Table~\ref{tab:radius}.
In other words, station-specific parameters account for a large portion of the variation in bike usage, which supports modeling the data at the individual station level.
The high degree of heterogeneity across the BSS can be attributed to the fact that 
each station is associated with multiple aspects of usages which is natural in a large metropolitan city.
This indicates that partitioning stations into a handful of clusters may ignore the complex nodal features that drive the usage of bikes at each station.


\subsubsection{Parameter estimates}
\label{sec:coef}

While some inferential methods exist for high-dimensional generalized linear models \citep{belloni2016post}, they do not easily apply to our setting because of the presence of a fusion penalty.
Instead, we examine the estimates capturing the overall trend and the covariate effects obtained from the 6 folds used in the CV step, along with those obtained from the full training data.
We focus on the results obtained with $r = 1500$ in the main text and report the rest of the results in Appendix~\ref{sec:radius} where analogous conclusions are drawn. 

Table~\ref{tab:coef} shows that, while the values of the estimates vary slightly from one fold to another, their signs and overall magnitude do not change, confirming that as noted in~\ref{c:trend} and~\ref{c:cov}, the variables have meaningful effects on overall bike usage across the system.
Also, Table~\ref{tab:coef} demonstrates the invariance of parameter estimates with respect to the choice of penalization, which indicates that the superior performance of the proposed fused Lasso method over alternative approaches, is attributed to how well the station-specific parameters (i.e.\ $\theta^{\hod}_{s, h}$ and $\theta^{\dow}_{s, d}$) are estimated, upon which the penalization is imposed.

\begin{table}[htb]
\centering
\caption{Seoul: Estimated coefficients for the trend and the covariate effects by the proposed fused Lasso regression method from each fold used in the $6$-fold CV and from the full training data when $r = 1500$.
For comparison, we also report the estimates obtained with  Lasso-only, full-interaction and no-interaction methods.}
\label{tab:coef}
\resizebox{0.9\columnwidth}{!}
{\small 
\begin{tabular}{c cccccc c c c c }
\toprule
& \multicolumn{6}{c}{Fold} & \multicolumn{4}{c}{Full data}  \\
\cmidrule(lr){2-7} \cmidrule(lr){8-11}
& 1 & 2 & 3 & 4 & 5 & 6 & Fused  & Lasso-only & Full & No \\
\cmidrule(lr){1-1} \cmidrule(lr){2-7} \cmidrule(lr){8-8} \cmidrule(lr){9-9} \cmidrule(lr){10-10} \cmidrule(lr){11-11}
$\alpha$ & 0.065 & 0.063 & 0.061 & 0.059 & 0.06 & 0.062  & 0.062 & 0.062 & 0.062 & 0.062 \\
$\beta^{\rain}$ &  -2.180 & -2.359 & -2.501 & -2.298 & -2.221 & -2.454 & -2.343  & -2.343 & -2.344 & -2.332 \\
$\beta^{\air}_1$ & 0.115 & 0.105 & 0.064 & 0.072 & 0.135 & 0.138 & 0.103  & 0.102 & 0.103 & 0.100 \\
$\beta^{\air}_2$ &  0.135 & 0.132 & 0.068 & 0.088 & 0.130 & 0.162 & 0.116 & 0.115 & 0.113 & 0.118 \\
$\beta^{\air}_3$ & 0.256 & 0.272 & 0.036 & 0.220 & 0.273 & 0.262 & 0.227   & 0.226 & 0.228 & 0.209 \\
\bottomrule
\end{tabular}}
\end{table}

\begin{figure}[htb]
\centering
\subfloat{\includegraphics[width = .4\textwidth]{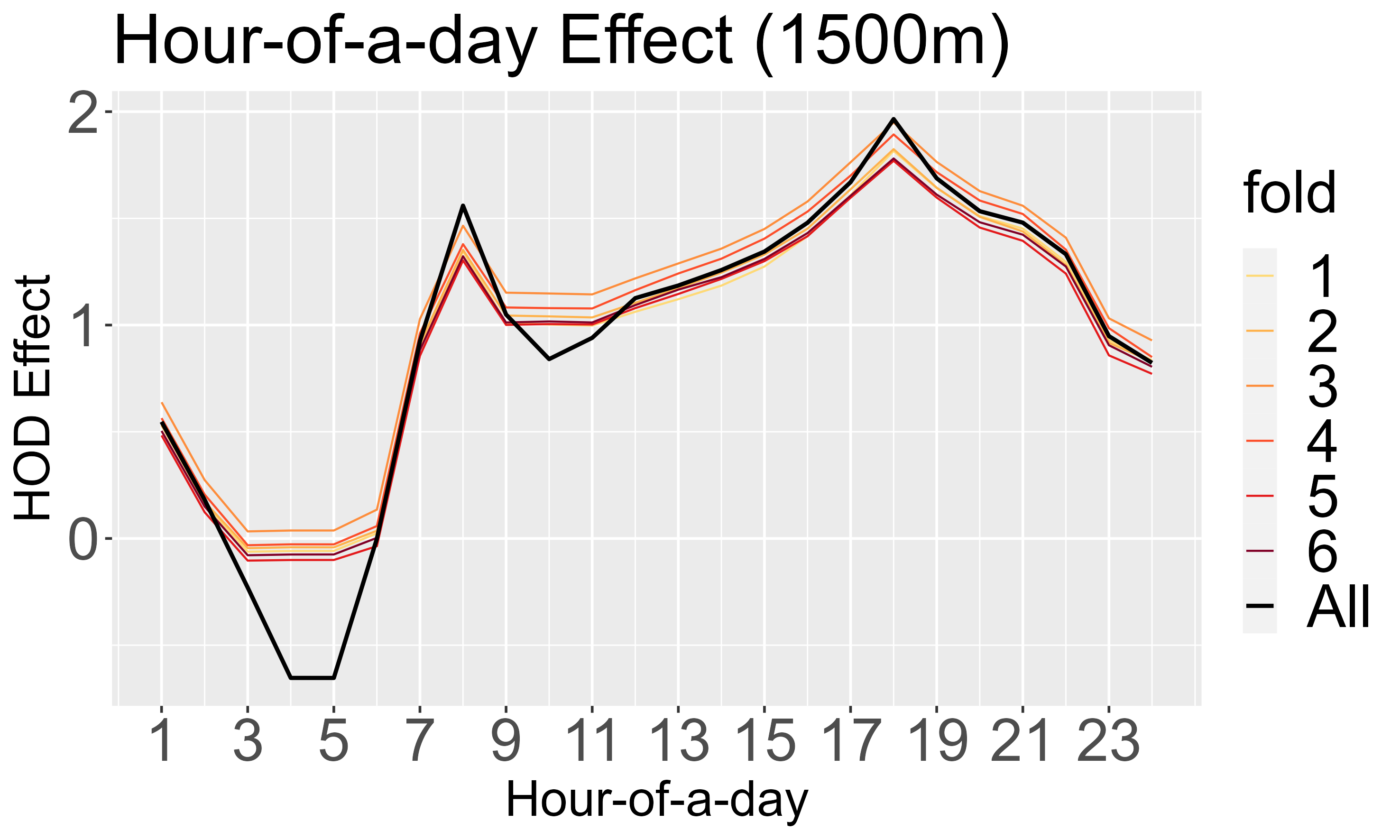}}
\subfloat{\includegraphics[width = .4\textwidth]{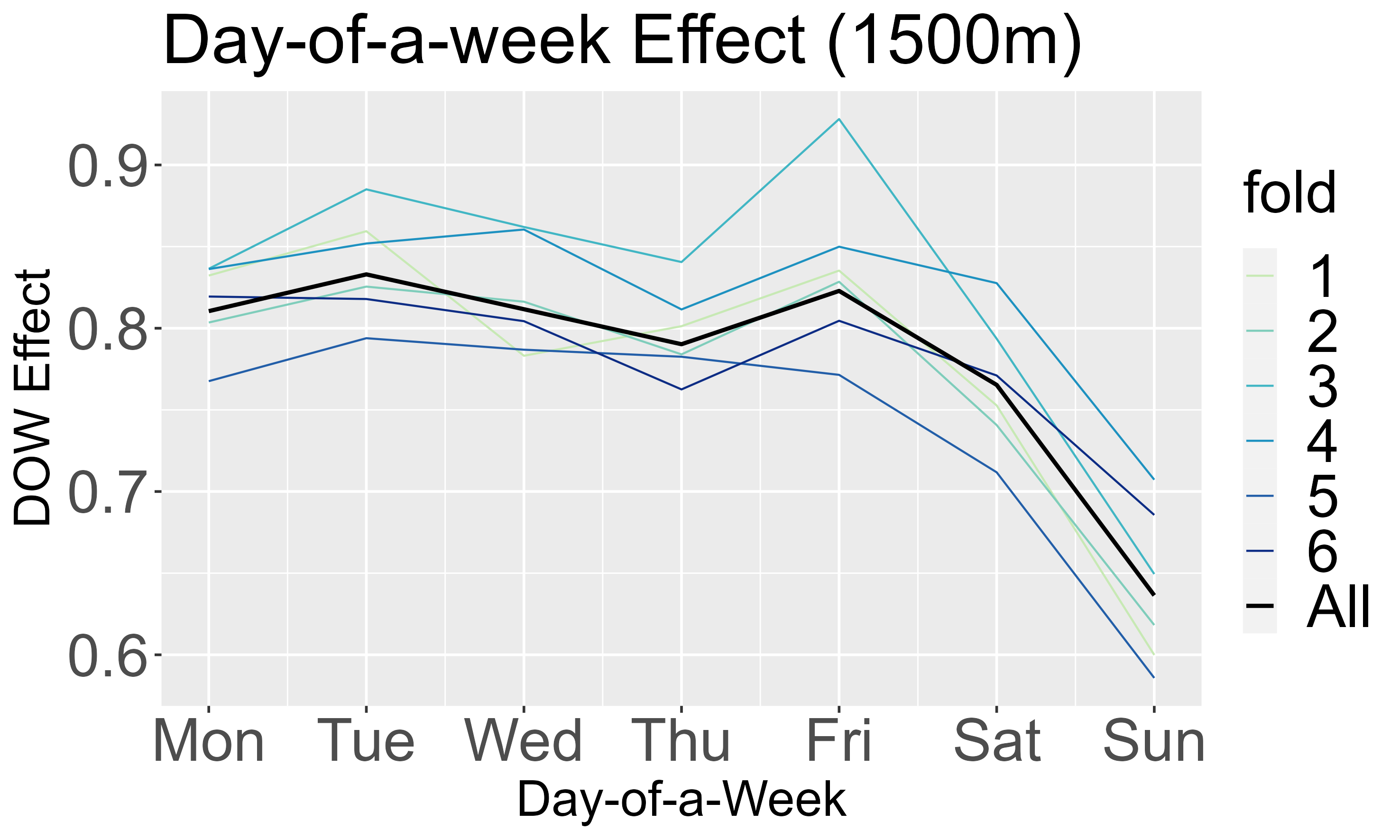}}
\caption{Seoul: Parameter estimates for $\theta^{\hod}_h, \, h \in \mc H$ (left) and $\theta^{\dow}_d, \, d \in \mc D$ (right) from each fold used in the $6$-fold CV and from the full data when $r = 1500$.}
\label{fig:thetas}
\end{figure}

\begin{figure}[htb]
\centering
\subfloat{\resizebox*{4.5cm}{!}{\includegraphics{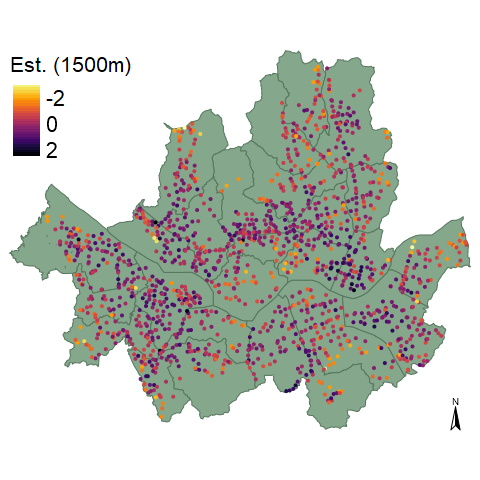}}}
\subfloat{\resizebox*{4.5cm}{!}{\includegraphics{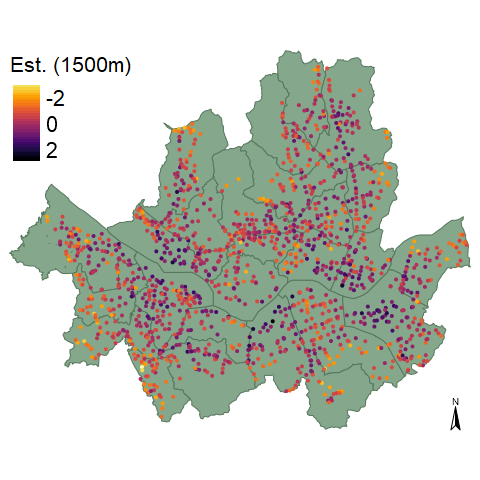}}}
\subfloat{\resizebox*{4.5cm}{!}{\includegraphics{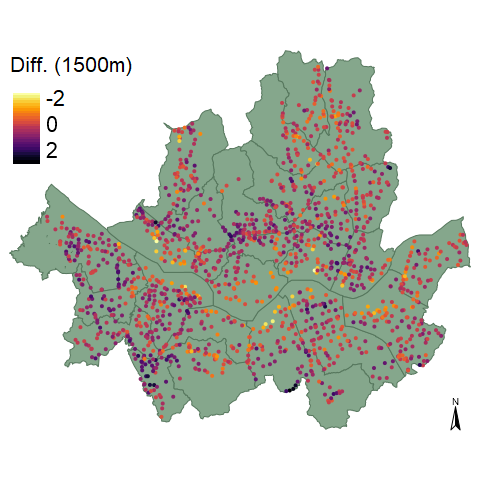}}}
\caption{Seoul: Estimated station-specific bike demands in log-scale (given by $\wh\theta_s + \wh\theta^{\hod}_h + \wh\theta^{\dow}_d + \wh\theta^{\hod}_{s, h} + \wh\theta^{\dow}_{s, d}$) from the model fitted with $r = 1500$ at 8am on Tuesdays (left), at 8pm on Sundays (middle) and their differences (right).} 
\label{fig:fitted}
\end{figure}

Figure~\ref{fig:thetas} plots the estimates of $\theta^{\hod}_h$ and $\theta^{\dow}_d$ which are shared by all the stations.
We observe that the smooth transition over the course of a day and a week is well captured across the $6$ folds, along with the peaks corresponding to the high demand by commuters.
In addition, we plot the combined effects of temporal variables on the mean bike demand (in log-sale), namely $\wh\theta_s + \wh\theta^{\hod}_h + \wh\theta^{\dow}_d + \wh\theta^{\hod}_{s, h} + \wh\theta^{\dow}_{s, d}$ under~\eqref{eq:model}, for all stations $s \in \mc S$ when $(h, d) = (8, \text{Tu})$ (8am on Tuesday) and $(20, \text{Su})$ (8pm on Sunday), see Figure~\ref{fig:fitted}.
As expected, the spatial distribution of bike usage concentration is markedly different when $(h, d) = (8, \text{Tu})$ (when bikes are primarily used for commuting) and $(h, d) = (20, \text{Su})$ (bikes tend to be used for leisure activities).

\subsection{Predictive performance}
\label{sec:forecasting}

In this section, we perform a prediction exercise on the three datasets collected in Seoul, New York and San Francisco.
For each dataset, we produce the predictions of the hourly bike usage of the entire BSS on the last $7$ days which is set aside as the test data (denoted by $\mc T_{\text{test}})$, based on the model trained on the remaining data.
Denoting by $Y_t(s)$ the usage at a time point $t \in \mc T_{\text{test}}$ and $\wh{\mu}_t(s)$ its prediction, we measure the prediction error (PE) by
\begin{align}
\label{eq:mspr:pe}
\text{PE} = \frac{1}{S \vert \mc T_{\text{test}} \vert} \sum_{s = 1}^S \sum_{t \in \mc T_{\text{test}}} \frac{ (Y_t(s) - \wh{\mu}_t(s))^2}{\max(\wh{\mu}_t(s), \epsilon)}.
\end{align}
The small constant~$\epsilon = 0.01$ is introduced to the denominator to avoid inflation of PE when $\wh\mu_t(s) = 0$ while $Y_t(s) > 0$; the proposed fused Lasso does not get affected by the introduction of $\epsilon$ as it rarely returns near-zero predictions when $Y_t(s) > 0$.
For the complete results of the prediction exercise including the errors via alternative measures, see Appendix~\ref{sec:add:pred}.

In addition to the penalized regression methods considered in the previous section, we include competitors that are trained to model bike usage data collected from each station individually, which are: (i)~Integer-valued generalized autoregressive conditional heteroscedasticity model proposed for univariate count time series (INGARCH, \citealp{liboschik2017}), (ii) Random Forest (RF; Chapter~15 of \citealp{hastie2009elements}), and (iii) XGBoost \citep{chen2016xgboost}, see Appendix~\ref{sec:ingarch} for full information on their implementations.
All methods are supplied with the same information on the covariate and temporal effects in~\eqref{eq:model}.
In principle, RF and XGBoost can be applied to jointly model the entire system but due to the high dimensionality relative to the training sample size, they encounter computational issues; for this reason, we choose to apply them to individual stations.

The dataset from New York contains some irregularities as some stations were introduced during the period in consideration such that the training data mostly contain zeros, see Appendix~\ref{sec:irregular}. 
These irregularities pose a convergence issue for INGARCH and to remedy this, we remove the top $0.5\%$ largest errors when reporting the average PE from INGARCH on the data.
On the other hand, the proposed fusion penalty effectively handles such anomalous observations through pooling information from the neighbors.

\begin{figure}[h!t!]
\centering
\begin{tabular}{c}
\includegraphics[width = 0.8\textwidth]{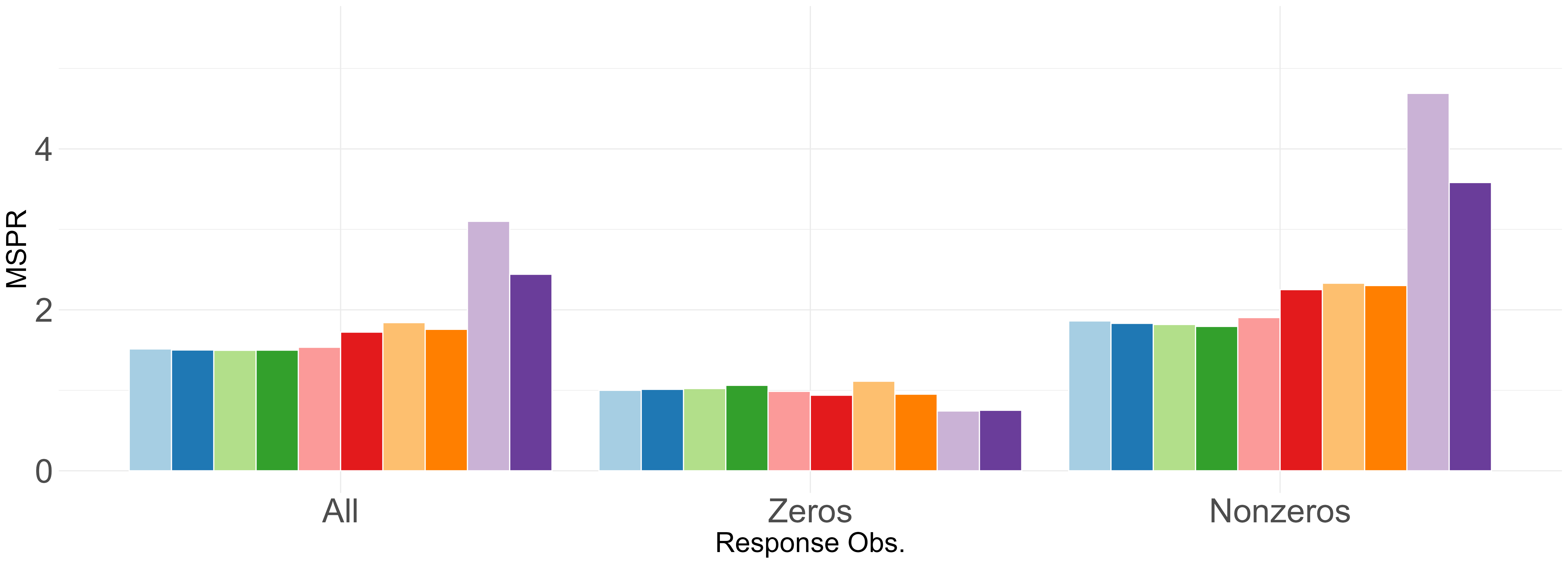} \\
\includegraphics[width = 0.8\textwidth]{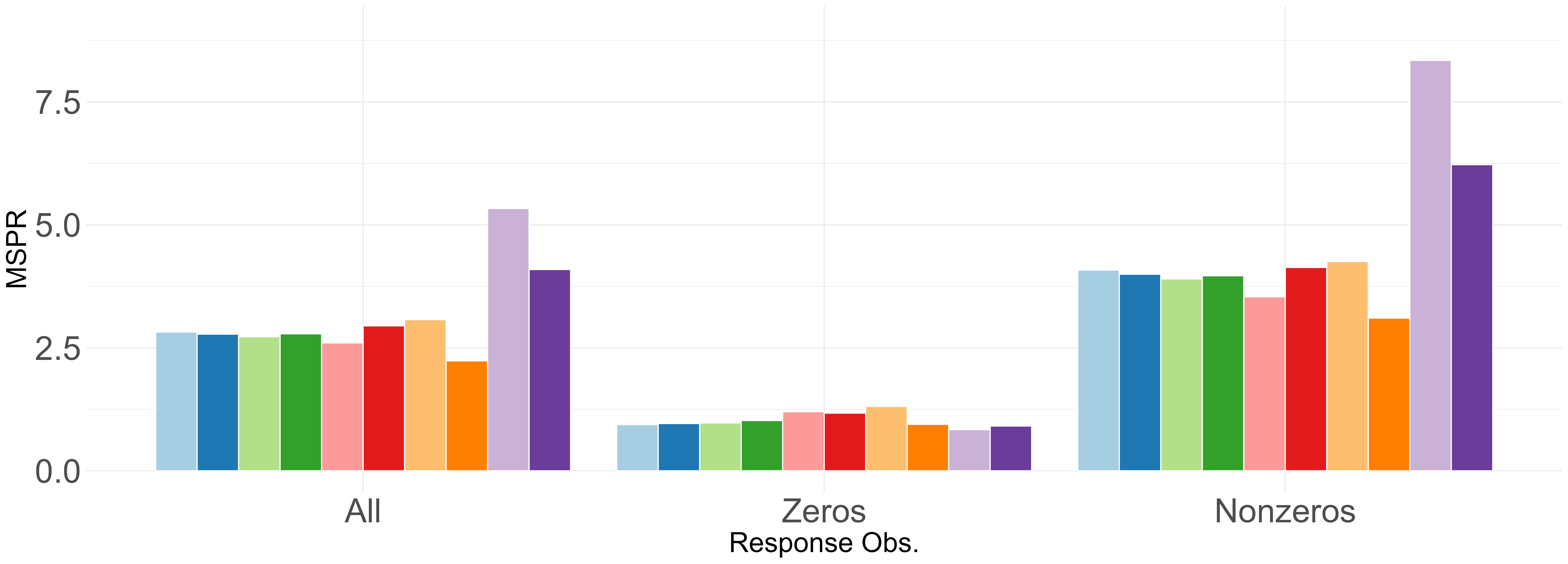} \\
\includegraphics[width = 0.8\textwidth]{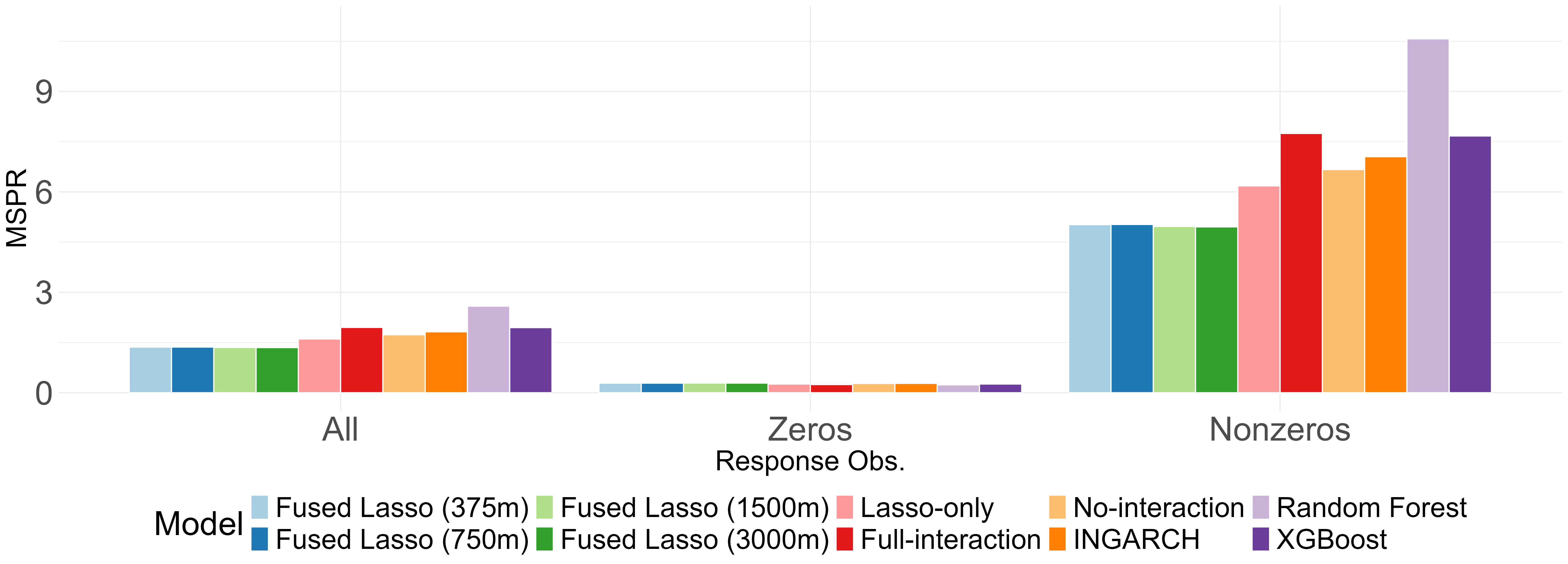}
\end{tabular}
\caption{
Modified MSPRs on the test data (PE) from Seoul (top), New York (middle) and San Francisco (bottom). The left column displays the PE computed from the entire test set $\mc T_{\text{test}}$, and the middle and right columns display the PE computed with the subsets of $\mc T_{\text{test}}$ where $Y_t(s) = 0$ and $Y_t(s) > 0$, respectively.}
\label{fig:mspr_per_cases}
\end{figure}

Figure~\ref{fig:mspr_per_cases} shows that overall, the proposed fused Lasso method performs competitively regardless of the choice of $r$.
Competitors exhibit slightly better prediction accuracy for the entries with $Y_t(s) = 0$, particularly on the data from Seoul and San Francisco, but their performance deteriorate for the situations with $Y_t(s) > 0$, which results in the large overall PE. 
Our modeling approach does not address possible zero-inflation in the data yet it produces predictions with good accuracy when the demand is large. 
We attribute its overall good performance to that, by modeling the entire BSS simultaneously via the proposed fusion penalty, it is able to better capture the stylized features of the bike usage patterns; this in turn leads to better predictive performance across multiple datasets. 

\section{Conclusions}
\label{sec:conc}

In this study, we address the problem of modeling bike usage patterns for the entire BSS spanning a large metropolitan city.
We model the data at the granularity of individual stations by incorporating covariate effects as well as spatial and temporal characteristics commonly observed in bike usage data.
The proposed multilayer fused Lasso penalty is imposed on the networks encoding the geographical proximity of the stations over hourly layers, and successfully captures the spatial and temporal homogeneity.
Combined with the data-driven choice of penalty parameters, our penalized regression approach strikes a good balance between a simplistic model that does not allow for station-specific behavior, and a complex model possibly suffering from over-parameterization, and its good predictive performance is demonstrated on datasets collected in Seoul, New York and San Francisco.

We envision that the proposed method is applicable to different datasets with network-like features, such as those collected from large transportation, communication or logistic systems.
In particular, when information about the factors driving link homophily (such as the nodal features related to land use, slope of terrain, nearby landmarks, and other modes of transportation in the case of BSSs) is not readily available,
our penalized regression approach enables learning of the linkages in an unobservable network from the fusion of parameters induced by the penalties defined on an observable network. 

\bibliographystyle{apalike}
\bibliography{fbib}

\clearpage

\appendix
\numberwithin{equation}{section}
\numberwithin{figure}{section}
\numberwithin{table}{section}

\section{Computational considerations}
\label{sec:comp}

\subsection{ADMM algorithm for penalized MLE}
\label{sec:admm}

\subsubsection{ADMM framework}

We adopt the alternating direction method of multipliers (ADMM) algorithm for fitting~\eqref{eq:org_obj_admm} based on \cite{wahlberg2012admm},
where the optimization of objective functions with fusion penalties is discussed. 
We recall that the fused Lasso penalties lead to the objective function of the form
\begin{footnotesize}
\begin{align}
\min_{\Theta, \Theta_H, \Theta_D, \Delta} & \sum_{i=1}^{n} \mu_i(\Theta, \Theta_H, \Theta_D, \Delta) - \sum_{i=1}^{n} y_i \log(\mu_i(\Theta, \Theta_H, \Theta_D, \Delta)) \nn \\
&+ \lambda 
\sum_{s \in \mc S} \l( \sum_{h \in \mc H} \vert \theta^{\hod}_{s, h} \vert
+ \sum_{d \in \mc D} \vert \theta^{\dow}_{s, d} \vert \r)
\nn \\
&+ \lambda_N \sum_{s \in \mc S} \sqrt{ \vert \mc N_r(s) \vert \, \sum_{s^\prime \in \mc N_r(s)} \l[ 2(\theta_s - \theta_{s^\prime})^2 + \sum_{h = 1}^{23} (\phi^{\hod}_{s, h} - \phi^{\hod}_{s', h})^2 +
\sum_{d \in \mc D \setminus \{\text{Mo}\}} (\phi^{\dow}_{s, d} - \phi^{\dow}_{s^\prime, d})^2 \r]}  \nn \\
&+ \lambda_H \sum_{s \in \mc S} 
\sum_{h = 0}^{23} \l\vert \phi^{\hod}_{s, h} - \phi^{\hod}_{s, h + 1} \r\vert.
\label{admmeq:org_obj_admm}
\end{align}
\end{footnotesize}
In this section, we treat $\Theta$, $\Theta_{H}$ and $\Theta_{D}$ as row-vectors without confusion:
\begin{align*}
    \Theta =& (\theta_s, \, s \in {\mc S}), \\
    \Theta_H =& \left((\theta_{s,1}^{\hod}, \, s\in {\mc S}), \ldots, (\theta_{s, 23}^{\hod}, \, s\in {\mc S}), \, (\theta_1^{\hod}, \ldots, \theta_{23}^{\hod}) \right), \\
    \Theta_D =& \left((\theta_{s, \text{Mo}}^{\dow}, \, s\in {\mc S}), \ldots, (\theta_{s, \text{Su}}^{\dow}, \, s\in {\mc S}), \, (\theta_{\text{Mo}}^{\dow}, \ldots, \theta_{\text{Su}}^{\dow}) \right).
\end{align*}
By re-parametrizing the fused lasso penalty terms, we re-write~\eqref{admmeq:org_obj_admm} as,
\begin{align}
\min_{\Theta, \Theta_H, \Theta_D, \Delta, \Gamma, \Psi} & \sum_{i=1}^{n} \mu_i(\Theta, \Theta_H, \Theta_D, \Delta) - \sum_{i=1}^{n} y_i \log(\mu_i(\Theta, \Theta_H, \Theta_D, \Delta)) \nn \\
&+ \lambda 
\sum_{s \in \mc S} \l( \sum_{h \in \mc H} \vert \theta^{\hod}_{s, h} \vert
+ \sum_{d \in \mc D} \vert \theta^{\dow}_{s, d} \vert \r)
\nn \\
&+ \lambda_N \cdot \widetilde{P}_{N} (\Gamma) + \widetilde{P}_H(\Psi) 
+ \mbf I_{\mathcal{C}}\left( \Theta, \Theta_{H}, \Theta_{D}, \Gamma, \Psi \right).
\label{admm:obj2} 
\end{align}
Here, the penalty functions are defines as
\begin{align*}
&\widetilde{P}_N(\Gamma) =
 \sum_{s \in \mc S} \sqrt{ \vert \mc N_r(s) \vert \, \sum_{s^\prime \in \mc N_r(s)} \l[ 
\left(\gamma_{s, s'}\right)^2 + \sum_{h=1}^{23} \left(\gamma_{s, s',h}^{\hod}\right)^2 + \sum_{d \in \mc D \setminus \{\text{Mo}\}} \left(\gamma_{s, s',d}^{\dow}\right)^2  
\r]}, 
\\
&\text{where \ }
\Gamma = (\bm\gamma_s, \, s \in \mc S) \text{ \ with \ }
\bm\gamma_s = \l( \gamma_{s, s'}, (\gamma^{\hod}_{s, s', h}, \, h \in \mc H \setminus \{0\}),
(\gamma^{\dow}_{s, s', d}, \, d \in \mc D \setminus \{\text{Mo}\})), \, s' \in \mc N_r(s) \r),
\end{align*}
and
\begin{align*}
\widetilde{P}_H(\Psi) &= \sum_{s \in \mc S}\sum_{h \in \{0, \ldots, 23 \}} \vert \psi_{s,h} \vert
\quad \text{with} \quad 
\Psi = \l( \bm\psi_s = (\psi_{s, 0}, \ldots, \psi_{s, 23}), \, s \in \mc S \r).
\end{align*}
In addition, $I_{\mathcal{C}}$ denotes an indicator function on the constraint set $\mathcal{C}$ which forces the newly defined penalty functions $\widetilde{P}_{N}(\Gamma)$ (resp.\  $\widetilde{P}_H(\Psi)$) to be the same as ${P}_N(\Theta, \Theta_{H}, \Theta_{D})$ (resp.\ $P_H(\Theta, \Theta_{H}, \Theta_{D})$) such that
\begin{align*}
I_\mathcal{C}(\mbf z) =
\begin{cases}
 0 & \text{ if  } \mbf z \in \mathcal{C} \\
 \infty & \text{ otherwise}.
\end{cases}
\end{align*}

The constraint set $\mathcal{C}$ is a subset of $\vert (\Theta, \Theta_H, \Theta_D, \Gamma, \Psi) \vert$-dimensional space whose elements fulfil
\begin{align}
 \gamma_{s, s'} &= \sqrt{2} (\theta_{s} - \theta_{s'}),  \nn \\
 \gamma_{s, s',h}^{\hod} &= \theta_{s} - \theta_{s'} + \theta_{s,h}^{\hod} - \theta_{s', h}^{\hod} \: \text{ for } h \in \mathcal{H}\setminus \{0\}, \nn  \\
 \gamma_{s, s',d}^{\dow} & =  \theta_{s} - \theta_{s'} + \theta_{s,d}^{\dow} - \theta_{s', d}^{\dow} \: \text{ for } d \in \mathcal{D}\setminus \{\text{Mo}\}
 \label{admm:para1}
\end{align}
for all $s, s' \in \mc S$, and
\begin{align}
\label{admm:para2}
  \psi_{s,h} &= \theta_{h+1}^{\hod}-\theta_{h}^{\hod} + \theta_{s,h+1}^{\hod} - \theta_{s,h}^{\hod} \: \text{ for } h \in \{1, \ldots 22\}, \nn \\
  \psi_{s, 0} &= \theta_{1}^{\hod} + \theta_{s, 1}^{\hod} 
  \quad \text{and} \quad 
  \psi_{s, 23} = -\theta_{23}^{\hod} -\theta_{s, 23}^{\hod},
\end{align}
for all $s \in \mc S$,
where~\eqref{admm:para1} and \eqref{admm:para2} account for that the baseline parameters are set to be zero. 

To utilize ADMM, we rewrite the objective function \eqref{admm:obj2} as follows:
\begin{align}
\label{admm:obj3} 
\min_{ \substack {\Theta, \Theta_H, \Theta_D, \Delta, \Gamma, \Psi, \\
Z_{\Theta}, Z_{\Theta_{H}}, Z_{\Theta_{D}}, S_{\Gamma}, S_{\Psi}
}} & \sum_{i=1}^{n} \mu_i(\Theta, \Theta_H, \Theta_D, \Delta) - \sum_{i=1}^{n} y_i \log(\mu_i((\Theta, \Theta_H, \Theta_D, \Delta))) \nn \\
&+ \lambda 
\sum_{s \in \mc S} \l( \sum_{h \in \mc H} \vert \theta^{\hod}_{s, h} \vert
+ \sum_{d \in \mc D} \vert \theta^{\dow}_{s, d} \vert \r)
\nn \\
&+ \lambda_N \cdot \widetilde{P}_{N} (\Gamma) + \widetilde{P}_H(\Psi) + \mbf I_{\mathcal{C}}\left( Z_{\Theta}, Z_{\Theta_{H}}, Z_{\Theta_{D}}, S_{\Gamma}, S_{\Psi} \right) 
\\
\text{subject to} &  \quad (\Theta, \Theta_{H}, \Theta_{D}) = (Z_{\Theta}, Z_{\Theta_H}, Z_{\Theta_D}) 
\text{ and } (\Gamma, \Psi) = (S_{\Gamma}, S_{\Psi}). \nn
\end{align}
The ADMM optimizes~\eqref{admm:obj3} in three steps as follows:
\begin{enumerate}[wide, labelwidth=!, labelindent=0pt, label = {\bf Step~\arabic*:}]
\item  Update the primal variables as
\begin{small}
\begin{align}
    (\Theta^{k+1}, \Theta_{H}^{k+1}, \Theta_{D}^{k+1}, \Delta^{k+1})  =& \argmin_{\Theta, \Theta_H, \Theta_D, \Delta} 
    \sum_{i = 1}^{n} \mu_i(\Theta, \Theta_H, \Theta_D, \Delta) - \sum_{i=1}^{n} y_i \log(\mu_i((\Theta, \Theta_H, \Theta_D, \Delta))) \nn \\
&+ \lambda 
\sum_{s \in \mc S} \l( \sum_{h \in \mc H} \vert \theta^{\hod}_{s, h} \vert
+ \sum_{d \in \mc D} \vert \theta^{\dow}_{s, d} \vert \r)
\nn \\
&+ \frac{\rho}{2} \l\Vert   (\Theta, \Theta_{H}, \Theta_{D}) -   (Z_{\Theta}^k, Z_{\Theta_{H}}^k, Z_{\Theta_{D}}^k) +  (U_{\Theta}^k, U_{\Theta_{H}}^k, U_{\Theta_{D}}^k) \r\Vert^2, 
\label{admm:step:one:one} \\
\Gamma^{k+1} =& \, \argmin_{\Gamma} \lambda_N \widetilde{P}_N(\Gamma) + \frac{\rho}{2} \l\Vert \Gamma - S_{\Gamma}^{k} + T_{\Gamma}^{k} \r\Vert^{2},  
\label{admm:step:one:two} \\
\Psi^{k+1} =& \, \argmin_{\Psi} \lambda_H \widetilde{P}_N(\Psi) + \frac{\rho}{2} \l\Vert \Psi - S_{\Psi}^{k} + T_{\Psi}^{k} \r\Vert^{2}, 
\label{admm:step:one:three} 
\end{align}
\end{small} 

where $(U_{\Theta}, U_{\Theta_H}, U_{\Theta_D})$, $T_{\Gamma}$, and $T_{\Psi}$ are dual variables associated with the constraints $(\Theta, \Theta_H, \Theta_D)=(Z_{\Theta}, Z_{\Theta_H}, Z_{\Theta_D})$, $\Gamma = S_{\Gamma}$ and $\Psi = S_{\Psi}$, respectively.

\item Update $( Z_{\Theta}, Z_{\Theta_{H}}, Z_{\Theta_{D}}, S_{\Gamma}, S_{\Psi})$ by projecting $(\Theta^{k+1} + U_{\Theta}^{k}, \Theta_H^{k+1} + U_{\Theta_H}^{k}, \Theta_{D}^{k+1} + U_{\Theta_D}^{k}, \Gamma^{k+1} + T_{\Gamma}^{k}, \Psi^{k+1} + T_{\Psi}^{k})$ onto the constraint set $\mathcal{C}$, as
\begin{small}
\begin{align}
       & \left( Z_{\Theta}^{k+1}, Z_{\Theta_{H}}^{k+1}, Z_{\Theta_{D}}^{k+1}, S_{\Gamma}^{k+1}, S_{\Psi}^{k+1} \right) 
       \nn \\
       =&  \prod_{\mathcal{C}} (\Theta^{k+1}+U_{\Theta}^{k}, \Theta_H^{k+1}+U_{\Theta_H}^{k}, \Theta_{D}^{k+1}+U_{\Theta_D}^{k}, \Gamma^{k+1}+T_{\Gamma}^{k}, \Psi^{k+1}+T_{\Psi}^{k}),
       \label{admm:step:two}
\end{align}
\end{small}
with $\prod_{\mc C}$ denoting the projection operator.

\item Update the dual variable as
    \begin{small}
    \begin{align*}
        (U_{\Theta}^{k+1},\; U_{\Theta_H}^{k+1}, \; U_{\Theta_D}^{k+1}) =& \, (U_{\Theta}^k, U_{\Theta_H}^k, U_{\Theta_D}^k) + (\Theta^{k + 1}, \Theta_H^{k + 1}, \Theta_D^{k + 1}) - 
        (Z_{\Theta}^{k+1}, Z_{\Theta_H}^{k+1}, Z_{\Theta_D}^{k+1}), \\
        (T_{\Gamma}^{k+1}, T_{\Psi}^{k+1}) =& \,  (T_{\Gamma}^k, T_{\Psi}^k) +  (\Gamma^{k+1}, \Psi^{k+1})
        -  (S_\Gamma^{k+1}, S_\Psi^{k+1}).
    \end{align*}
        \end{small}
\end{enumerate}

While Step~3 is straightforward, Steps~1 and~2 involve relatively heavy computations. The detail of Step~1 and Step~2 are discussed in the following subsections.

\subsubsection{Computational details of Step~1}
\label{subsec:glmnet}

Let ${\mc P} = (\Theta, \Theta_{H}, \Theta_{D})$.

\begin{enumerate}[wide, labelwidth=!, labelindent=0pt, label = {\bf Step~1.\arabic*:}]
\item 
    We update $(\Theta, \Theta_{H}, \Theta_{D}, \Delta)$ by minimizing the objective function in~\eqref{admm:step:one:one}.
It in turn can be optimized
via Iteratively Reweighted Least Square (IRLS) method with the Lasso penalty as below,
at some given fixed values for $(Z_{\Theta}^k, Z_{\Theta_H}^k, Z_{\Theta_D}^k)$ and 
$(U_{\Theta}^k, U_{\Theta_H}^k, U_{\Theta_D}^k)$. The $j+1$th iteration of IRLS is as follows:
\begin{footnotesize}
\begin{align}
\label{admm:irls1}
& (\Theta^{j + 1}, \Theta_{H}^{j + 1}, \Theta_{D}^{j + 1}, \Delta^{j + 1}) =  \argmin_{\Theta, \Theta_{H}, \Theta_{D}, \Delta} 
\left(\mbf X
\begin{pmatrix}
\Theta^\top \\
\Theta_{H}^\top \\
\Theta_{D}^\top \\
\Delta^\top
\end{pmatrix} 
- {\mbf z}^j \right)^\top
{\mbf W}^j
\left(\mbf X
\begin{pmatrix}
\Theta^\top \\
\Theta_{H}^\top \\
\Theta_{D}^\top \\
\Delta^\top
\end{pmatrix} 
- {\mbf z}^j \right) \nn \\
& + \lambda 
\sum_{s \in \mc S} \l( \sum_{h \in \mc H} \vert \theta^{\hod}_{s, h} \vert
+ \sum_{d \in \mc D} \vert \theta^{\dow}_{s, d} \vert \r)
+ \frac{\rho}{2} \l\Vert (\Theta, \Theta_{H}, \Theta_{D}) -   (Z_{\Theta}^{k}, Z_{\Theta_{H}}^{k}, Z_{\Theta_{D}}^{k}) +  (U_{\Theta}^{k}, U_{\Theta_{H}}^{k}, U_{\Theta_{D}}^{k}) \r\Vert^2,
\end{align} 
\end{footnotesize}
where $\mbf X$ is a data matrix of dimension $n \times (\vert\Delta\vert + (\vert \mc D \vert + \vert \mc H\vert -1)\cdot\vert\mc S\vert)$. 
Also, ${\mbf W}^j$ is an $n \times n$ diagonal matrix with its $i$th entry being $\wh\mu_i(\Theta^j, {\Theta_H}^j, {\Theta_D}^j, \Delta^j)$, the fitted value of the $i$th observation after the $j$th iteration, and ${\mbf z}^{j}$ is a length-$n$ vector defined as follows:
\begin{align*}
{\mbf z}^{j} = {\mbf X} 
(\Theta^j, \Theta_{H}^j, \Theta_{D}^j, \Delta^j)^\top
+ \left({\mbf W}^{j}\right)^{-1}{\mbf y} - \mathbbm {1}_{n}.
\end{align*}
Here, $\mbf y$ denotes a length-$n$ vector, the $i$th entry of which is the $i$th response observation, and $\mathbbm{1}_n$ denotes an all-one vector of length $n$. The objective function in~\eqref{admm:irls1} can further be written as the $\ell_1$-penalized least squares estimation problem as follows:
\begin{footnotesize}
\begin{align}
& (\Theta^{j+1}, \Theta^{j+1}_{H}, \Theta^{j+1}_{D}, \Delta^{j+1}) = 
\nn \\
& \argmin_{\Theta, \Theta_{H}, \Theta_{D}, \Delta} 
\left\Vert {\mbf X}_{\text{ext}}^{j}
(\Theta, \Theta_H, \Theta_D, \Delta)^\top
- {\mbf z}_{\text{ext}}^j\right\Vert^{2}  + \lambda 
\sum_{s \in \mc S} \l( \sum_{h \in \mc H} \vert \theta^{\hod}_{s, h} \vert
+ \sum_{d \in \mc D} \vert \theta^{\dow}_{s, d} \vert \r),
\label{admm:lassoform}
\end{align} 
\end{footnotesize}
where ${\mbf X}_{\text{ext}}^{j}$ and ${\mbf z}_{\text{ext}}^{j}$ are a matrix of size
$(n + \vert \mc P \vert) \times (\vert \mc P \vert + \vert\Delta\vert)$ and a vector of length $ (n + \vert \mc P \vert)$, respectively, such that
\begin{align*}
    \mbf{X}_{\text{ext}}^{j}=& 
\left(
\begin{array}{  c  c }
\multicolumn{2}{c}{ \left({\mbf W}^{j}\right)^{1/2} {\mbf X} } \\
\cdashline{1-2} 
\multicolumn{1}{ p{10mm} : }{$\sqrt{\rho} I_{\vert \mc P \vert} $}  & \multicolumn{1}{ p{10mm} }{$\mathbf{0}_{\vert \mc P\vert \times \vert\Delta\vert} $}   \\ 
\end{array}
\right), \\
{\mbf z}^{j}_{\text{ext}}=&
\left(
\begin{array}{c}
     \left({\mbf W}^{j}\right)^{1/2} {\mbf z}^{j} \\
      \sqrt{\rho} (Z_{\Theta}^k, Z_{\Theta_{H}}^k, Z_{\Theta_{D}}^k)^\top - \sqrt{\rho} (U_{\Theta}^k, U_{\Theta_{H}}^k, U_{\Theta_{D}}^k)^\top
\end{array}
\right).
\end{align*}
We evaluated~\eqref{admm:lassoform} using the R package 
\texttt{glmnet} \citep{glmnet}.

\item For~\eqref{admm:step:one:two}--\eqref{admm:step:one:three},
we obtain $\Gamma^{k + 1}$ and $\Psi^{k + 1}$ using a soft-thershold operator $\mathbb{S}_{\lambda}$ that takes an input vector and outputs $\mathbb{S}_{\lambda}(\mbf v) = \left(1 - \lambda/\Vert \mbf v \Vert \right)_{+} \cdot \mbf v$ with $\mathbb{S}_{\lambda}(\mathbf 0) = \mbf 0$ and $c_{+} = \max\{0, c\}$. 

\begin{align*}
\bm\gamma_s^{k+1} &= \mathbb{S}_{\rho^{-1} \sqrt{{\mc N}_r(s)}\lambda_{N}}  \l(S_{\bm\gamma_s}^{k} - T_{\bm\gamma_s}^{k}\r) \text{ \ for each \ } s \in {\mc S}, \\
\psi_{s, h}^{k+1} &= \mathbb{S}_{\rho^{-1}\lambda_{H}}(S_{\Psi, s, h}^k - T_{\Psi, s, h}^k)   \: \text{ for } \: s \in {\mc S} \text{, } \: h \in {\mc H}.
\end{align*}
\end{enumerate}

\subsubsection{Computational details of Step~2}
\label{sec:comp_bottle}

In Step~2, the update of $\left( Z_{\Theta}, Z_{\Theta_{D}}, Z_{\Theta_{H}}, S_{\Gamma}, S_{\Psi} \right)$ is achieved via projection in~\eqref{admm:step:two},
where ${\mc C}$ is the constraint set specified in~\eqref{admm:para1} and \eqref{admm:para2}. 
This step is the bottleneck of the computation due to the large number of variables to be projected onto a constraint set \citep{wahlberg2012admm}. 
Denoting the number of stations as $|\mathcal{S}|$, we carefully utilize the sparsity of the large matrices involved in this step, which reduces the computational complexity from $O( \vert \mc S \vert^{3} \vert (\mc D \vert + \vert \mc H \vert)^{3} )$ to $O(\vert \mc S \vert^2 (\mc D \vert + \vert \mc H \vert) )$.
In this section, we illustrate the details of the projection procedure.
Throughout, we denote by $\mbf I$ and $\mbf 0$ an identity matrix and
a matrix of zeros, respectively, and their dimensions are determined by the context unless specified. 

The projection $(A_1, A_2, A_3, B_1, B_2)= \prod_{\mc C} (E_1, E_2, E_3, F_1, F_2)$ is equivalent to the following minimization problem
\begin{align*}
    \min_{A_1,A_2,A_3,B_1, B_2} & \Vert A_1 -E_1\Vert^2+\Vert A_2 -E_2\Vert^2 +\Vert A_3 -E_3\Vert^2+\Vert B_1 -F_1\Vert^2+\Vert B_2 -F_2\Vert^2 \\
    \text{subject to} & \quad B_1 = (A_1, A_2, A_3) \mbf D_\Theta^\top \text{ and } 
    B_2 = (A_1, A_2, A_3) \mbf D_{\mc H}^\top,
\end{align*}
where $\mbf D_\Theta$ and $\mbf D_{\mc H}$ are matrices encoding the constraints~\eqref{admm:para1} and ~\eqref{admm:para2}, respectively. Then, the above optimization problem can be re-written as
\begin{footnotesize}
\begin{align*}
\min_{A_1,A_2,A_3} & \Vert A_1 -E_1\Vert^2+\Vert A_2 -E_2\Vert^2 +\Vert A_3 -E_3\Vert^2+\Vert (A_1,A_2,A_3) \mbf D_\Theta^\top - F_1\Vert^2+\Vert (A_1, A_2, A_3) \mbf D_{\mc H}^\top - F_2 \Vert^2,
\end{align*}
\end{footnotesize}
and its optimizer is the solution of its normal equation
\begin{footnotesize}
\begin{align}
\label{projsol}
    (A_1, A_2, A_3) \underbrace{\left( \mbf I + \mbf D_\Theta^\top \mbf D_\Theta + \mbf D_{\mc H}^\top \mbf D_{\mc H}
    \right)}_{\mbf P} 
    = (E_1, E_2, E_3) + F_1 \mbf D_\Theta + F_2 \mbf D_{\mc H}.
\end{align}
\end{footnotesize}
Once the inverse of $\mbf P$ is available, 
the solution $(A_1, A_2, A_3)$ of~\eqref{projsol} can be calculated in a straightforward manner. Also, the inverse matrix remains the same throughout the iterations and thus no re-computation is required. 
In our problem, however, as the size of the matrix $\mbf P$ is huge, its dimension reaching approximately $47,000 \times 47,000$, and inverting this matrix can be very demanding with the computational complexity of $\mathcal{O}(10^{11})$. Additionally, even if we can compute the inverse matrix, it is huge in size and occupies a large portion of memory space which hinders efficient computation. 
Given the situation, we avoid direct computation of $\mbf P^{-1}$ and find the solution of~\eqref{projsol} by utilizing the specific structure of the matrix $\mbf P$. 
    
We start by defining an $M \times \vert\mc S\vert$ matrix $\dnet$ to be a matrix that represents the network constructed from the neighborhood relations so that each row is associated with two connected stations where $M = \sum_{s \in {\mc S}} \vert{\mc N}_r(s)\vert$. Defining $\mc R_{\text{ind}}: \mc S \times \mc S \rightarrow \{1, \ldots, M\}$ to be the mapping that returns the row index of $s-s'
    $ connection for $s' \in {\mc N}_r(s)$, the $\mc R_{\text{ind}}(s, s')$-th row of $\dnet$ is given by $\mbf e_s - \mbf e_{s'}$,
    where $\mbf e_i$ is a standard basis vector of length $\vert\mc S\vert$. 
    That is, each row of $\dnet$ is composed of $\{-1,0,1\}$ with exactly one $1$ and one $-1$ and the rest of the entries are all $0$s. By construction, the Laplacian matrix, say $\lnet$, of the network can be represented by $\dnet$ as follows:
    \begin{align*}
    \lnet = \frac{1}{2}\dnet^\top \dnet.    
    \end{align*}
Then, we have 
    \begin{align*}
        \dnet \Theta^\top &= (\theta_s - \theta_{s'}, \,  s' \in \mc N_r(s), \,  s \in {\mc S})^\top, \\
        \dnet (\Theta^{\circ}_{H, h})^\top &= (\theta^{\hod}_{s, h} - \theta^{\hod}_{s', h}, \, s' \in \mc N_r(s), \, s \in \mc S)^\top, \\
        \dnet (\Theta^{\circ}_{D, d})^\top &= (\theta^{\dow}_{s, d} - \theta^{\hod}_{s', d}, \, s' \in \mc N_r(s), \, s \in \mc S)^\top, \\
    \end{align*} 
    where $\Theta_{H, h}^{\circ} = (\theta^{\hod}_{s, h}, \, s \in \mc S)$ and $\Theta_{D, d}^{\circ} = (\theta^{\dow}_{s, d}, \, d \in \mc D)$.
    
    The matrix $\mbf D_\Theta$ can be written with $\dnet$ as
    
\begin{tikzpicture}
\matrix [matrix of math nodes,left delimiter=(,right delimiter=),row sep=0.5cm,column sep=0.5cm] (m) {
\dnet & {\mbf 0} &  {\mbf 0}  &  {\mbf 0} &  -\dnet &  {\mbf 0}&  {\mbf 0} & {\mbf 0} &  {\mbf 0}\\
 {\mbf 0} & \ddots &  {\mbf 0} &  {\mbf 0} & \vdots &  {\mbf 0} &   {\mbf 0} &  {\mbf 0} &  {\mbf 0}  \\
 {\mbf 0} &  {\mbf 0} & \dnet &  {\mbf 0} &-\dnet &  {\mbf 0} &  {\mbf 0} &  {\mbf 0} &  {\mbf 0}   \\
  {\mbf 0} &  {\mbf 0} &  {\mbf 0} &  {\mbf 0} & -\dnet & \dnet &   {\mbf 0} &  {\mbf 0} &  {\mbf 0}  \\
 {\mbf 0} &  {\mbf 0} &  {\mbf 0} &  {\mbf 0} & \vdots &  {\mbf 0} &  \ddots &  {\mbf 0} &  {\mbf 0}  \\
  {\mbf 0} &  {\mbf 0} &  {\mbf 0} &  {\mbf 0} &-\dnet &  {\mbf 0} & {\mbf 0} &  \dnet &  {\mbf 0}  \\
 {\mbf 0} &  {\mbf 0} &  {\mbf 0} &  {\mbf 0} &-\sqrt{2}\dnet &  {\mbf 0} &  {\mbf 0} &  {\mbf 0} &  {\mbf 0} \\};
\draw[dashed] ($0.5*(m-1-4.north east)+0.5*(m-1-3.north west)$) --
     ($0.5*(m-7-4.south east)+0.5*(m-7-3.south west)$);

\draw[dashed] ($0.5*(m-1-4.north east)+0.5*(m-1-5.north west)$) --
     ($0.3*(m-7-4.south east)+0.7*(m-7-5.south west)$);

\draw[dashed] ($0.5*(m-1-5.north east)+0.5*(m-1-6.north west)$) --
     ($0.8*(m-7-5.south east)+.2*(m-7-6.south west)$);
\draw[dashed] ($0.5*(m-1-8.north east)+0.5*(m-1-9.north west)$) --
     ($0.5*(m-7-8.south east)+0.5*(m-7-9.south west)$);
\draw[dashed] ($0.5*(m-3-1.south west)+0.5*(m-4-1.north west)$) --
 ($0.5*(m-3-9.south east)+0.5*(m-4-9.north east)$);
\draw[dashed] ($0.5*(m-6-1.south west)+0.5*(m-7-1.north west)$) --
 ($0.5*(m-6-9.south east)+0.5*(m-7-9.north east)$);

\node[above=3pt of m-1-1] (top-1) {};
\node[above=3pt of m-1-2] (top-2) {};
\node[above=3pt of m-1-3] (top-3) {};
\node[above=3pt of m-1-4] (top-4) {};
\node[above=3pt of m-1-5] (top-5) {};
\node[above=3pt of m-1-6] (top-6) {};
\node[above=3pt of m-1-7] (top-7) {};
\node[above=3pt of m-1-8] (top-8) {};
\node[above=3pt of m-1-9] (top-9) {};

\node[left=13pt of m-1-1] (left-1) {};
\node[left=13pt of m-2-1] (left-2) {};
\node[left=13pt of m-3-1] (left-3) {};
\node[left=13pt of m-4-1] (left-4) {};
\node[left=13pt of m-5-1] (left-5) {};
\node[left=13pt of m-6-1] (left-6) {};
\node[left=17pt of m-7-1] (left-7) {};
\node[rectangle,above delimiter=\{] (del-top-1) at ($0.5*(top-1.south) +0.5*(top-3.south)$) {\tikz{\path (0,0) rectangle (3.5,0);}};
\node[above=10pt] at (del-top-1.north) {$\scard \cdot \dcard$};

\node[rectangle,above delimiter=\{] (del-top-2) at ($1.0*(top-4.south)$) {\tikz{\path (0,0) rectangle (0.7,0);}};
\node[above=10pt] at (del-top-2.north) {$\dcard$};

\node[rectangle,above delimiter=\{] (del-top-3) at ($1.0*(top-5.south)$) {\tikz{\path (0,0) rectangle (1.6,0);}};
\node[above=10pt] at (del-top-3.north) {$\scard$};

\node[rectangle,above delimiter=\{] (del-top-4) at ($1.0*(top-7.south)$) {\tikz{\path (0,0) rectangle (3.5,0);}};
\node[above=10pt] at (del-top-4.north) {$\scard \cdot \hcard $};

\node[rectangle,above delimiter=\{] (del-top-5) at ($1.0*(top-9.south)-(0.15,0)$) {\tikz{\path (0,0) rectangle (0.6,0);}};
\node[above=10pt] at (del-top-5.north) {$\hcard$};
\node[rectangle,left delimiter=\{] (del-left-1) at ($0.5*(left-1.east) +0.5*(left-3.east)$) {\tikz{\path (left-1.north east) rectangle (left-3.south west);}};
\node[left=10pt] at (del-left-1.west) {$M \cdot \dcard$};

\node[rectangle,left delimiter=\{] (del-left-2) at ($0.5*(left-4.east) +0.5*(left-6.east)$) {\tikz{\path (left-4.north east) rectangle (left-6.south west);}};
\node[left=10pt] at (del-left-2.west) {$M \cdot \hcard$};

\node[rectangle,left delimiter=\{] (del-left-2) at ($1.0*(left-7.east)$) {\tikz{\path (0,.5) rectangle (0,1.1);}};
\node[left=10pt] at (del-left-2.west) {$M$};
\end{tikzpicture}

where ${\mc D}^{\circ}={\mc D}\setminus\{\text{Mo}\}$ and ${\mc H}^{\circ}={\mc H}\setminus\{0\}$ so that the baseline parameters are removed. 

Denoting $\frac{1}{2}\lnet$ by $\wtlnet$,  the form of $\mbf D_\Theta^\top \mbf D_\Theta$ is as below:

\begin{tikzpicture}
\matrix [matrix of math nodes,left delimiter=(,right delimiter=),row sep=0.5cm,column sep=0.5cm] (m) {
\wtlnet & {\mbf 0} &  {\mbf 0}  &  {\mbf 0} &  -\wtlnet &  {\mbf 0}&  {\mbf 0} & {\mbf 0} &  {\mbf 0}\\
 {\mbf 0} & \ddots &  {\mbf 0} &  {\mbf 0} & \vdots &  {\mbf 0} &   {\mbf 0} &  {\mbf 0} &  {\mbf 0}  \\
 {\mbf 0} &  {\mbf 0} & \wtlnet &  {\mbf 0} &-\wtlnet &  {\mbf 0} &  {\mbf 0} &  {\mbf 0} &  {\mbf 0}   \\
  {\mbf 0} &  {\mbf 0} & {\mbf 0} &  {\mbf 0} & {\mbf 0} &  {\mbf 0} &  {\mbf 0} &  {\mbf 0} &  {\mbf 0}   \\
-\wtlnet &  \cdots &  -\wtlnet &    {\mbf 0} & (\dcard+\hcard)\wtlnet& -\wtlnet &  \cdots &  -\wtlnet &  {\mbf 0}   \\
  {\mbf 0} &  {\mbf 0} &  {\mbf 0} &  {\mbf 0} & -\wtlnet & \wtlnet &   {\mbf 0} &  {\mbf 0} &  {\mbf 0}  \\
 {\mbf 0} &  {\mbf 0} &  {\mbf 0} &  {\mbf 0} & \vdots &  {\mbf 0} &  \ddots &  {\mbf 0} &  {\mbf 0}  \\
  {\mbf 0} &  {\mbf 0} &  {\mbf 0} &  {\mbf 0} &-\wtlnet &  {\mbf 0} & {\mbf 0} &  \wtlnet &  {\mbf 0}  \\
 {\mbf 0} &  {\mbf 0} &  {\mbf 0} &  {\mbf 0} & {\mbf 0} &  {\mbf 0} &  {\mbf 0} &  {\mbf 0} &  {\mbf 0} \\};
\draw[dashed] ($0.5*(m-1-4.north east)+0.5*(m-1-3.north west)$) --
     ($0.5*(m-9-4.south east)+0.5*(m-9-3.south west)$);

\draw[dashed] ($0.6*(m-1-4.north east)+0.4*(m-1-5.north west)$) --
     ($0.7*(m-9-4.south east)+0.3*(m-9-5.south west)$);

\draw[dashed] ($0.5*(m-1-5.north east)+0.5*(m-1-6.north west)$) --
     ($0.4*(m-9-5.south east)+.6*(m-9-6.south west)$);
     
\draw[dashed] ($0.5*(m-1-8.north east)+0.5*(m-1-9.north west)$) --
     ($0.5*(m-9-8.south east)+0.5*(m-9-9.south west)$);
\draw[dashed] ($0.5*(m-3-1.south west)+0.5*(m-4-1.north west)$) --
 ($0.5*(m-3-9.south east)+0.5*(m-4-9.north east)$);
\draw[dashed] ($0.5*(m-4-1.south west)+0.5*(m-5-1.north west)$) --
 ($0.5*(m-4-9.south east)+0.5*(m-5-9.north east)$);
 \draw[dashed] ($0.5*(m-5-1.south west)+0.5*(m-6-1.north west)$) --
 ($0.5*(m-5-9.south east)+0.5*(m-6-9.north east)$);
  \draw[dashed] ($0.5*(m-8-1.south west)+0.5*(m-9-1.north west)$) --
 ($0.5*(m-8-9.south east)+0.5*(m-9-9.north east)$);

\node[above=3pt of m-1-1] (top-1) {};
\node[above=3pt of m-1-2] (top-2) {};
\node[above=3pt of m-1-3] (top-3) {};
\node[above=3pt of m-1-4] (top-4) {};
\node[above=3pt of m-1-5] (top-5) {};
\node[above=3pt of m-1-6] (top-6) {};
\node[above=3pt of m-1-7] (top-7) {};
\node[above=3pt of m-1-8] (top-8) {};
\node[above=3pt of m-1-9] (top-9) {};

\node[rectangle,above delimiter=\{] (del-top-1) at ($0.5*(top-1.south) +0.5*(top-3.south)$) {\tikz{\path (0,0) rectangle (3.5,0);}};
\node[above=10pt] at (del-top-1.north) {$\scard \cdot \dcard$};

\node[rectangle,above delimiter=\{] (del-top-2) at ($1.0*(top-4.south)$) {\tikz{\path (0,0) rectangle (0.7,0);}};
\node[above=10pt] at (del-top-2.north) {$\dcard$};

\node[rectangle,above delimiter=\{] (del-top-3) at ($1.0*(top-5.south)$) {\tikz{\path (0,0) rectangle (1.6,0);}};
\node[above=10pt] at (del-top-3.north) {$\scard$};

\node[rectangle,above delimiter=\{] (del-top-4) at ($1.0*(top-7.south)$) {\tikz{\path (0,0) rectangle (3.5,0);}};
\node[above=10pt] at (del-top-4.north) {$\scard \cdot \hcard $};

\node[rectangle,above delimiter=\{] (del-top-5) at ($1.0*(top-9.south)-(0.15,0)$) {\tikz{\path (0,0) rectangle (0.6,0);}};
\node[above=10pt] at (del-top-5.north) {$\hcard$};

\end{tikzpicture}

Now, we define $\mbf D_{\mc H}$, the difference matrix that addresses the association between consecutive hours of each of the stations so that
\begin{align*}
    \mbf D_{\mc H} (\Theta, \Theta_D, \Theta_{H})^\top = \left(
    \begin{array}{c}
          (- (\theta_{1}^{\hod} + \theta_{s,1}^{\hod}), \,  s \in {\mc S} )^\top\\
          \cdashline{1-1}
          ((\theta_{1}^{\hod} + \theta_{s,1}^{\hod})- (\theta_{2}^{\hod} + \theta_{s,2}^{\hod}), \,  s \in {\mc S} )^\top \\
          \vdots \\
           ((\theta_{22}^{\hod} + \theta_{s,22}^{\hod})- (\theta_{23}^{\hod} + \theta_{s,23}^{\hod}), \,  s \in {\mc S} )^\top \\ 
           \cdashline{1-1}
           ((\theta_{23}^{\hod} + \theta_{s,23}^{\hod})^\top, \,  s \in {\mc S} )^\top
    \end{array}
    \right).
\end{align*}

The specific form of $\mbf D_{\mc H}$ is as follows:

\begin{footnotesize}
\begin{tikzpicture}
\hspace{-1.3cm}
\matrix [matrix of math nodes,left delimiter=(,right delimiter=),row sep=0.5cm,column sep=0.4cm] (m) {
{\mbf 0} & {\mbf 0} & {\mbf 0} & {\mbf 0} &  {\mbf 0}  &  {\mbf 0} &  {\mbf 0} &  {\mbf 0}& {\mbf 0} & -1 &  0 & 0 & 0  \\
{\mbf 0} & {\mbf 0} &  {\mbf 0} & {\mbf 0} &  {\mbf 0}  &  {\mbf 0} &  {\mbf 0} &  {\mbf 0}&  {\mbf 0} & 1 & -1 & 0 & 0  \\
{\mbf 0} & {\mbf 0} &{\mbf 0} & {\mbf 0} &  {\mbf 0}  &  {\mbf 0} &  {\mbf 0} &  {\mbf 0}&  {\mbf 0} &   0 & \ddots & \ddots & 0 \\
{\mbf 0} & {\mbf 0} & {\mbf 0} & {\mbf 0} &  {\mbf 0}  &  {\mbf 0} &  {\mbf 0} &  {\mbf 0}&  {\mbf 0} & 0 & 0 & 1 & -1  \\
{\mbf 0} & {\mbf 0} & {\mbf 0} & {\mbf 0} &  {\mbf 0}  &  {\mbf 0} &  {\mbf 0} &  {\mbf 0} & {\mbf 0} &0 & 0 & 0 & 1  \\
{\mbf 0} & {\mbf 0} & {\mbf 0} & {\mbf 0} &  {\mbf 0}  &  -\mbf I_{\scard} &  {\mbf 0} &  {\mbf 0}&  {\mbf 0} & -\mathbbm{1}_{\scard}  & {\mbf 0} & {\mbf 0} & {\mbf 0} \\
{\mbf 0} & {\mbf 0} & {\mbf 0} & {\mbf 0} &  {\mbf 0}  &  \mbf I_{\scard} &  -\mbf I_{\scard} &  {\mbf 0}&  {\mbf 0} & \mathbbm{1}_{\scard} & -\mathbbm{1}_{\scard} & {\mbf 0} & {\mbf 0}  \\
{\mbf 0} & {\mbf 0} & {\mbf 0} & {\mbf 0} &  {\mbf 0}  & {\mbf 0} &  \ddots & \ddots&  {\mbf 0} & {\mbf 0} & \ddots & \ddots & {\mbf 0}  \\
{\mbf 0} & {\mbf 0} & {\mbf 0} & {\mbf 0} &  {\mbf 0}  & {\mbf 0} &  {\mbf 0} & \mbf I_{\scard} &  -\mbf I_{\scard} & {\mbf 0} & {\mbf 0} &\mathbbm{1}_{\scard} & -\mathbbm{1}_{\scard}   \\
{\mbf 0} & {\mbf 0} & {\mbf 0} & {\mbf 0} &  {\mbf 0}  & {\mbf 0} &  {\mbf 0} & {\mbf 0} &  \mbf I_{\scard} & {\mbf 0} &{\mbf 0} & {\mbf 0} & \mathbbm{1}_{\scard}  \\
 };
\draw[dashed] ($0.5*(m-1-4.north east)+0.5*(m-1-3.north west)$) --
     ($0.5*(m-10-4.south east)+0.5*(m-10-3.south west)$);
\draw[dashed] ($0.5*(m-1-5.north east)+0.5*(m-1-4.north west)$) --
     ($0.5*(m-10-5.south east)+0.5*(m-10-4.south west)$);
\draw[dashed] ($0.5*(m-1-6.north east)+0.5*(m-1-5.north west)$) --
     ($0.5*(m-10-6.south east)+0.5*(m-10-5.south west)$);
     
\draw[dashed] ($0.5*(m-1-10.north east)+0.5*(m-1-9.north west)$) --
     ($0.6*(m-10-10.south east)+0.4*(m-10-9.south west)$);

\draw[dashed] ($0.5*(m-5-1.south west)+0.5*(m-6-1.north west)$) --
 ($0.5*(m-5-13.south east)+0.5*(m-6-13.north east)$);

\node[above=3pt of m-1-1] (top-1) {};
\node[above=3pt of m-1-2] (top-2) {};
\node[above=3pt of m-1-3] (top-3) {};
\node[above=3pt of m-1-4] (top-4) {};
\node[above=3pt of m-1-5] (top-5) {};
\node[above=3pt of m-1-6] (top-6) {};
\node[above=3pt of m-1-7] (top-7) {};
\node[above=3pt of m-1-8] (top-8) {};
\node[above=3pt of m-1-9] (top-9) {};
\node[above=3pt of m-1-10] (top-10) {};
\node[above=3pt of m-1-11] (top-11) {};
\node[above=3pt of m-1-12] (top-12) {};
\node[above=3pt of m-1-13] (top-13) {};

\node[left=13pt of m-1-1] (left-1) {};
\node[left=13pt of m-2-1] (left-2) {};
\node[left=13pt of m-3-1] (left-3) {};
\node[left=13pt of m-4-1] (left-4) {};
\node[left=13pt of m-5-1] (left-5) {};
\node[left=13pt of m-6-1] (left-6) {};
\node[left=13pt of m-7-1] (left-7) {};
\node[left=13pt of m-8-1] (left-8) {};
\node[left=13pt of m-9-1] (left-9) {};
\node[left=13pt of m-10-1] (left-10) {};
\node[rectangle,above delimiter=\{] (del-top-1) at ($1.0*(top-2.south)$) {\tikz{\path (0,0) rectangle (2.2,0);}};
\node[above=10pt] at (del-top-1.north) {$\scard \cdot \dcard$};

\node[rectangle,above delimiter=\{] (del-top-2) at ($1.0*(top-4.south)$) {\tikz{\path (0,0) rectangle (0.5,0);}};
\node[above=10pt] at (del-top-2.north) {$\dcard$};

\node[rectangle,above delimiter=\{] (del-top-3) at ($1.0*(top-5.south)$) {\tikz{\path (0,0) rectangle (0.5,0);}};
\node[above=10pt] at (del-top-3.north) {$\scard$};

\node[rectangle,above delimiter=\{] (del-top-4) at ($0.5*(top-7.south)+0.5*(top-8.south)$) {\tikz{\path (0,0) rectangle (4.7,0);}};
\node[above=10pt] at (del-top-4.north) {$\scard \cdot \hcard $};

\node[rectangle,above delimiter=\{] (del-top-5) at ($0.5*(top-11.south)+0.5*(top-12.south)$) {\tikz{\path (0,0) rectangle (4.7,0);}};
\node[above=10pt] at (del-top-5.north) {$ \hcard $};

\node[rectangle,left delimiter=\{] (del-left-1) at ($0.5*(left-1.east) +0.5*(left-5.east)$) {\tikz{\path (left-6.north east) rectangle (left-10.south west);}};
\node[left=10pt] at (del-left-1.west) {$ \vert \mc H \vert$};

\node[rectangle,left delimiter=\{] (del-left-1) at ($0.5*(left-6.east) +0.5*(left-10.east)$) {\tikz{\path (left-1.north east) rectangle (left-5.south west);}};
\node[left=10pt] at (del-left-1.west) {$ \vert \mc H \vert \cdot \scard$};

\end{tikzpicture}. \\
\end{footnotesize}

Thus, $\mbf D_{\mc H}^\top \mbf D_{\mc H}$ has the form
\begin{align}
 D_{\mc H}^\top D_{\mc H} =&   
 \left( \begin{array}{c : c : c}
      {\mbf 0} & {\mbf 0} & {\mbf 0}\\
     \cdashline{1-3}
     {\mbf 0}^\top & \mbf M_1 & \mbf M_2 \\
      {\mbf 0}^\top & \mbf M_2^\top & \mbf M_3
    \end{array} \right), \quad \text{where}
    \nn \\
    \mbf M_1=& \begin{pmatrix}
    2 \mbf I_{\scard} & -\mbf I_{\scard} & {\mbf 0} & {\mbf 0} & {\mbf 0} \\
    -\mbf I_{\scard} &  2 \mbf I_{\scard} & -\mbf I_{\scard} & {\mbf 0} & {\mbf 0} \\
    {\mbf 0} & \ddots & \ddots & \ddots  &  {\mbf 0}\\
        {\mbf 0} &  {\mbf 0} &   -\mbf I_{\scard} &  2 \mbf I_{\scard} & -\mbf I_{\scard} \\
       {\mbf 0} &    {\mbf 0} &  {\mbf 0} &   -\mbf I_{\scard} &  2 \mbf I_{\scard}
    \end{pmatrix}, \label{proj_def_m1} \\
    \mbf M_2= & \begin{pmatrix}
    2 \mathbbm{1}_{\scard} & -\mathbbm{1}_{\scard} & {\mbf 0} & {\mbf 0} & {\mbf 0} \\
    -\mathbbm{1}_{\scard} &  2 \mathbbm{1}_{\scard} & -\mathbbm{1}_{\scard} & {\mbf 0} & {\mbf 0} \\
     {\mbf 0} & \ddots & \ddots & \ddots  &  {\mbf 0}\\
        {\mbf 0} &  {\mbf 0} &   -\mathbbm{1}_{\scard} &  2 \mathbbm{1}_{\scard} & -\mathbbm{1}_{\scard} \\
       {\mbf 0} &    {\mbf 0} &  {\mbf 0} &   -\mathbbm{1}_{\scard} &  2 \mathbbm{1}_{\scard}
    \end{pmatrix}, \nn \\
     \mbf M_3= & \begin{pmatrix}
    2 (\scard + 1) & -(\scard + 1) & 0 & 0 & 0 \\
    -(\scard + 1) &  2 (\scard + 1) & -(\scard + 1) & 0 & 0 \\
    0 &   \ddots & \ddots & \ddots & 0 \\
        0 &  0 &   -(\scard + 1) &  2 (\scard + 1) & -(\scard + 1) \\
       0 &    0 &  0 &   -(\scard + 1) &  2 (\scard + 1)
    \end{pmatrix}. \nn
\end{align}

From these, we can illustrate the structure of $\mbf P$. 
One notable thing is that both $\mbf D_{\Theta}$ and $\mbf D_{\mc H}$ have
their columns corresponding to $(\theta_{\text{Tu}}^{\dow}, \ldots, \theta_{\text{Su}}^{\dow})$ set exactly to be zero.
Thus, in computing $\mbf P^{-1}$, we remove their corresponding columns so that the inverse matrix is applied only to $(\Theta, \Theta_{D^{\circ}}, \Theta_{H})$. The structure of
$\bar{\mbf P}$, a sub-matrix of $\mbf P = \mbf I + \mbf D_{\Theta}^\top \mbf D_{\Theta}+ \mbf D_{\mc H}^\top \mbf D_{\mc H}$ without the columns corresponding to the daily parameters, is as follows:
\begin{footnotesize}
\begin{align}
    & \bar{\mbf P} = \nn \\
    &\left(
    \begin{array}{ccc:c:ccccc:c}
    \wtlnet + \mbf I & {\mbf 0} & {\mbf 0} & -\wtlnet & \multicolumn{5}{c:}{{\mbf 0} } & {\mbf 0}\\
    {\mbf 0}  & \ddots & {\mbf 0} & \vdots & \multicolumn{5}{c:}{ \vdots } & {\mbf 0} \\
    {\mbf 0}  &  {\mbf 0} &\wtlnet + \mbf I & -\wtlnet & \multicolumn{5}{c:}{{\mbf 0}} &  {\mbf 0} \\
    \cdashline{1-10}
     -\wtlnet & \cdots & -\wtlnet & (\vert \mc D \vert + \vert \mc H\vert)\wtlnet + \mbf I & -\wtlnet & \multicolumn{3}{c}{{\cdots }} & -\wtlnet & {\mbf 0}  \\
      \cdashline{1-10}
      {\mbf 0} &   \cdots &   {\mbf 0} & -\wtlnet & \wtlnet + 3\mbf I & -\mbf I &   {\mbf 0} & {\mbf 0} & {\mbf 0} & \multirow{5}{*}{$\mbf M_2$} \\
      \multirow{3}{*}{$\vdots$} &  \multirow{3}{*}{$\ddots$} &  \multirow{3}{*}{$\vdots$} &  \multirow{3}{*}{$\vdots$} & -\mbf I & \wtlnet + 3\mbf I & -\mbf I &  {\mbf 0} & \\
      & & &  &   {\mbf 0} & \ddots & \ddots & \ddots &  {\mbf 0} & \\
       &  &  &  &  {\mbf 0} &  {\mbf 0} & -\mbf I & \wtlnet +3\mbf I & -\mbf I & \\
       {\mbf 0} & \cdots & {\mbf 0} & -\wtlnet & {\mbf 0} & {\mbf 0} & {\mbf 0} & -\mbf I & \wtlnet + 3\mbf I & \\
        \cdashline{1-10}
        {\mbf 0} & \cdots & {\mbf 0} & {\mbf 0} & \multicolumn{5}{c:}{\mbf M_2^\top} & \mbf M_3+\mbf I_{\hcard}
       \end{array}
    \right) \nn
\end{align}
\end{footnotesize}
where $\mbf M_2$ and $\mbf M_3$ are as in~\eqref{proj_def_m1}. Here, the size of each block is $(\scard \cdot \dcard)$, $\scard$, $(\scard \cdot \hcard)$, and $\hcard$ from left to right and top to bottom, which match that of $\mbf D_{\Theta}^\top \mbf D_{\Theta}$ and $\mbf D_{\mc H}^\top \mbf D_{\mc H}$ specified above after removing corresponding columns of $(\theta_{\text{Tu}}^{\dow}, \ldots, \theta_{\text{Su}}^{\dow})$.

Let $\wtlnet = \mbf E \bm\Lambda \mbf E^\top$ denote the eigenvalue decomposition of $\wtlnet$ with $\bm\Lambda$ denoting a diagonal matrix having the eigenvalues as its diagonal entries. 
Then,
\begin{footnotesize}
\begin{align}
    \label{inverse2}
    & {\mbf W}^\top \bar{\mbf P} {\mbf W}= \nn \\
    &\left(
    \begin{array}{ccc:c:ccccc:c}
    \bm\Lambda + \mbf I & {\mbf 0} & {\mbf 0} & -\bm\Lambda & \multicolumn{5}{c:}{{\mbf 0} } & {\mbf 0}\\
    {\mbf 0}  & \ddots & {\mbf 0} & \vdots & \multicolumn{5}{c:}{ \vdots } & {\mbf 0} \\
    {\mbf 0}  &  {\mbf 0} &\bm\Lambda + \mbf I & -\bm\Lambda & \multicolumn{5}{c:}{{\mbf 0}} &  {\mbf 0} \\
    \cdashline{1-10}
     -\bm\Lambda & \cdots & -\bm\Lambda & (\vert \mc D \vert + \vert \mc H\vert)\bm\Lambda + \mbf I & -\bm\Lambda & \multicolumn{3}{c}{{\cdots }} & -\bm\Lambda & {\mbf 0}  \\
      \cdashline{1-10}
      {\mbf 0} &   \cdots &   {\mbf 0} & -\bm\Lambda & \bm\Lambda + 3\mbf I & - \mbf I &   {\mbf 0} & {\mbf 0} & {\mbf 0} & \multirow{5}{*}{$\wt{\mbf M}_2$} \\
      \multirow{3}{*}{$\vdots$} &  \multirow{3}{*}{$\ddots$} &  \multirow{3}{*}{$\vdots$} &  \multirow{3}{*}{$\vdots$} & - \mbf I & \bm\Lambda + 3\mbf I & - \mbf I &  {\mbf 0} &  {\mbf 0} \\
      & & &  &   {\mbf 0} & \ddots & \ddots & \ddots &  {\mbf 0} & \\
       &  &  &  &  {\mbf 0} &  {\mbf 0} & - \mbf I & \bm\Lambda +3\mbf I & - \mbf I & \\
       {\mbf 0} & \cdots & {\mbf 0} & -\bm\Lambda & {\mbf 0} & {\mbf 0} & {\mbf 0} & - \mbf I & \bm\Lambda + 3\mbf I & \\
        \cdashline{1-10}
        {\mbf 0} & \cdots & {\mbf 0} & {\mbf 0} & \multicolumn{5}{c:}{\wt{\mbf M}_2^\top} & \mbf M_3 + \mbf I_{ \hcard }
       \end{array}
    \right)
\end{align}
where $\wt{\mbf M}_2$ is
\begin{align*}
    \wt{\mbf M}_2= & \begin{pmatrix}
    2 \mbf E^{(\text{r})} & -\mbf E^{(\text{r})} & {\mbf 0} & {\mbf 0} & {\mbf 0} \\
    -\mbf E^{(\text{r})} &  2 \mbf E^{(\text{r})} & -\mbf E^{(\text{r})} & {\mbf 0} & {\mbf 0} \\
     {\mbf 0} & \ddots & \ddots & \ddots  &  {\mbf 0}\\
        {\mbf 0} &  {\mbf 0} &   -\mbf E^{(\text{r})} &  2 \mbf E^{(\text{r})} & -\mbf E^{(\text{r})} \\
       {\mbf 0} &    {\mbf 0} &  {\mbf 0} &   -\mbf E^{(\text{r})} &  2 \mbf E^{(\text{r})}
    \end{pmatrix} \text{ \ with \ } \mbf E^{(\text{r})} = \mbf E \mathbbm{1}_{\scard}, \text{ \ and \ }
    {\mbf W} =& \left(\begin{array}{ccc :c }
    \mbf E & {\mbf 0} &  {\mbf 0} & {\mbf 0} \\
    {\mbf 0} & \ddots & {\mbf 0} &  \vdots \\
    {\mbf 0} & {\mbf 0} & \mbf E & {\mbf 0} \\
    \cdashline{1-4}
    \multicolumn{3}{c:}{ {\mbf 0}} & \mbf I_{\hcard}
    \end{array}
    \right).
\end{align*}
\end{footnotesize}
Thus, the solution of $\bar{\mbf P} {\mbf z} = {\mbf b}$ can be achieved by solving
\begin{align}
    \label{inverse3}
   \left( {\mbf W}^\top \bar{\mbf P} {\mbf W}\right)  {\widetilde {\mbf z}} = {\wt{\mbf b}}
\end{align}
with $\wt{\mbf z} = {\mbf W}^\top {\mbf z}$ and $ {\widetilde {\mbf b}} = {\mbf W}^\top {\mbf b}$, and then finally setting $\mbf z = {\mbf W} \widetilde{\mbf z}$.
Solving~\eqref{inverse3} can be achieved efficiently by taking into account the structure of ${\mbf W}$ and ${\mbf W}^\top \bar{\mbf P} {\mbf W}$. 
Specifically, $\mbf W$ is block diagonal with repeated blocks which reduces the matrix multiplication complexity. The matrix ${\mbf W}^\top \bar{\mbf P} {\mbf W}$ is very sparse as described in~\eqref{inverse2} with all blocks involving $\bm\Lambda$ being diagonal, which facilitates efficient computation. 
Therefore, \eqref{inverse3} can be solved in an iterative manner using LU decomposition. The details of this procedure is illustrated later in this section. Having the full illustration of the structures, our suggested method for the projection step is described in Algorithm~\ref{algo:prod}.

\begin{algorithm}
  \caption{Projection in Step~2 of ADMM}
  \label{algo:prod}
  \begin{algorithmic}[1]
    \Inputs{ $
    \begin{array}{l}
         \left(\Theta^{k+1}, \; \Theta_{D}^{k+1}, \; \Theta_{H}^{k+1},  \;\Gamma^{k+1}, \;\Psi^{k+1}\right), \:
    \left( U_{\Theta}^{k}, \;, U_{\Theta_{D}}^{k},\; U_{\Theta_{H}}^{k}, \; T_{\Gamma}^{k}, \;, T_{\Psi}^{k}\right), \\    
    {\mbf D}_{\Theta},\; {\mbf D}_{\mc H}, \; {\mbf W}
    \end{array}
    $ } \\
    Compute ${\mbf b}_1 \leftarrow   \left(\Theta^{k+1 }, \; \Theta_{D}^{k+1}, \; \Theta_{H}^{k+1}\right)^\top +  \left( U_{\Theta}^{k}, \;, U_{\Theta_{D}}^{k},\; U_{\Theta_{H}}^{k}\right)^\top$
    \\
    Compute $ {\mbf b}_2 \leftarrow \mbf D_{\Theta} \left( \Gamma^{k+1} + T_{\Gamma}^{k}\right)^\top + \mbf D_{\mc H} \left( \Psi^{k+1} + T_{\Psi}^{k}\right)^\top$ \\
    Compute $\widetilde {\mbf b}  \leftarrow {\mbf W}^\top \left({\mbf b}_1 + {\mbf b}_2 \right) $ \\
    Solve   $\left( {\mbf W}^\top\left( \mbf I + \mbf D_{\Theta}^\top \mbf D_{\Theta}+ \mbf D_{\mc H}^\top \mbf D_{\mc H} \right){\mbf W}\right)^{-1}  {\widetilde {\mbf z}} = {\widetilde {\mbf b}}$ for $\widetilde {\mbf z}$\\
    Compute ${\mbf z} \leftarrow  {\mbf W} \widetilde{{\mbf z}}$ \\
    Set $(Z_{\Theta}^{k+1},  Z_{\Theta_{D}}^{k+1},  Z_{\Theta_{H}}^{k+1}) \leftarrow ({\mbf z}_{\Theta}, {\mbf z}_{\Theta_{D}}, {\mbf z}_{\Theta_{H}})$ where ${\mbf z} = ({\mbf z}_{\Theta}, {\mbf z}_{\Theta_{D}}, {\mbf z}_{\Theta_H})$ \\ 
    Set $(S_{\Gamma}^{k+1}, S_{\Psi}^{k+1}) \leftarrow (\mbf D_{\Theta} {\mbf z}, \; \mbf D_{\mc H} {\mbf z}_{\Theta_H} )$ 
    \Outputs{
    $(Z_{\Theta}^{k+1},  Z_{\Theta_{D}}^{k+1},  Z_{\Theta_{H}}^{k+1}, S_{\Gamma}^{k+1}, S_{\Psi}^{k+1})$
    }
  \end{algorithmic}
\end{algorithm}

In order to solve \eqref{inverse3} in line 5 of Algorithm~\ref{algo:prod}, we can utilize the LU decomposition of the matrix in~\eqref{inverse2}. Writing the LU decomposition of the matrix~\eqref{inverse2} by $\mbf L\mbf L^\top$, the matrix $\mbf L$ has the banded block diagonal structure as follows:
\begin{align}
    \label{Lmatrix} 
    \mbf L = \left(\begin{array}{c :c :c :c }
        \mbf A_{11} & {\mbf 0} &{\mbf 0} & {\mbf 0}\\
        \cdashline{1-4}
        \mbf A_{21} & \mbf A_{22}& {\mbf 0} & {\mbf 0} \\
         \cdashline{1-4}
        {\mbf 0} & \mbf A_{32} & \mbf A_{33} & {\mbf 0} \\
         \cdashline{1-4}
        {\mbf 0} & {\mbf 0} & \mbf A_{43} & \mbf A_{44}
    \end{array}
    \right).
\end{align}
The blocks match the size of their counterparts in~\eqref{inverse2}. Here, the relatively large matrices $\mbf A_{11}$, $\mbf A_{21}$, $\mbf A_{22}$, $\mbf A_{32}$ and $\mbf A_{33}$ have specific repetitive sparse structures which facilitate memory saving and efficient computation. 
Specifically, these contain repeated sub-matrices of dimension $\scard \times \scard$ which are of the following forms:
\begin{align*}
    \mbf A_{11}=& \, \left( 
    \begin{array}{c c c }
    \mathbb{D}_{11} &  {\mbf 0} &  {\mbf 0}\\
      {\mbf 0} & \ddots &  {\mbf 0}\\
     {\mbf 0} & {\mbf 0} & \mathbb{D}_{11}
    \end{array}
    \right), \quad
    \mbf A_{21}= \, \left( \mathbb{D}_{21} \ldots \mathbb{D}_{21} \right), \quad
    \mbf A_{22}= \, \mathbb{D}_{22}, \quad
    \mbf A_{32}= \, \left(
    \mathbb{D}_{32}
    \cdots 
    \mathbb{D}_{32}
    \right)^\top, \\
    \mbf A_{33} = & \, \left(
    \begin{array}{cccccc}
    d_{1,1} &  {\mbf 0} &  {\mbf 0} &  {\mbf 0} &  {\mbf 0} & {\mbf 0}\\
    g_{2,1}  & d_{2,2} &  {\mbf 0}  & \vdots  & \multirow{2}{*}{$\vdots$}  & \multirow{2}{*}{$\vdots$}\\ 
     f_1 & g_{3,2} & d_{3,3} & {\mbf 0}  &   &  \\
     \vdots & f_2 & \ddots & \ddots &  {\mbf 0}  &  \\
        \vdots & \vdots &  & \ddots & \ddots  & {\mbf 0} \\
     f_1 & f_2 & \cdots & f_{\hcard -1 } & g_{\hcard,\hcard-1} & d_{\hcard,\hcard} \\
    \end{array}
    \right).
\end{align*}

Using this banded diagonal structure of the LU-decomposition, solving~\eqref{inverse3} can be done in two steps, (i) solving $\mbf L \mbf y = \widetilde{\mbf b}$, and then (ii) solving $\mbf L^\top \widetilde{\mbf z} = {\mbf y}$. The details are presented below.

\begin{enumerate}[label = (\roman*)]
    \item Solve $\mbf L \mbf y = \widetilde{\mbf b}$ with $\mbf y = (\mbf y^{(1)}, \ldots, \mbf y^{(4)})^\top$:
    \begin{enumerate}
        \item Solve $\mbf A_{11}{\mbf y}^{(1)} = {\widetilde{\mbf b}}^{(1)}$: \\
        $ {\mbf y}^{(1)}_{i}=  \mathbb{D}_{11}^{-1} {\widetilde{\mbf b}}^{(1)}_i \text{ for } i=1, \ldots, \dcard. $
        \item Solve $\mbf A_{22}{\mbf y}^{(2)} = {\widetilde{\mbf b}}^{(2)} - \mbf A_{21}{\mbf y}^{(1)}$: \\
        $
    {\mbf y}^{(2)}=  \mathbb{D}_{22}^{-1} \left( {\widetilde{\mbf b}}^{(2)} - \mathbb{D}_{21} \sum_{i=1}^{\dcard} {\mbf y}^{(1)}_{i} \right).$
        \item Solve $\mbf A_{33} {\mbf y}^{(3)} = {\widetilde{\mbf b}}^{(3)} - \mbf A_{32}{\mbf y}^{(2)}$: 
        \begin{align*}
            {\mbf y}^{(3)}_1 &= d_{1,1}^{-1}\left( {\widetilde{\mbf b}}^{(3)}_1  - \mathbb{D}_{32} {\mbf y}^{(3)}_1 \right), \\
            {\mbf y}^{(3)}_2 &= d_{2,2}^{-1} \left(\left( {\widetilde{\mbf b}}^{(3)}_2  - \mathbb{D}_{32} {\mbf y}^{(3)}_2 \right) - g_{2,1}  {\mbf y}^{(3)}_1 \right), \\
                {\mbf y}^{(3)}_i &=d_{i,i}^{-1} \left(\left( {\widetilde{\mbf b}}^{(3)}_i  - \mathbb{D}_{32} {\mbf y}^{(3)}_i \right) - g_{i,i-1}  {\mbf y}^{(3)}_{i-1} - \sum_{j=1}^{i-2} f_{j} {\mbf y}^{(3)}_{j} \right) 
         \text{ for } i =3, \ldots, \hcard.
        \end{align*}
    \item  ${\mbf y}^{(4)} = \mbf A_{44}^{-1} \left({\widetilde{\mbf b}}^{(4)} - \mbf A_{43}{\mbf y}^{(3)}\right)$.
    \end{enumerate}
    \item Solve $\mbf L^\top \widetilde{\mbf z} = \mbf y$ with $\wt{\mbf z} = (\wt{\mbf z}^{(1)}, \ldots, \wt{\mbf z}^{(4)})^\top$:
    \begin{enumerate}
        \item ${\widetilde{\mbf z}}^{(4)} = \left(\mbf A_{44}^\top\right)^{-1} {\mbf y}^{(4)}$ 
        \item Solve $\mbf A_{33}^\top {\widetilde{\mbf z}}^{(3)} = {\mbf y}^{(3)} - \mbf A_{43}^\top {\widetilde{\mbf z}}^{(4)}$: 
        \begin{align*}
            {\widetilde{\mbf z}}^{(3)}_{\hcard} &= d_{ \hcard, \hcard}^{-1} \left(  {\mbf y}^{(3)}_{\hcard} - \left(\mbf A_{43}^\top\right)_{\hcard} {\widetilde{\mbf z}}^{(4)} \right), \\
            {\widetilde{\mbf z}}^{(3)}_{\hcard-1} &= d_{ \hcard-1, \hcard-1}^{-1} \left(  {\mbf y}^{(3)}_{\hcard-1}  - \left(\mbf A_{43}^\top\right)_{\hcard-1} {\widetilde{\mbf z}}^{(4)} -g_{\hcard, \hcard-1} {\widetilde{\mbf z}}^{(3)}_{\hcard} \right),
            \\
            {\widetilde{\mbf z}}^{(3)}_{i} &= d_{i, i}^{-1} \left(  {\mbf y}^{(3)}_{i}  - \left(\mbf A_{43}^\top\right)_{i} {\widetilde{\mbf z}}^{(4)} -g_{i+1, i} {\widetilde{\mbf z}}^{(3)}_{i+1} - (\hcard -i - 1)f_{i} \sum_{j=i+2}^{\hcard} {\widetilde{\mbf z}}^{(4)}_{j}\right) 
            \end{align*}
for $i = \hcard -2, \ldots, 1$.
        \item Solve $\mbf A_{22}^\top {\widetilde{\mbf z}}^{(2)} = {\mbf y}^{(2)} - \mbf A_{32}^\top {\widetilde{\mbf z}}^{(3)}$: \\ 
        ${\widetilde{\mbf z}}^{(2)} = \mathbb{D}_{22}^{-1}\left({\mbf y}^{(2)} - \mathbb{D}_{32} \left( \sum_{i=1}^{\hcard} {\widetilde{\mbf z}}^{(3)}_i \right)\right)$
        \item Solve $\mbf A_{11}^\top {\widetilde{\mbf z}}^{(1)} = {\mbf y}^{(1)} - \mbf A_{21}^\top {\widetilde{\mbf z}}^{(2)}$: \\ 
        ${\widetilde{\mbf z}}^{(1)}_i = \mathbb{D}_{11}^{-1}\left({\mbf y}^{(1)}_i - \mathbb{D}_{21} {\widetilde{\mbf z}}^{(2)}\right)$ \text{ for } $i = 1, \ldots, \dcard$.
    \end{enumerate}
\end{enumerate}
Here, $(\mbf A_{43})_i$ denotes its $i$th row. The vector arguments are partitioned as
\begin{align*}
{\mbf x}^\top = \l(   (\mbf x^{(1)}_1)^\top , \ldots , (\mbf x^{(1)}_{\dcard})^\top,  \: (\mbf x^{(2)})^\top, \:  (\mbf x^{(3)}_{1})^\top, \ldots,  (\mbf x^{(3)}_{\hcard})^\top, \: (\mbf x^{(4)})^\top \r)^\top,
\end{align*}
where ${\mbf x}^{(j)}_i$ for $j=1, 3$ and ${\mbf x}^{(2)}$ are of length $\scard$ and ${\mbf x}^{(4)}$ is of length $\hcard$.

\subsubsection{Computation time}
\label{subsubsec:comput_time}

The computing time for training the proposed model varies significantly based on several factors, including data size, the number of neighbors (determined by the radius parameter $r$) and the precision of the Lasso sub-optimization process described in Step~1.1 of Section~\ref{subsec:glmnet}.

For a given set of tuning parameters, training times were as follows: approximately 2 to 24 hours for the Seoul dataset (1,806,000 samples), 2 to 48 hours for the New York dataset (1,011,240 samples), and 10 minutes to 2 hours for the SF dataset (90,312 samples). These computations were performed on a system with two Intel(R) Xeon(R) Gold 6226 CPUs @ 2.70GHz and 128GB of RAM. In our experiment, the primary computational bottleneck occurred during the execution of \texttt{glmnet} in Step~1.1 of Section~\ref{subsec:glmnet}. Although computational bottlenecks typically arise in Step~2 \citep{wahlberg2012admm}, we have observed that, due to our approach described in Appendix~\ref{subsubsec:comput_time}, the empirical bottleneck in our case is Step~1.1.
Each \texttt{glmnet} execution can take up to 2 to 3 minutes for large scale data (e.g.\ Seoul and New York datasets) with the default precision setting of $10^{-11}$, while the remaining steps take about 10 seconds combined. The runtime of \texttt{glmnet} is highly dependent on its precision setting. There is a trade-off between precision and the number of \texttt{glmnet} calls required for Step~1.1. Specifically, higher precision reduces the number of iterations (and corresponding \texttt{glmnet} calls) needed for the convergence of Step~1.1.

In our experiments, we found it advantageous to set a moderate precision level in the early stages of training, gradually increasing it as the process progresses. This strategy is effective because the early stages of sub-optimization require many \texttt{glmnet} iterations for Step~1.1 to converge. As training advances, the number of \texttt{glmnet} calls decreases significantly, making higher precision more critical for achieving overall convergence.

\subsection{Computation of model complexity}
\label{sec:mc}

The numerator of MC in~\eqref{eq:mc} counts the number of connected components of the given graphs. Typically, the number of connected components of a graph can be found by counting the number of zero-eigenvalue of the graph's Laplacian matrix.
The Laplacian matrix of $\mathfrak{N}_{\text{single}}(r) \cap \wh{\mathfrak{N}}_{D, d}$ has the dimension of $\scard \times \scard$ for each $d$, and it is feasible to compute its eigenvalues and calculate $\mc C (\mathfrak{N}_{\text{single}}(r) \cap \wh{\mathfrak{N}}_{D, d})$.

However, computing $\mc C (\mathfrak{N}_{\text{multi}}(r) \cap \wh{\mathfrak{N}}_{H})$ is not so straightforward since the Laplacian matrix of the graph $\mathfrak{N}_{\text{multi}}(r) \cap \wh{\mathfrak{N}}_H $ has the dimension of $\scard \vert \mc H \vert \times \scard \vert \mc H \vert $, it is not practical to compute its eigendecomposition.
Instead, we propose to obtain $\mc C (\mathfrak{N}_{\text{multi}}(r) \cap \wh{\mathfrak{N}}_{H})$ by first finding the layer-specific connected components, and then coalescing the components of two consecutive layers if they are connected transversely over the layers. 

Before describing the proposed method, we introduce some notations relevant to $\mathfrak{N}_{\text{single}}(r) \cap \wh{\mathfrak{N}}_{H, h}$, a single-layer network for each $h \in \mc H$: 
\begin{align*}
\wh{\mathfrak{N}}_{H, h} =& \, \l(\mc S, \, \mc E_{H, h}\r) \text{ \ with \ } \mc E_{H, h} = \l\{(s, s'), \, s \ne s': \, \wh\phi^{\hod}_{s, h} = \wh\phi^{\hod}_{s', h} \r\},
\\
\mathfrak{G}_h =& \, \mathfrak{N}_{\text{single}}(r) \cap \wh{\mathfrak{N}}_{H, h}, 
\\
N_{h} =& \, \mc C (\mathfrak{G}_h), 
\\
\mbf{M}^{(0)} =& \, \text{$\vert \mc H \vert \times \scard $ matrix that encodes the layer-specific connected components of }\mathfrak{G}_h, 
\\
\mbf L_h =& \, \text{the Laplacian matrix of }\mathfrak{G}_h 
\\
\mbf{c}_{h,k}(\mbf M) =& \, \text{the index vector of entries that corresponds to the elements in the $k$th cluster} 
\\
&\text{of the $h$th row of a cluster label matrix $\mbf M$} .
\end{align*}
Specifically, for each connected component, the corresponding $(h, s)$ elements of $\mbf M^{(0)}$ take a unique value, and thus the number of unique values in $\mbf M^{(0)}$ agrees with the number of total connected components $\sum_{h=0}^{23} N_h$.

We begin by describing how to construct $\mbf M^{(0)}$. Recall that the number of zero eigenvalues of $\mbf L_h$ corresponds to $N_h$. The eigenvector associated with the zero eigenvalue provides some information of the connected components. 
Namely, it is a linear combination of the connected component indicator vectors $\mbf c_{h, k}(\mbf M^{(0)})$. 
For an arbitrary index vector $\mbf g = (g_i: \, g_i \in \{1, \ldots, \scard\})$, we denote the indicator vector of $\mbf g$ by $\mbf e_{\mbf g}$:
It has its $g_i$th entry to be one for $i = 1, \ldots, \vert \mbf g \vert$, and all the rest are zeros. 
Denote by $\mbf v^{(0)}_{h, i}$ the $i$th eigenvector associated with the zero eigenvalues of $\mbf L_h$ for $i = 1, \ldots, N_h$.
Then, we have $\mbf v^{(0)}_{h,i} = \sum_{k = 1}^{N_h} a^{(h,i)}_{k} \mbf e_{\mbf c_{h, k}(\mbf M^{(0)})}$ 
for some constants $a^{(h,i)}_k$s. Thus, we utilize the eigenvectors to identify the cluster label of each stations and construct $\mbf M^{(0)}$. Precisely, we find the partition $\mc P$ of the index set $\{1, \ldots, \scard\}$ with the smallest cardinality
such that for each $P$ belonging to the partition,
all the elements of $\mbf v^{(0)}_{h, i}$ located at $P$ take the same value, for all $i = 1, \ldots, N_h$.
This procedure is described in Algorithm~\ref{algo:layer_wise}.

\begin{algorithm}
  \caption{Construction of $\mbf M^{(0)}$.}
  \label{algo:layer_wise}
  \begin{algorithmic}[1]
    \Inputs{
    The Laplacian matrix $\mbf L_h$ for $h \in \mc H$
     } 
    \Initialize{
    $\mbf M^{(0)} \leftarrow$ a $\vert {\mc H} \vert \times \scard$-matrix of zeros \\
    $\texttt{cluster\_label} \leftarrow 1$
    }
    \For{$h \in \mc H$}
    \State Perform eigenvalue decomposition of $\mbf L_h$ 
    \State $N_h \leftarrow$ the number of zero eigenvalues
    \State $\mbf V \leftarrow$ a $\scard \times N_h$ matrix having the eigenvector $\mbf v^{(0)}_{h,i}$ as its $i$th column
    \State $\texttt{unlabeled} \leftarrow \{1, 2, \ldots, \scard \}$
    \For{$s \in \{1, \ldots, \scard\}$}
    \If{$s \notin \texttt{unlabeled}$}
    \State $\texttt{cluster} \leftarrow \{1, \ldots, \scard\}$
    \For{$i \in \{1, \ldots, N_h\}$}
    \State $ \texttt{value} \leftarrow V_{s,i}$, the $s$th entry of the $i$th eigenvector $\mbf v^{(0)}_{h,i}$
    \State $\texttt{value\_set} \leftarrow \{s': V_{s',i} = \texttt{value}\}$  
    \State $\texttt{cluster} \leftarrow \texttt{cluster} \cap \texttt{value\_set}$
    \EndFor
    \State $M^{(0)}_{h, s'}\leftarrow \texttt{cluster\_label}$ for $s' \in \texttt{cluster}$
    \State $\texttt{unlabeled} \leftarrow \texttt{unlabeled}\setminus \texttt{cluster}$
    \State $\texttt{cluster\_label} \leftarrow \texttt{cluster\_label}+1$
    \EndIf
    \EndFor
    \EndFor
    \Outputs{$\mbf M^{(0)}$}
  \end{algorithmic}
\end{algorithm}

Once the layer-specific cluster label matrix $\mbf M^{(0)}$ is provided, we link the connected components throughout the hourly layers as described in Algorithm~\ref{algo:comp}. Within the procedure, the sub-routine given in Algorithm~\ref{algo:link} is utilized, which sequentially links connected components lying in two consecutive layers. In each run, Algorithm~\ref{algo:link} links a component in evaluation to exactly one component in another layer while there can be more than one component that are supposed to be linked. Thus, Algorithm~\ref{algo:comp} keeps running the sub-algorithm until there exists no more component left to be linked. One notable feature is that the sub-algorithm is executed twice with different order vector $\mbf{r}$ each time. This is to account for the circular feature of hour-of-a-day. The function $\texttt{is}(\cdot)$ in the sub-algorithm~\ref{algo:link} is defined to return the Boolean of the input statement.

\begin{algorithm}
  \caption{Counting the connected components of $\mathfrak{N}_{\text{multi}}(r) \times \wh{\mathfrak{N}}_H$.}
  \label{algo:comp}
  \begin{algorithmic}[1]
    \Inputs{The layer-specific cluster label matrix $\mbf M^{(0)}$ \\
    The estimates $\wh\phi^{\hod}_{s, h}$ for $s \in \mc S$ and $h \in \mc H$}
    \State $\mbf M \leftarrow \mbf M^{(0)}$
    \State $\texttt{changed} \leftarrow \texttt{true}$
    \While{\texttt{changed}}
    \State $\texttt{changed} \leftarrow \texttt{false}$
    \State Make links across the layers in a forward manner by running Algorithm~\ref{algo:link}
    \State with $\mbf r = (0,1,\ldots,23)$ 
    \State Connect the layers of $h=0$ and $h=23$ in a forward manner by running Algorithm~\ref{algo:link} 
    \State with $\mbf r = (23,0,1,\ldots,22)$ 
    \EndWhile
    \Outputs{the number of unique values in $\mbf M$}
  \end{algorithmic}
\end{algorithm}

\begin{algorithm}
  \caption{Link across ordered layers.}
  \label{algo:link}
  \begin{algorithmic}[1]
    \Inputs{ 
    The estimates $\wh\phi^{\hod}_{s, h}$ for $s \in \mc S$ and $h \in \mc H$ \\
    An order vector $\mbf{r}$ of length $\vert \mc H \vert$ \\
    An $\vert \mbf r \vert \times \vert \mc S \vert$ cluster label matrix $\mbf{M}$ \\
    A Boolean variable \texttt{changed}
     }
    \State $\mbf M^{\text{old}} \leftarrow \mbf M$
    \For{$i \in \{1, \ldots, \vert\mbf{r} \vert\}$}
    \State \textbf{// link the $r_i$th and the $r_{i+1}$th layers:}
    \State $\mbf M^{\text{new}} \leftarrow \mbf M^{\text{old}}$
    \For{$k \in \{1, \ldots, N_{r_i}\}$}
    \State \textbf{// investigate whether each component is subject to further connection}
    \State $\text{ind} \leftarrow \mbf{c}_{r_i,k}(\mbf M^{\text{old}})$
    \State $j \leftarrow 0$
    \State $\texttt{matched} \leftarrow \texttt{false}$
    \State $\texttt{exhausted} \leftarrow \texttt{is}(k \ge \vert \mbf{c}_{r_i,k}(\mbf M^{\text{old}}) \vert)$
    \While{ \texttt{matched} $=$ \texttt{false} and \texttt{exhausted} $=$ \texttt{false}}
    \State $j \leftarrow j+1$
    \State $c_1 \leftarrow \wh{\phi}^{\hod}_{\text{ind}_j, r_i}$
    \State $c_2 \leftarrow \wh{\phi}^{\hod}_{\text{ind}_{j}, r_{i+1}}$
    \State $\text{ind}^{(j,r_{i+1})} \leftarrow$  the index vector of the cluster to which the $j$th entry of \texttt{ind} belongs, in the $r_{i+1}$th row of $\mbf M^{\text{new}}$ i.e.\ $\mbf c_{r_{i+1}, {k'}}(\mbf M^{\text{ new}})$ for some $k'$ such that $\texttt{ind}_j \in c_{r_{i+1}, {k'}}(\mbf M^{\text{new}})$
    \State $m_1 \leftarrow \min\; {\mbf M^{\text{old}}}_{r_i,\text{ind}}$
    \State $m_2 \leftarrow \min\; {\mbf M}^{\text{new}}_{r_{i+1}, \text{ind}^{(j,r_{i+1})}}$
    \State $\texttt{same\_estimate}\leftarrow \texttt{is}(c_1=c_2)$
    \State $\texttt{same\_label} \leftarrow \texttt{is}(m_1=m_2)$
    \State $\texttt{should\_connect} \leftarrow \texttt{is}(\texttt{same\_estimate} \text{ and not } \texttt{same\_label})$
    \If{$\texttt{should\_connect}$}
    \State \textbf{// connect the to components and update their labels}
    \State $m = \min \{m_1, m_2\}$ 
    \State ${\mbf M}^{\text{new}}_{r_i, \text{ind}} \leftarrow m$ 
      \State ${\mbf M}^{\text{new}}_{r_{i+1}, \text{ind}^{(j,r_{i+1})} } \leftarrow m$ 
    \State $\texttt{changed} \leftarrow \texttt{true}$ 
    \State $\texttt{matched} \leftarrow \texttt{true}$ 
    \State $\texttt{exhausted} \leftarrow \texttt{is}(k \ge \vert \mbf{c}_{r_i,k}(\mbf M^{\text{old}}) \vert)$
    \EndIf
    \EndWhile
    \EndFor
    \State $\mbf M^{\text{old}} \leftarrow \mbf M^{\text{new}}$
    \EndFor
    \State $\mbf M \leftarrow \mbf M^{\text{new}}$
    \Outputs{The updated cluster label matrix $\mbf M$ \\
    The updated Boolean variable \texttt{changed}
    }
  \end{algorithmic}
\end{algorithm}

\clearpage

\section{Data descriptions and exploratory analysis}
\label{sec:data}

\subsection{Bike-sharing system in Seoul}

In this section, we provide basic information about the BSS in Seoul, South Korea, followed by in-depth exploratory analysis of the usage data.

\subsubsection{Basic information}
\label{sec:suppinfo}

In this paper, we analyze the public bike share usage in Seoul, the capital city in South Korea. 
The dataset is available at \url{https://data.seoul.go.kr/}

Seoul metropolitan government implemented a pilot of `Ddareungi', the public bike-sharing system, between November 2010 and April 2015 with 440 bikes. After its second round of pilot with 967 bikes, the official Ddareungi system launched on March 1, 2016, covering 11 districts with 2000 bikes.


         
         
         
         
         

Both subscribed and casual users can rent a bike through a mobile app. The difference is that while only 1-Day pass is available to casual users, subscribed users can choose among 1-Day, 1-Week, 1-Month, 6-Months and 1-Year passes. Upon subscription, the registered user can choose between a 1-hour pass or a 2-hour pass. The information on pricing is provided in Table~\ref{tab:price}. 

\begin{table}[h!]
\caption{Usage price of the bike-sharing system in Seoul, South Korea in $2019$.}
    \label{tab:price}
\footnotesize
    \centering
    \begin{tabular}{c c c c c c c}
\toprule
        \multirow{2}{*}{Hours of use} & \multirow{2}{*}{1-Day} & \multirow{2}{*}{7-Day} & \multirow{2}{*}{30-Day} & \multirow{2}{*}{180-Day} & \multirow{2}{*}{365-Day} & Group ticket\\
        &&&&&&(2~5people)\\
        
\cmidrule(lr){1-1} \cmidrule(lr){2-7}
        \multirow{2}{*}{1-hour pass} & \$0.89 & \$2.68 & \$4.47 & \$13.41 & \$26.82 & Per person \$0.89\\
        &(₩1,000) & (₩3,000) & (₩5,000) & (₩15,000) &(₩1,000) & (Per person ₩1,000)\\
        
\cmidrule(lr){1-1} \cmidrule(lr){2-7}
        \multirow{2}{*}{2-hour pass}& \$1.79 & \$3.58 & \$6.26 & \$17.88 & \$35.77 & Per person \$1.79\\
        & (₩2,000) & (₩4,000) & (₩7,000) & (₩20,000) & (₩40,000) & (Per person ₩2,000)\\
        
\bottomrule
    \end{tabular}
\end{table}

The stations undergo rebalancing three times a day during the following three time windows: 8 AM to 4 PM, 4 PM to 11 PM, and 11 PM to 4 AM the following day.
While a particular station may temporarily run out of bikes, users can check real-time bike availability at all stations through a mobile app in advance. This feature enables them to make informed decisions, allowing users to select a nearby station with bikes readily available for immediate use.

\subsubsection{Exploratory data analysis}
\label{sec:exploratory}

In this section, we perform the Seoul BSS data to observe usage patterns commonly shared in BSS and motivate the regression model proposed in Section~\ref{sec:method}.
We use the hourly rental records at individual stations between April 1, 2019 and May 30, 2019. 
By selecting the temperate months of April and May, and excluding three public holidays falling in this period, we avoid dealing with seasonality or possibly abnormal observations due to extreme weather conditions.
Thus, our final dataset comprises observations from $T = 57$ days and $S = 1505$ stations.

\begin{figure}[h!]
\centering
\includegraphics[width = .5\textwidth]{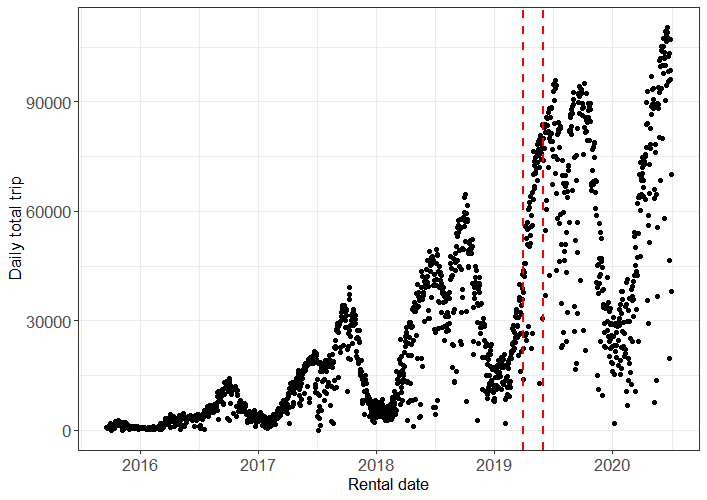}
\caption{Daily trip frequencies between 2015--2020.}
\label{fig:tripfreq}
\end{figure}

Since the launch of the BSS in 2010, the numbers of subscribed users, stations and available bikes have steadily increased, as has the number of trips.
In Figure~\ref{fig:tripfreq}, we plot the daily trip frequencies between 2015 and 2019. The start and the end of the analysis sample period is denoted by the two vertical lines.

\begin{figure}[h!t!b!]
\centering
\subfloat{\includegraphics[width = .48\textwidth]{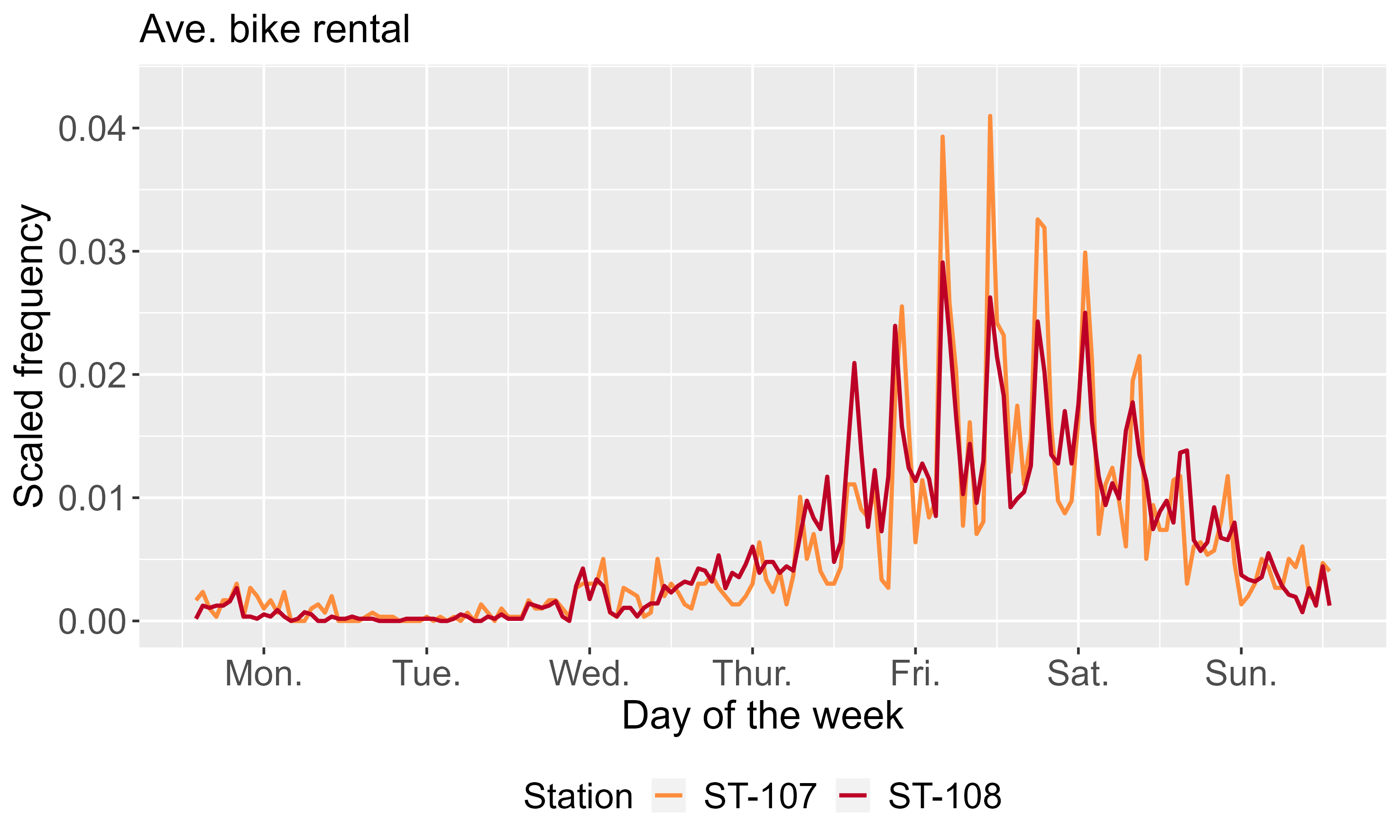}} 
\subfloat{\includegraphics[width = .48\textwidth]{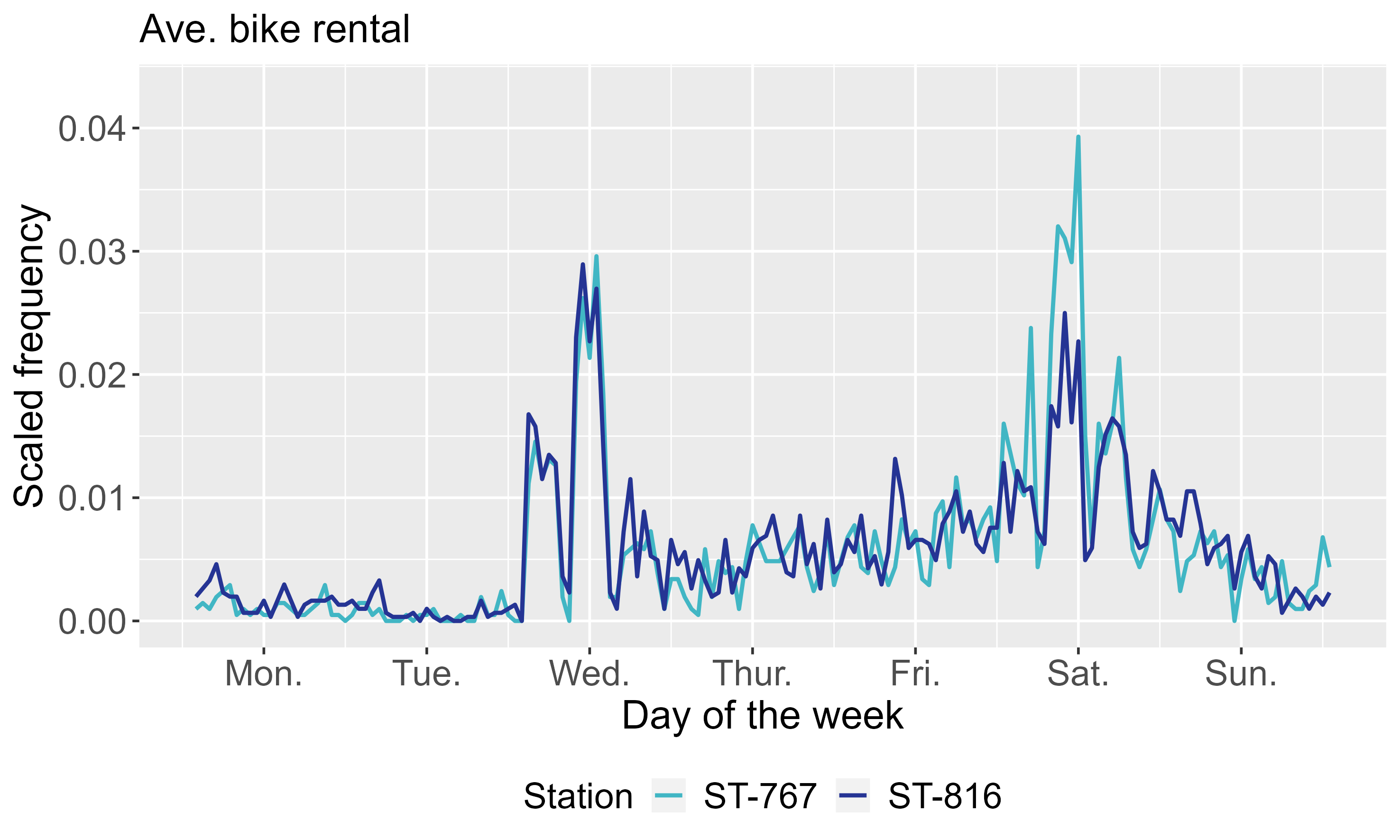}} 
\caption{Relative hourly bike rental frequencies from four selected stations averaged over eight week of individual stations. ST-107 and ST-108 (left), and ST-767 and ST-816 (right) are adjacent to each other, respectively.} 
\label{fig:temporal}
\end{figure}

Bike rental patterns exhibit substantial heterogeneity across stations. 
In Figure~\ref{fig:temporal}, we present a plot depicting the relative hourly bike rental  frequencies averaged over the observation period for selected pairs of adjacent stations.
There are clear station-specific patterns over the course of a day and a week, and nearby stations tend to display similar usage patterns. 
Specifically, stations ST-107 and ST-108 are located close to a large riverside park in Seoul and thus are used for leisure activities, as evidenced by the large number of rentals concentrated on weekends. 
On the other hand, ST-767 and ST-816 are located in a commercial district with government agencies and large firms, and exhibit spikes in bike rental frequency during weekday commuting-time.
Additionally, although daily peaks and troughs can be observed, these temporal patterns do not undergo abrupt changes in the sense that the number of bikes rented out between 9am and 10am is reasonably close to that between 10am and 11am.

\begin{figure}[h!t!b!p!]
\centering
\subfloat{\resizebox*{4.5cm}{!}{\includegraphics{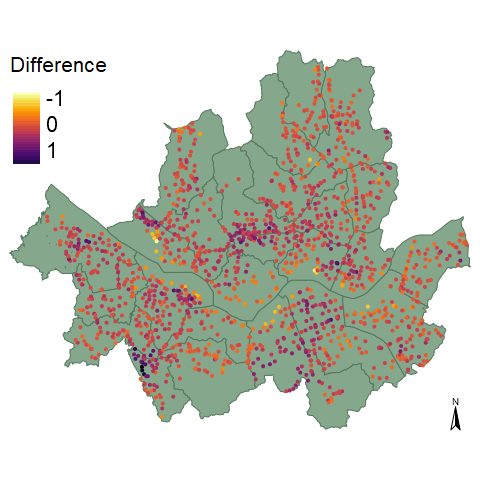}}}
\subfloat{\resizebox*{4.5cm}{!}{\includegraphics{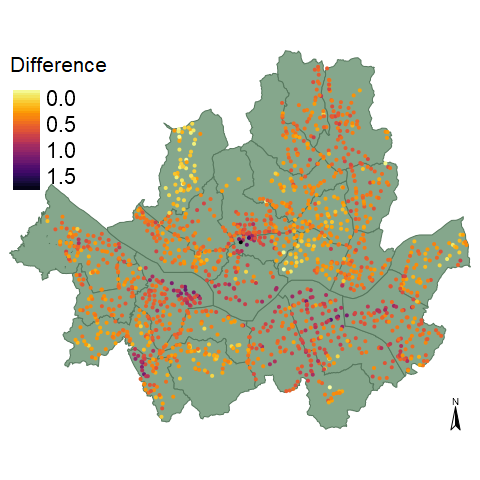}}}
\subfloat{\resizebox*{4.5cm}{!}{\includegraphics{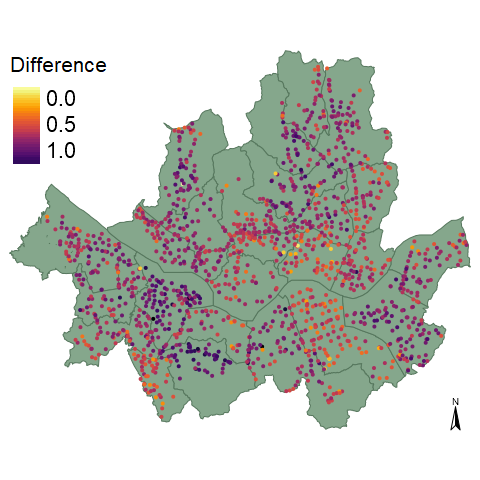}}}
\caption{Difference in average daily bike rentals in log-scale at individual stations between weekday and weekend (left), and due to precipitation (middle) and air quality (right).}
\label{fig:raindust}
\end{figure}



The left panel of Figure~\ref{fig:raindust} shows that bike usage behavior differs between weekdays and weekends, and that the degree of variation differs across stations.
It further exhibits the presence of local clusters that share similar weekday/weekend variations.
The middle and right panels of Figure~\ref{fig:raindust} show that bike usage depends on the weather condition and air quality.
Recording each day as ``rainy'' if positive precipitation is recorded in any part of the city at any time of the day, we observe that precipitation reduces bike usage as expected.
Similar observations are made with respect to the air quality: Adopting the qualitative categorization into ``good'', ``average'', ``bad,'' and ``very bad'' following the classification system of the Korean Ministry of Environment based on PM10 and PM2.5 dust concentrations, the rental frequencies across the BSS decreases in volume when the air quality is ``very bad'' compared to when it is ``good''.

Based on the above observations, we conclude that capturing the characteristics \ref{c:trend}--\ref{c:cov} identified in Section~\ref{sec:method}, is crucial in the modeling and predicting the usage data collected from a BSS, which motivates our proposed model.

\subsection{Bike-sharing system in New York}

Launched in 2013, Citi Bike is a popular bike-sharing program in New York City, the United States, sponsored by Citibank. With docking stations spread throughout the five boroughs, users can rent bikes for short trips using a mobile app or membership card. The system is designed for quick rides, with pricing encouraging returns to any station within 30--45 minutes. 
We analyze the hourly rental records April 1, 2019 and May 30, 2019 ($T = 60$ days), collected from $S = 795$ stations.
The dataset is available from \url{https://s3.amazonaws.com/tripdata/index.html}.

\subsection{Bike-sharing system in San Francisco}

\begin{figure}[h!]
\centering
\subfloat{\includegraphics[width = .5\textwidth]{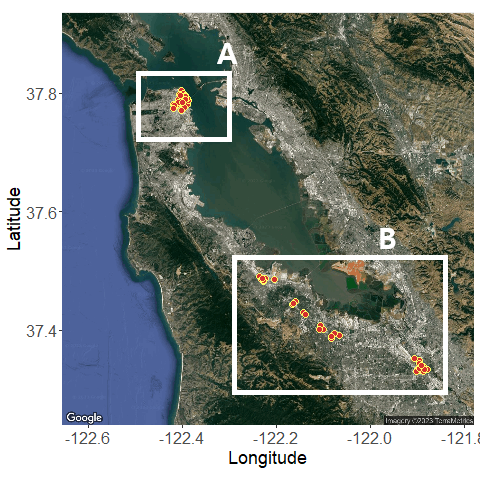}} 
\caption{Satellite map of bike rental stations in Bay area.} 
\label{fig:sf_area}
\end{figure}

Bay Wheels is a bike-sharing system serving the San Francisco Bay Area in California. It provides a network of docking stations distributed across five cities, Mountain View, Palo Alto, Redwood City, San Francisco and San Jose, and allows riders to rent bikes for short trips. 
We analyze the dataset is available in the R package \verb+bikeshare14+ \citep{bikeshare14}, which contains hourly usage of $S = 71$ stations situated throughout the Bay Area from April 1, 2014, to May 31, 2014 ($T = 61$ days). 
Figure~\ref{fig:sf_area} presents a satellite map displaying the bike rental stations, which are concentrated in two primary areas, denoted as A and B.

\subsection{Data irregularities}
\label{sec:irregular}

\begin{figure}[h!t!]
\centering
\subfloat{\includegraphics[width = .8\textwidth]{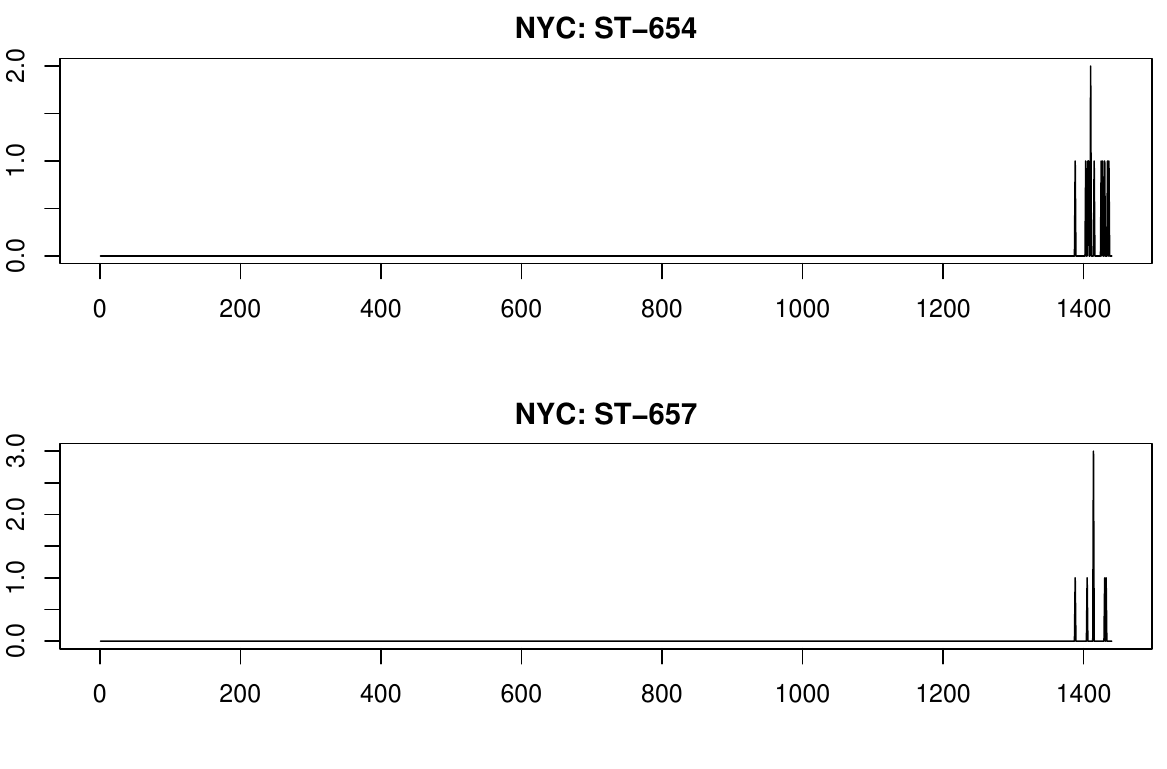}} 
\caption{Irregular usage patterns observed from bike stations in New York.} 
\label{fig:irregular}
\end{figure}

The dataset from New York contains some irregularities 
as some stations were introduced during the period in consideration (ST-654 and ST-657 in New York) such that the training data mostly contain zeros, see Figure~\ref{fig:irregular}. 
These irregularities pose a convergence issue for the INGARCH method considered in our prediction exercise (Section~\ref{sec:forecasting}).
On the other hand, the proposed fused Lasso method does not suffer from the presence of anomalous stations thanks to the fusion penalty that pools information from neighbors.

\clearpage

\section{Descriptions of the competitors}
\label{sec:ingarch}

In evaluating the forecasting performance of our proposal in Section~\ref{sec:forecasting}, we consider alternative approaches: INGARCH, 
Random Forest and XGBoost.
These methods individually model the data from each station $s \in \mc S$, namely $\{ Y_t(s), \, 1 \le t \le 24 T\}$, where $Y_t(s)$ denotes the usage at a time point $t$ and a station $s$, and $T$ the number of days in the (training) data.
On each dataset, we supply the same temporal (daily and weekly), trend and covariate effects, say $\wt{\mbf x}_t$, as those included in the model~\eqref{eq:model}, as an input to all competing methods.
Also, where cross validation is involved (random Forest and XGBoost below), we adopt the same $6$ folds as described in Section~\ref{sec:tuning} for fairness.
Further information on their implementation can be found below.

\paragraph{INGARCH.} The INGARCH approach models $\{ Y_t(s) \}_t$ as $Y_t(s) \vert \mc F_{t - 1}(s) \sim \text{Poisson}(\lambda_t(s))$ \citep{liboschik2017}.
Here, $\mc F_{t - 1}(s)$ denotes the history of the joint process $(Y_u(s), \lambda_u(s), \wt{\mbf x}_{u + 1})$ for $u \le t - 1$. 
We report the results from the INGARCH($1$, $1$) model with the log link function, i.e.\ 
\begin{align*}
\log(\lambda_t(s)) = \alpha(s) + \beta(s) \log(Y_{t - 1}(s) + 1) + \gamma(s) \log(\lambda_{t - 1}(s)) + \wt{\mbf x}_t^\top \bm\beta(s).
\end{align*}
We fit this model to each station $s = 1, \ldots, S$, using the R package {\tt tscount} \citep{tscount} on the training set, and forecast the usage on the remaining $7$ days using the {\tt predict} method provided in the package. 
Although not reported, we have attempted INGARCH models of larger orders and alternative conditional distributions such as Negative Binomial, which did not markedly improve the forecasting performance while sometimes, we observed numerical instabilities when applied to model a large number of time series.


\paragraph{Random Forest and XGBoost.} We adopt the packages \verb+randomForest+ \citep{rf} and \verb+xgboost+ \citep{xgb}, making use of the \verb+train+ and \verb+predict+ methods of the package \verb+caret+ \citep{caret}.
We have compared their performance with and without the trend, and found the former perform much worse than the latter.
For this reason, we report the results obtained without the covariate included for trend modeling in the paper.

\section{Additional empirical results}
\label{sec:empiric}

\subsection{Additional results from BSS in Seoul}
\label{sec:seoul}

\subsubsection{Fold-wise mean squared Pearson residuals}
\label{sec:fig_vals}

Table~\ref{tab:cv_err} and Table~\ref{tab:forcast_err} list precise mean squared Pearson residuals (MSPR) values reported in Figure~\ref{fig:mspr} and Figure~\ref{fig:mspr_per_cases}, respectively.

\begin{table}[h!t!]
\centering
\caption{Mean squared Pearson residuals (MSPRs) corresponding to Figure~\ref{fig:mspr}.
}
\label{tab:cv_err}
\resizebox{0.7\columnwidth}{!}
{
\begin{tabular}{c cccccc }
\toprule
& \multicolumn{6}{c}{Fold}   \\
Method & 1 & 2 & 3 & 4 & 5 & 6  \\
\cmidrule(lr){1-1} \cmidrule(lr){2-7} 
Fused Lasso (375m)  & 1.835 & 1.569 & 1.537 & 1.339 & 1.585 & 1.531 \\
Fused Lasso (750m)  & 1.822 & 1.564 & 1.53 & 1.334 & 1.58 & 1.525 \\
Fused Lasso (1500m)  & 1.819 & 1.563 & 1.528 & 1.332 & 1.578 & 1.524 \\
Fused Lasso (3000m)  & 1.825 & 1.583 & 1.538 & 1.347 & 1.61 & 1.543 \\
Lasso-only  & 1.882 & 1.603 & 1.585 & 1.371 & 1.627 & 1.556 \\
Full-interaction  & 2.035 & 1.739 & 1.763 & 1.472 & 1.804 & 1.691  \\
No-interaction  & 2.238 & 1.997 & 1.939 & 1.668 & 2.008 & 1.923 \\
\bottomrule
\end{tabular}}
\end{table}

\begin{table}[h!t!]
\centering
\caption{Modified mean Squared Pearson Residuals (PE) corresponding to Figure~\ref{fig:mspr_per_cases}.
}
\label{tab:forcast_err}
\resizebox{0.5\columnwidth}{!}
{
\begin{tabular}{c ccc }
\toprule
& \multicolumn{3}{c}{$Y_t(s)$}   \\
Method & All & Zeros & Non-zeros   \\
\cmidrule(lr){1-1} \cmidrule(lr){2-4} 
Fused Lasso (375m)  & 1.514 & 1.000 & 1.862\\
Fused Lasso (750m)  & 1.501 & 1.012 & 1.832 \\
Fused Lasso (1500m)  & 1.497 & 1.022 & 1.818\\
Fused Lasso (3000m)  & 1.499 & 1.062 & 1.794\\
Lasso-only  & 1.535 & 0.987 & 1.904 \\
Full-interaction   & 1.723 & 0.939 & 2.251\\
No-interaction  & 1.842 & 1.113 & 2.333\\
INGARCH & 1.758 & 0.952 & 2.302 \\
\bottomrule
\end{tabular}}
\end{table}

\subsubsection{Parameter estimates}
\label{sec:radius}

Table~\ref{tab:coef750} presents the estimated coefficients obtained with the radius $r \in \{ 375, 750, 3000\}$ used in defining the neighboring stations $\mc N_r(s)$, see Table~\ref{tab:coef} in the main text for the results obtained with $r = 1500$.

\begin{table}[h!t!]
\centering
\caption{Estimated coefficients for the covariate effects by the proposed fused Lasso regression method from each fold used in the $6$-fold CV and from the full training data when $r \in \{ 375, 750, 3000\}$.
}
\label{tab:coef750}
\resizebox{0.7\columnwidth}{!}
{
\begin{tabular}{cc ccccccc c c c c c}
\toprule
& & \multicolumn{6}{c}{Fold} &  \\
$r$ & & 1 & 2 & 3 & 4 & 5 & 6 & All \\
\cmidrule(lr){1-1} \cmidrule(lr){2-2} \cmidrule(lr){3-8} \cmidrule(lr){9-9} 
375m & $\alpha$ & 0.065 & 0.063 & 0.061 & 0.059 & 0.060 & 0.062 & 0.062  \\ 
& $\beta^{\rain}$ &  -2.181 & -2.359 & -2.501 & -2.297 & -2.222 & -2.454 & -2.343  \\ 
& $\beta^{\air}_1$ & 0.114 & 0.104 & 0.064 & 0.071 & 0.134 & 0.137 & 0.103  \\ 
& $\beta^{\air}_2$ &  0.134 & 0.132 & 0.067 & 0.087 & 0.130 & 0.161 & 0.115  \\ 
& $\beta^{\air}_3$ & 0.256 & 0.274 & 0.030 & 0.220 & 0.273 & 0.263 & 0.226  \\
\cmidrule(lr){1-1} \cmidrule(lr){2-2} \cmidrule(lr){3-8} \cmidrule(lr){9-9} 
750m & $\alpha$ & 0.065 & 0.063 & 0.061 & 0.059 & 0.060 & 0.062 & 0.062  \\ 
 & $\beta^{\rain}$ &  -2.180 & -2.359 & -2.501 & -2.297 & -2.221 & -2.454 & -2.343  \\ 
 & $\beta^{\air}_1$ & 0.114 & 0.104 & 0.064 & 0.071 & 0.134 & 0.137 & 0.103  \\ 
 & $\beta^{\air}_2$ &  0.134 & 0.131 & 0.067 & 0.087 & 0.129 & 0.161 & 0.115  \\ 
 & $\beta^{\air}_3$ & 0.256 & 0.272 & 0.033 & 0.220 & 0.273 & 0.262 & 0.227  \\
\cmidrule(lr){1-1} \cmidrule(lr){2-2} \cmidrule(lr){3-8} \cmidrule(lr){9-9} 
3000m & $\alpha$ & 0.065 & 0.063 & 0.061 & 0.059 & 0.060 & 0.061 & 0.062  \\ 
& $\beta^{\rain}$ &  -2.179 & -2.359 & -2.501 & -2.297 & -2.22 & -2.454 & -2.343  \\ 
& $\beta^{\air}_1$ & 0.116 & 0.106 & 0.066 & 0.073 & 0.136 & 0.138 & 0.105  \\ 
& $\beta^{\air}_2$ &  0.139 & 0.135 & 0.073 & 0.092 & 0.132 & 0.166 & 0.120  \\ 
& $\beta^{\air}_3$ & 0.260 & 0.274 & 0.051 & 0.221 & 0.278 & 0.266 & 0.232  \\
\bottomrule
\end{tabular}}
\end{table}

\begin{figure}[h!t!b!]
\centering
\subfloat{\includegraphics[width = .425\textwidth]{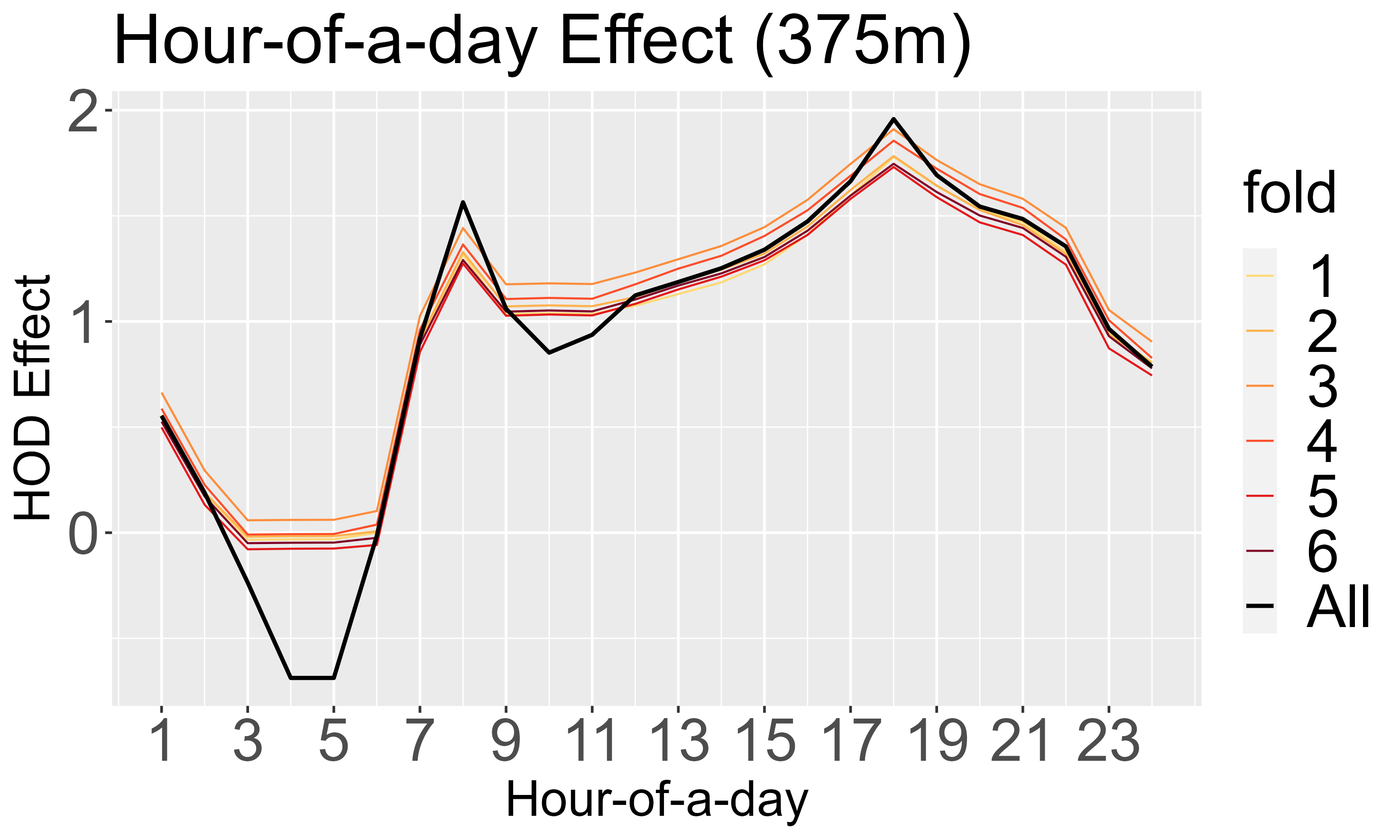}}
\subfloat{\includegraphics[width = .425\textwidth]{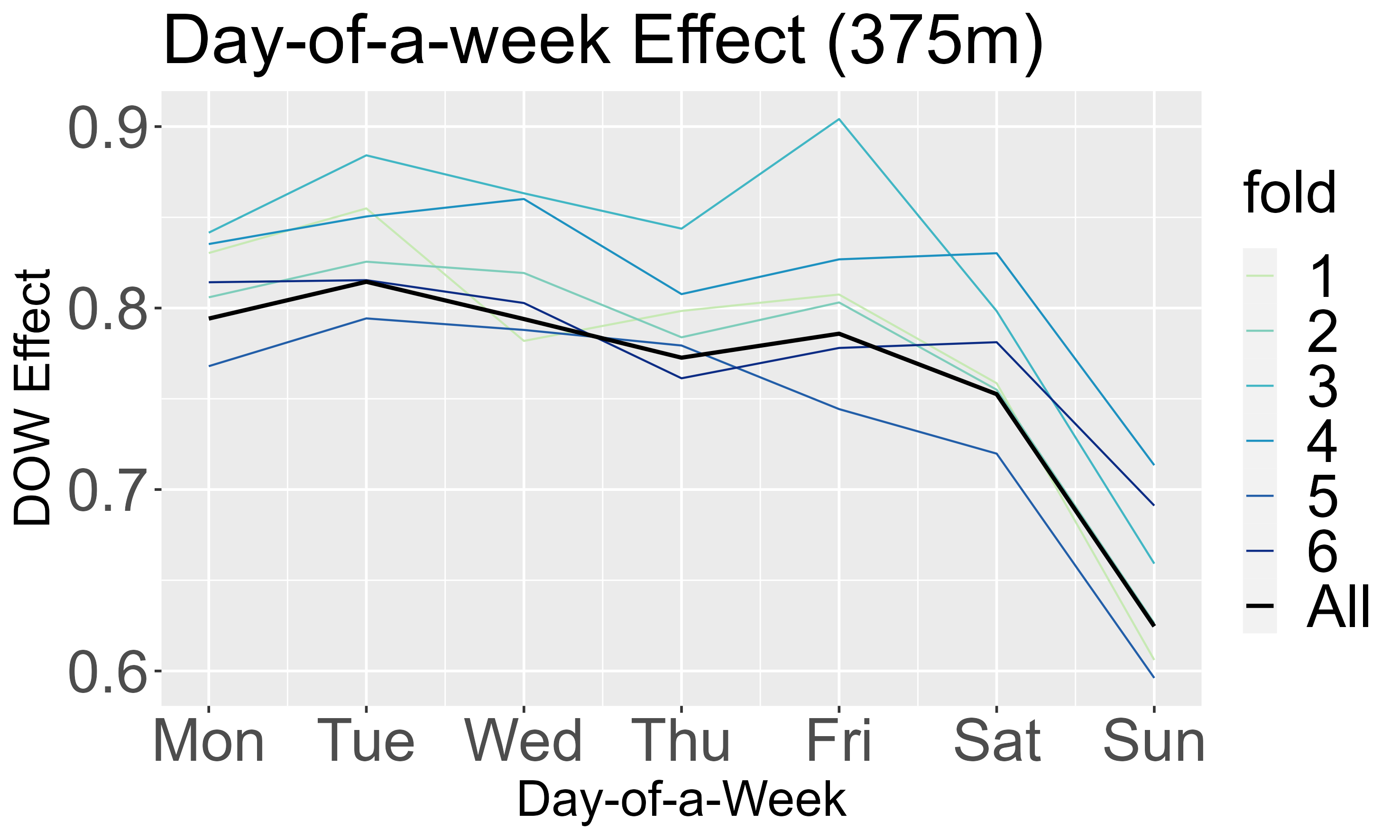}} \\
\subfloat{\includegraphics[width = .425\textwidth]{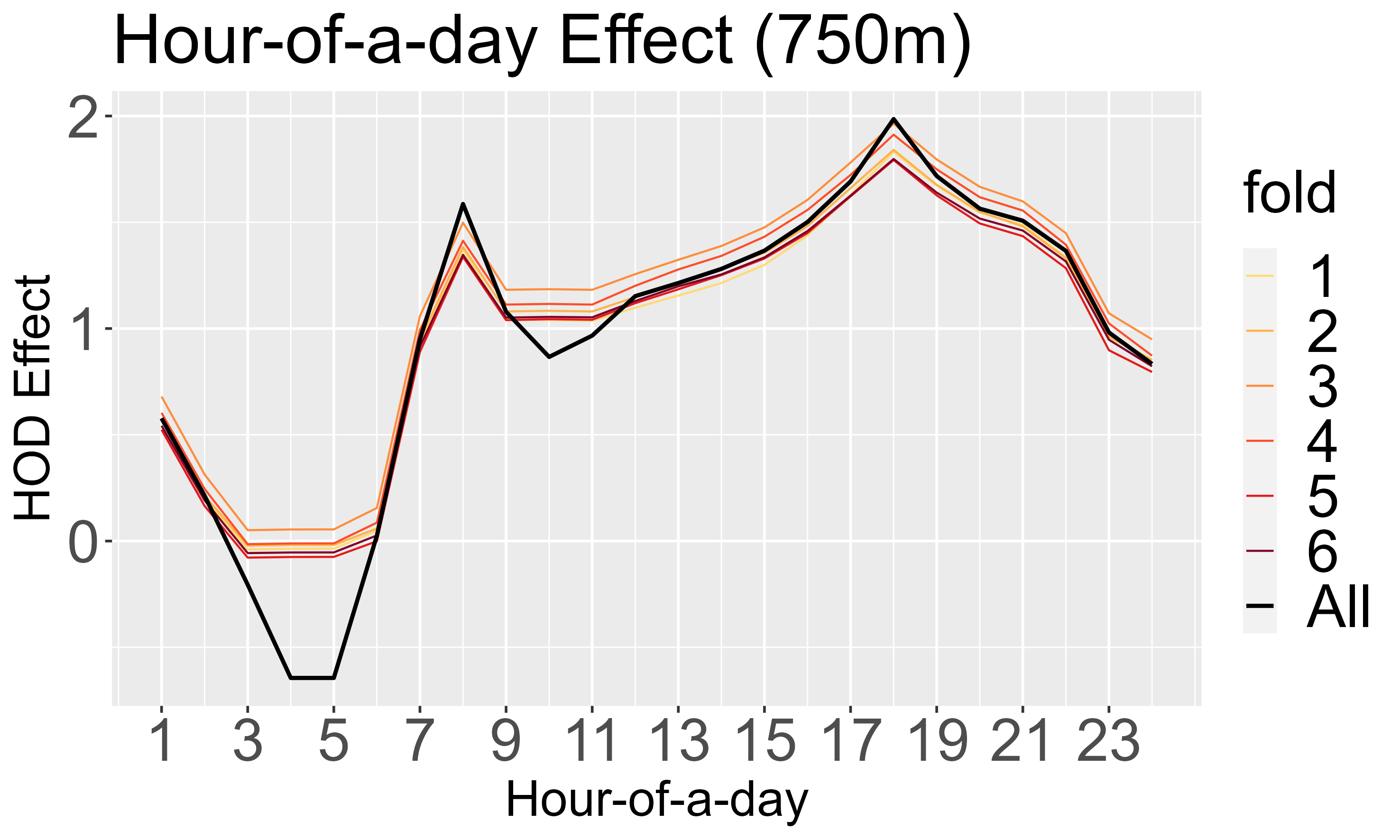}}
\subfloat{\includegraphics[width = .425\textwidth]{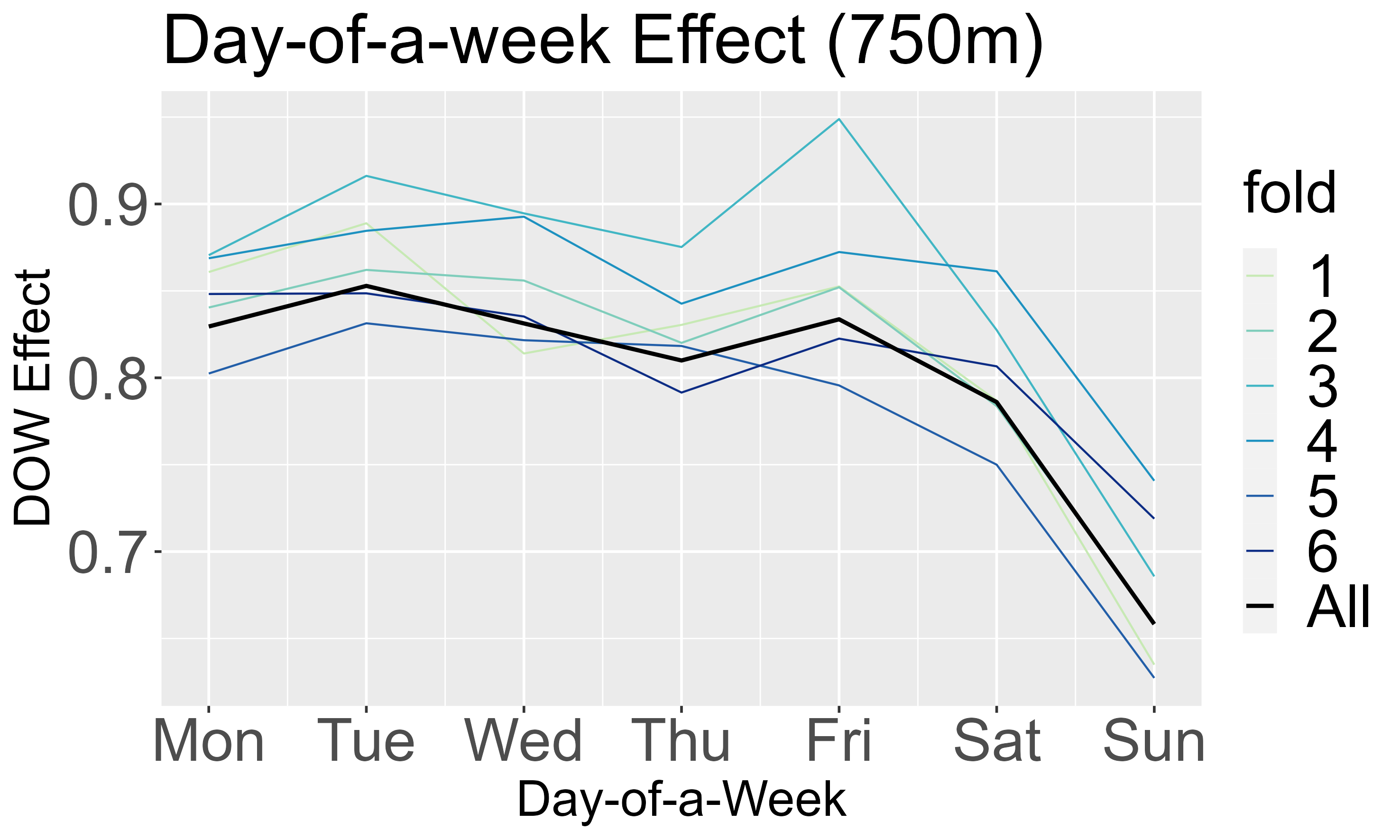}} \\
\subfloat{\includegraphics[width = .425\textwidth]{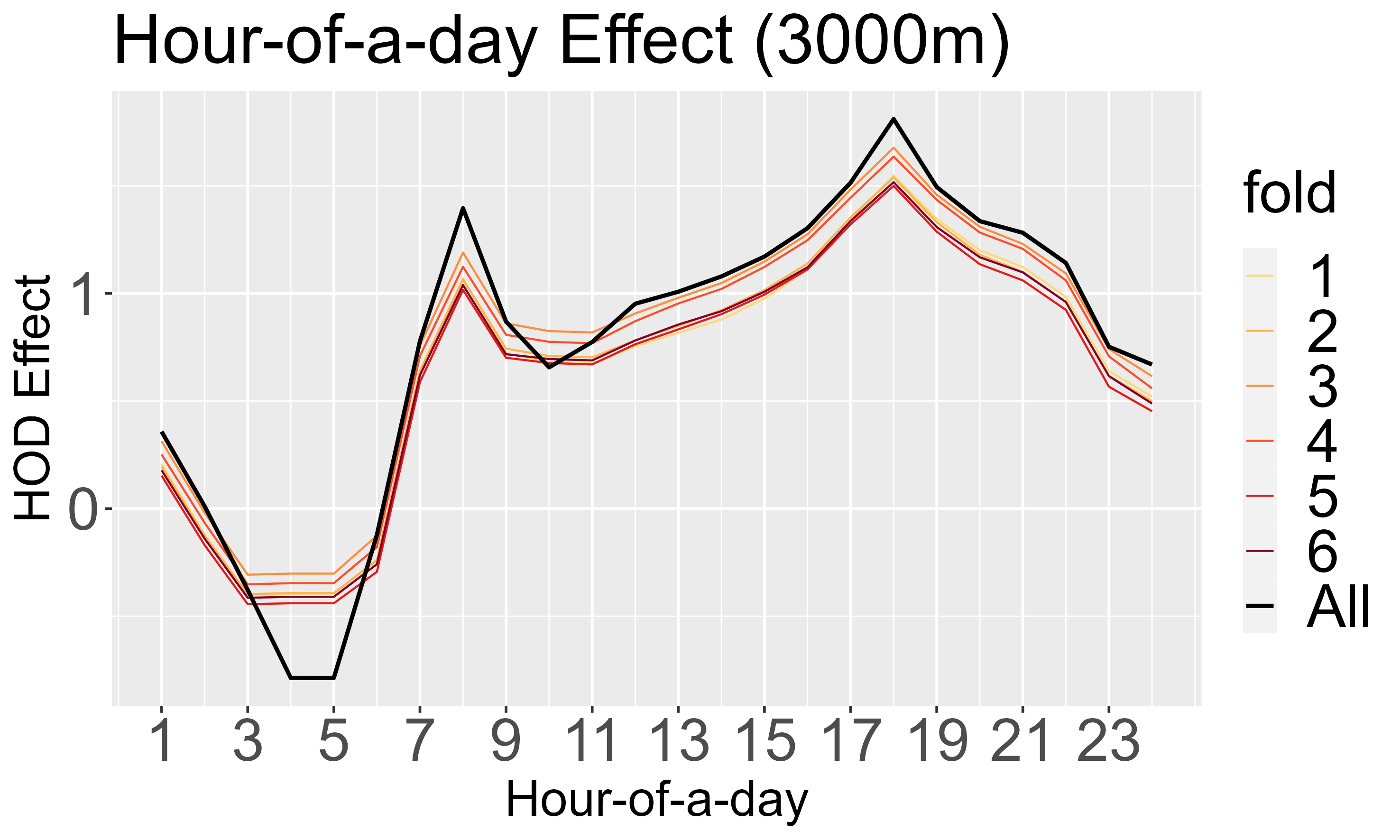}}
\subfloat{\includegraphics[width = .425\textwidth]{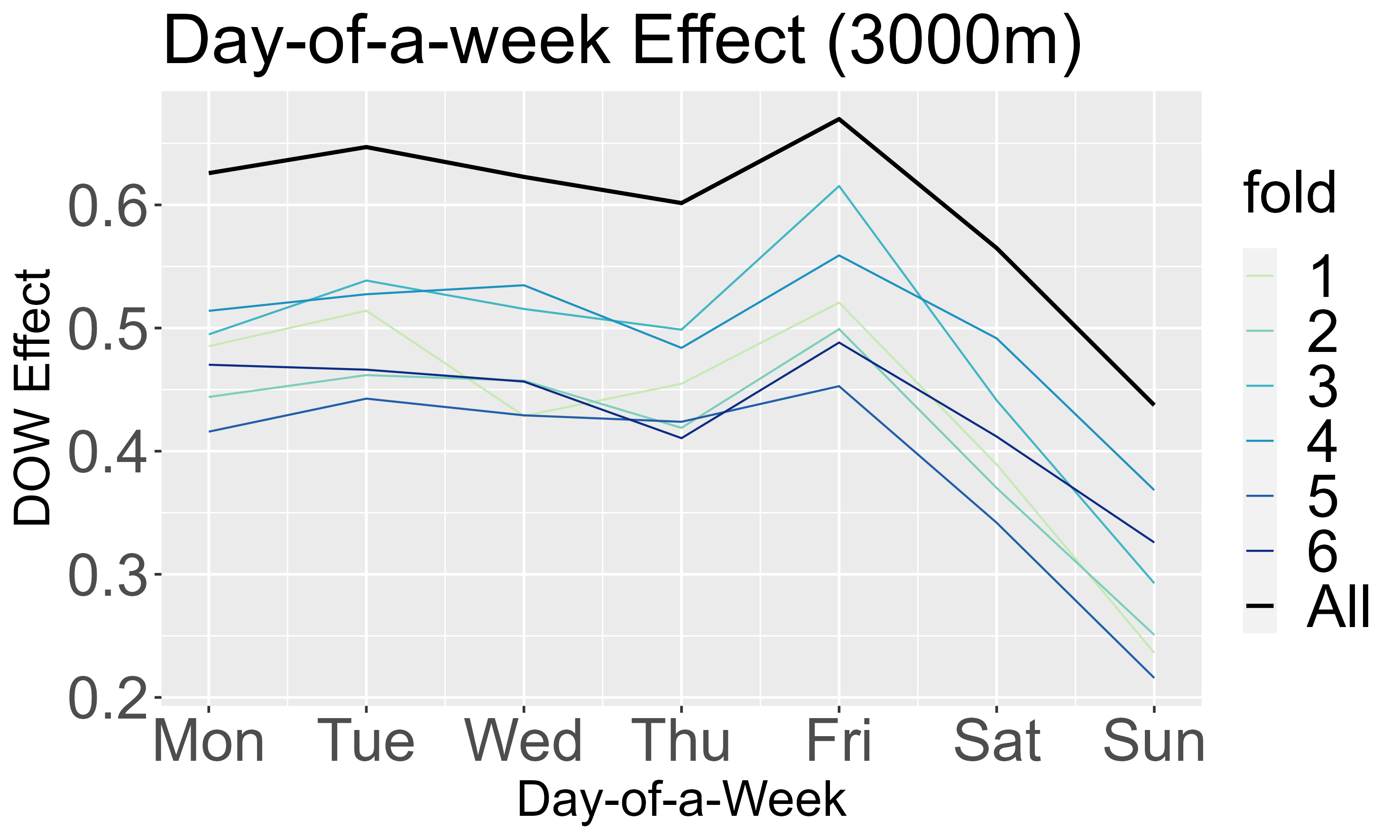}} \\
\caption{Parameter estimates for $\phi^{\hod}_h, \, h \in \mc H$ (left column) and $\phi^{\dow}_d, \, d \in \mc D$ (right column) from each fold used in the $6$-fold CV and from the full training data. The radius values in use are 375, 750 and 3000 meters, arranged from top to bottom.}
\label{fig:thetas:750}
\end{figure}

\begin{figure}[h!]
\centering
\subfloat{\resizebox*{5cm}{!}{\includegraphics{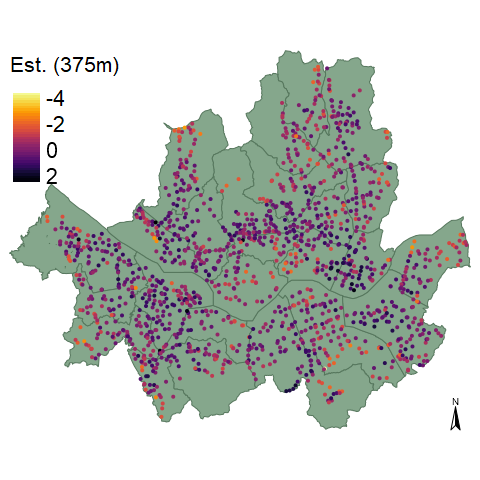}}}
\subfloat{\resizebox*{5cm}{!}{\includegraphics{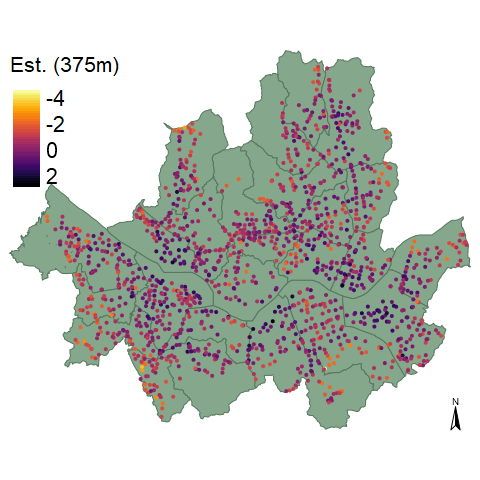}}}
\subfloat{\resizebox*{5cm}{!}{\includegraphics{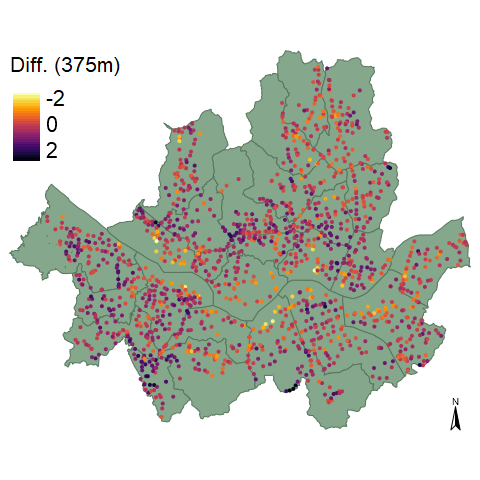}}} \\
\subfloat{\resizebox*{5cm}{!}{\includegraphics{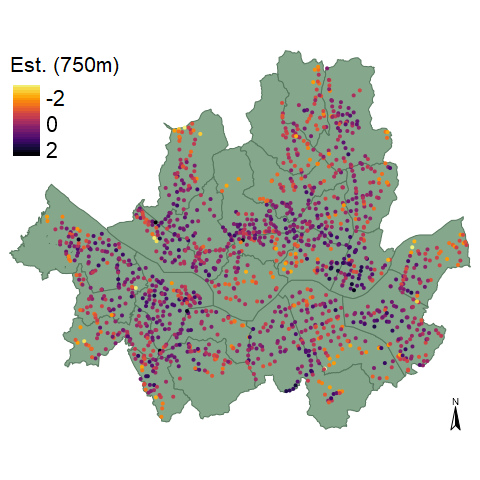}}}
\subfloat{\resizebox*{5cm}{!}{\includegraphics{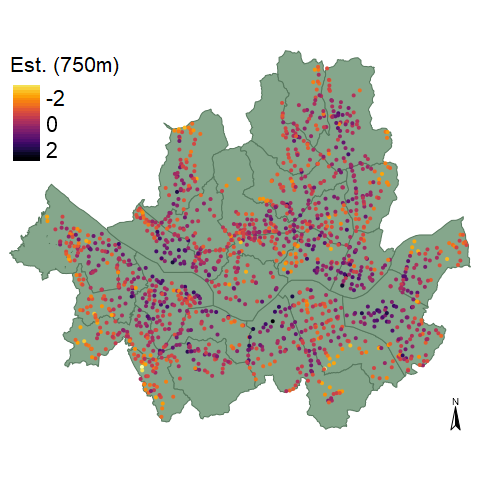}}}
\subfloat{\resizebox*{5cm}{!}{\includegraphics{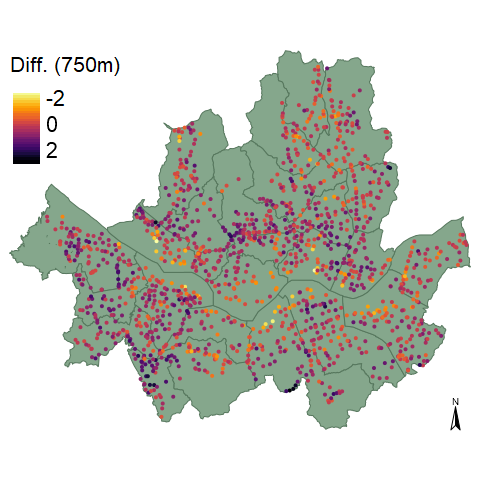}}} \\
\subfloat{\resizebox*{5cm}{!}{\includegraphics{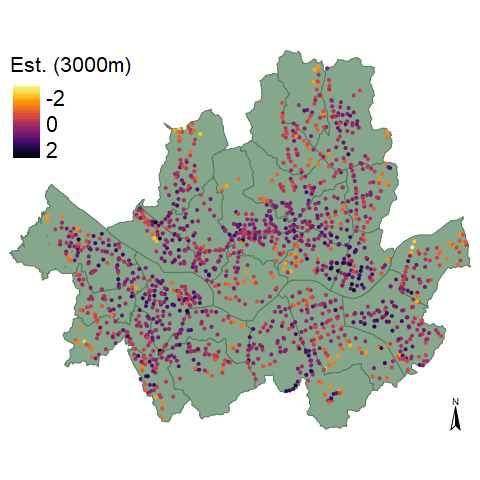}}}
\subfloat{\resizebox*{5cm}{!}{\includegraphics{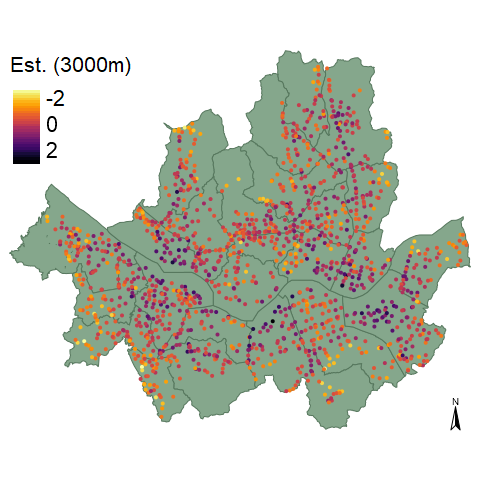}}}
\subfloat{\resizebox*{5cm}{!}{\includegraphics{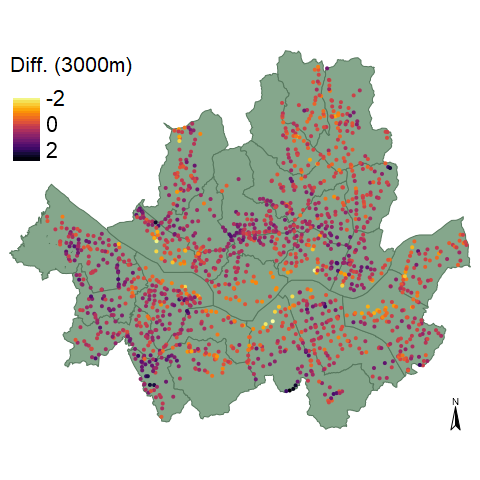}}}
\caption{Estimated station-specific bike demands in log-scale ($\wh{\theta}_s+\wh{\theta}_d + \wh{\theta}_{h} + \wh{\theta}_{s,d}+\wh{\theta}_{s,h} $) from the model fitted at 8am on Tuesdays (left column), at 8pm on Sundays (middle column) and their differences (right column). The radius values in use are 375, 750 and 3000 meters, arranged from top to bottom.} 
\label{fig:fitted:750}
\end{figure}

Figures~\ref{fig:thetas:750} and~\ref{fig:fitted:750} plot the results obtained with $r = 375, \ 750$ and $3000$, 
see Figures~\ref{fig:thetas} and~\ref{fig:fitted} in the main text for the results obtained with $r = 1500$.

\subsubsection{Cross validation for the selection of penalty parameters}
\label{sec:cv_table}

The proposed fused Lasso involves three tuning parameters, $\lambda$, $\lambda_N$ and $\lambda_H$ in~\eqref{eq:org_obj_admm}.
Based on preliminary numerical studies, we chose to fix $\lambda$ at $\lambda = 1.120 \times 10^{-5}$ which provided reasonable estimates. In search of $\lambda_N$ and $\lambda_H$, we performed grid-based search over a logarithmically equispaced grid ranging from $e^{-1}$ to $e^0$ with 10 grid points for $\lambda_N$, and that from 8 to 12 with 5 grid points for $\lambda_H$. Table~\ref{tab:cv} presents the MSPR in~\eqref{eq:cv} obtained from the $6$-fold CV with $r \in \{ 375, 750, 1500, 3000\}$ over the grids for $\lambda_H$ and $\lambda_N$.

\begin{table}[h!t!]
\centering
\caption{MSPR in~\eqref{eq:cv} from the $6$-fold CV with $r \in \{ 375, 750, 1500, 3000\}$ over varying $\lambda_H$ and $\lambda_N$.}
\label{tab:cv}
\resizebox{0.8\columnwidth}{!}
{
\begin{tabular}{c cccccccccc}
\toprule
\multicolumn{11}{c}{ $r=375$ } \\
\midrule
& \multicolumn{10}{c}{$\lambda_N$}   \\
$\lambda_H$ & 1.00  & 0.89 & 0.80 & 0.72 & 0.64 & 0.57 & 0.51 & 0.46 & 0.41 & 0.37 \\
\cmidrule(lr){1-1} \cmidrule(lr){2-11} 
12.00 &  1.567 & 1.567 & 1.568 & 1.568 & 1.568 & 1.568 & 1.568 & 1.569 & 1.569 & 1.569  \\ 
 10.84 &  1.567 & 1.567 & 1.567 & 1.567 & 1.568 & 1.568 & 1.568 & 1.568 & 1.569 & 1.569  \\ 
 9.89 &  1.566 & 1.567 & 1.567 & 1.567 & 1.567 & 1.568 & 1.568 & 1.568 & 1.568 & 1.569  \\ 
 8.85 &  1.566 & 1.567 & 1.567 & 1.567 & 1.568 & 1.568 & 1.568 & 1.568 & 1.569 & 1.569  \\ 
 8.00 &  1.567 & 1.567 & 1.568 & 1.568 & 1.568 & 1.568 & 1.569 & 1.569 & 1.569 & 1.569  \\
 \bottomrule
\multicolumn{11}{c}{ $r=750$ } \\
\midrule
& \multicolumn{10}{c}{$\lambda_N$}   \\
$\lambda_H$ & 1.00  & 0.89 & 0.80 & 0.72 & 0.64 & 0.57 & 0.51 & 0.46 & 0.41 & 0.37 \\
\cmidrule(lr){1-1} \cmidrule(lr){2-11} 
12.00 &  1.561 & 1.561 & 1.561 & 1.561 & 1.561 & 1.562 & 1.562 & 1.562 & 1.562 & 1.562  \\ 
 10.84 &  1.566 & 1.566 & 1.566 & 1.561 & 1.561 & 1.561 & 1.561 & 1.561 & 1.562 & 1.562  \\ 
 9.89 &  1.559 & 1.560 & 1.560 & 1.560 & 1.560 & 1.560 & 1.561 & 1.561 & 1.561 & 1.561  \\ 
 8.85 &  1.559 & 1.559 & 1.560 & 1.560 & 1.560 & 1.560 & 1.561 & 1.561 & 1.561 & 1.562  \\ 
 8.00 &  1.559 & 1.560 & 1.560 & 1.560 & 1.560 & 1.561 & 1.561 & 1.562 & 1.562 & 1.562  \\
 \bottomrule
\multicolumn{11}{c}{ $r=1500$ } \\
\midrule
& \multicolumn{10}{c}{$\lambda_N$}   \\
$\lambda_H$ & 1.00  & 0.89 & 0.80 & 0.72 & 0.64 & 0.57 & 0.51 & 0.46 & 0.41 & 0.37 \\
\cmidrule(lr){1-1} \cmidrule(lr){2-11} 
12.00 &  1.560 & 1.559 & 1.560 & 1.560 & 1.559 & 1.559 & 1.559 & 1.559 & 1.559 & 1.558  \\ 
 10.84 &  1.562 & 1.561 & 1.560 & 1.560 & 1.559 & 1.559 & 1.559 & 1.559 & 1.559 & 1.558  \\ 
 9.89 &  1.562 & 1.561 & 1.560 & 1.559 & 1.559 & 1.558 & 1.558 & 1.558 & 1.558 & 1.558  \\ 
 8.85 &  1.562 & 1.561 & 1.560 & 1.559 & 1.558 & 1.558 & 1.558 & 1.558 & 1.558 & 1.557  \\ 
 8.00 &  1.563 & 1.561 & 1.560 & 1.559 & 1.558 & 1.558 & 1.558 & 1.558 & 1.557 & 1.558  \\
 \bottomrule
\multicolumn{11}{c}{ $r=3000$ } \\
\midrule
& \multicolumn{10}{c}{$\lambda_N$}   \\
$\lambda_H$ & 1.00  & 0.89 & 0.80 & 0.72 & 0.64 & 0.57 & 0.51 & 0.46 & 0.41 & 0.37 \\
\cmidrule(lr){1-1} \cmidrule(lr){2-11} 
12.00 &  1.641 & 1.632 & 1.625 & 1.617 & 1.610 & 1.605 & 1.600 & 1.596 & 1.592 & 1.589  \\ 
 10.84 &  1.635 & 1.626 & 1.616 & 1.609 & 1.603 & 1.598 & 1.593 & 1.589 & 1.586 & 1.583  \\ 
 9.89 &  1.631 & 1.621 & 1.612 & 1.605 & 1.599 & 1.593 & 1.588 & 1.585 & 1.582 & 1.579  \\ 
 8.85 &  1.630 & 1.619 & 1.610 & 1.603 & 1.596 & 1.590 & 1.586 & 1.582 & 1.579 & 1.576  \\ 
 8.00 &  1.629 & 1.619 & 1.609 & 1.601 & 1.595 & 1.589 & 1.584 & 1.58 & 1.577 & 1.575  \\
 \bottomrule
\end{tabular}}
\end{table}

\subsubsection{Dispersion test}
\label{sec:overdisp}

We conduct a Wald-type hypothesis test against the alternative hypothesis of quasi-Poisson model with a positive dispersion parameter \citep{aer}.
Specifically, for the hourly rental frequency $Y_i$ with $\E(Y_i) = \mu_i$, we model its variance as $\Var(Y_i) = \mu_i + \alpha \mu_i$ and test the null hypothesis of $H_0: \alpha = 0$ against the alternative $H_1: \alpha > 0$; under $H_1$, the random variable $Y_i$ follows a quasi-Poisson distribution, see Table~\ref{tab:overdisp} for the results.
For all of the methods under consideration, including the newly added competitor based on a count time series model, the null hypothesis is rejected. 
This collective rejection provides evidence in favor of the presence of over-dispersion in the data, and the dispersion parameter is estimated at a similar value with the exception of `No-interaction' method.

\begin{table}[h!t!b!]
\centering
\caption{Results of the dispersion test. The top row presents the $p$-values from the Wald-type test and the bottom row gives the estimates of the dispersion parameter $\alpha$.}
\label{tab:overdisp}
\resizebox{0.95\columnwidth}{!}{
\begin{tabular}{c cccc cccc}
\toprule
& \multicolumn{4}{c}{Fused Lasso} & & & & \\
\cmidrule(lr){2-5}
& 375m & 750m & 1500m & 3000m & Lasso-only & Full-interaction & No-interaction & INGARCH \\
\cmidrule(lr){1-1} \cmidrule(lr){2-5} \cmidrule(lr){6-9}
$p$-value & $< 10^{-3}$  & $< 10^{-3}$ & $< 10^{-3}$ & $< 10^{-3}$ & $< 10^{-3}$  & $< 10^{-3}$ & $< 10^{-3}$ & $< 10^{-3}$ \\
$\wh{\alpha}$ & 1.495 & 1.497 & 1.499 & 1.529 & 1.491 & 1.451 & 1.875 & 1.397 \\
\bottomrule
\end{tabular}}
\end{table}

\clearpage

\subsection{Additional results from BSS in New York}
\label{sec:ny}

Similarly to the data analysis conducted in the main text on the bike-sharing system in Seoul, the first 53 days serves as a training set for selecting the penalization parameters.
The tuning parameters are chosen via $6$-fold cross validation, with each fold constructed to maintain a balanced representation of days of the week. The last 7 days are reserved for a forecasting exercise. 
Among three tuning parameters $\lambda$, $\lambda_N$ and $\lambda_H$, we fix $\lambda$ at $\lambda = 1.120 \times 10^{-5}$ as determined in the Seoul data analysis. We then performed a grid-based search over a logarithmically equispaced grid for $\lambda_N$ ranging from 
$e^{-2}$ to $e^1$ with 10 grid points, and for $\lambda_H$ ranging from $e^{1.386}$ to $e^{2.649}$ with 3 grid points\footnote{For forecasting exercises with $r = 3000$m, the tuning parameters chosen from $r = 1500$m were used.}.

The covariate information on air quality and the capacity of station is unavailable for this dataset and thus omitted in the data analysis. In all other respects, we employ the same methodology as outlined in Section~\ref{sec:real:penalty}.

\subsection{Additional results from BSS in San Francisco}
\label{sec:sf}

Consistent with the data analysis performed on the bike-sharing systems in Seoul and New York, the first 53 days are used as the training set to determine the tuning parameters through 6-fold cross-validation, ensuring that each fold maintains a balanced representation of the days of the week. The final 7 days are set aside for forecasting.

The explored grid values for $\lambda$, $\lambda_N$ and $\lambda_H$ are equispaced in log-scale ranging from $10^{-5}$ to $10^{-3}$ with five grid points, ranging from $0.1$ to $4.0$ with twelve grind points, and ranging from $12$ to $25$ with six grid points, respectively.
For this dataset, we do not have the covariate information on air quality which is omitted; otherwise, we take an approach identical to that in Section~\ref{sec:real:penalty}.

\begin{figure}[htb]
\centering
\includegraphics[width = 0.9\textwidth]{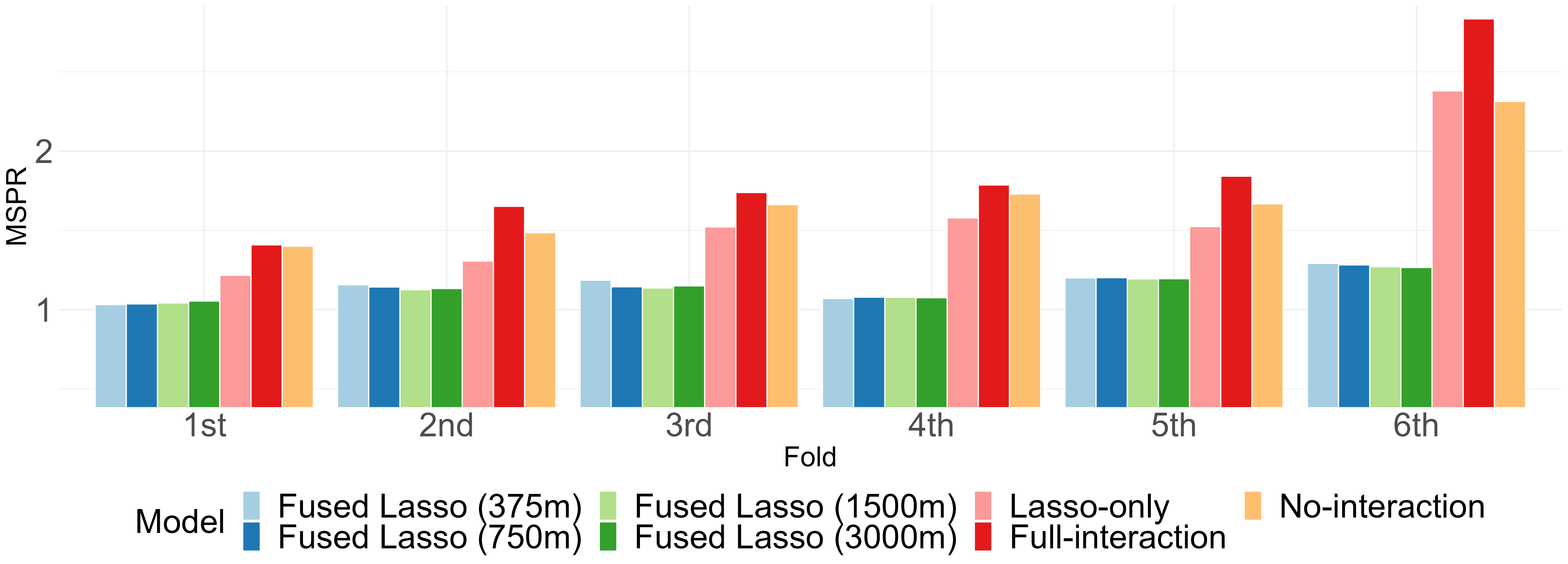}
\caption{San Francisco: MSPRs from each fold used in the $6$-fold CV.}
\label{fig:mspr_sf}
\end{figure}

\begin{figure}[htb]
\centering
\includegraphics[width = 0.9\textwidth]{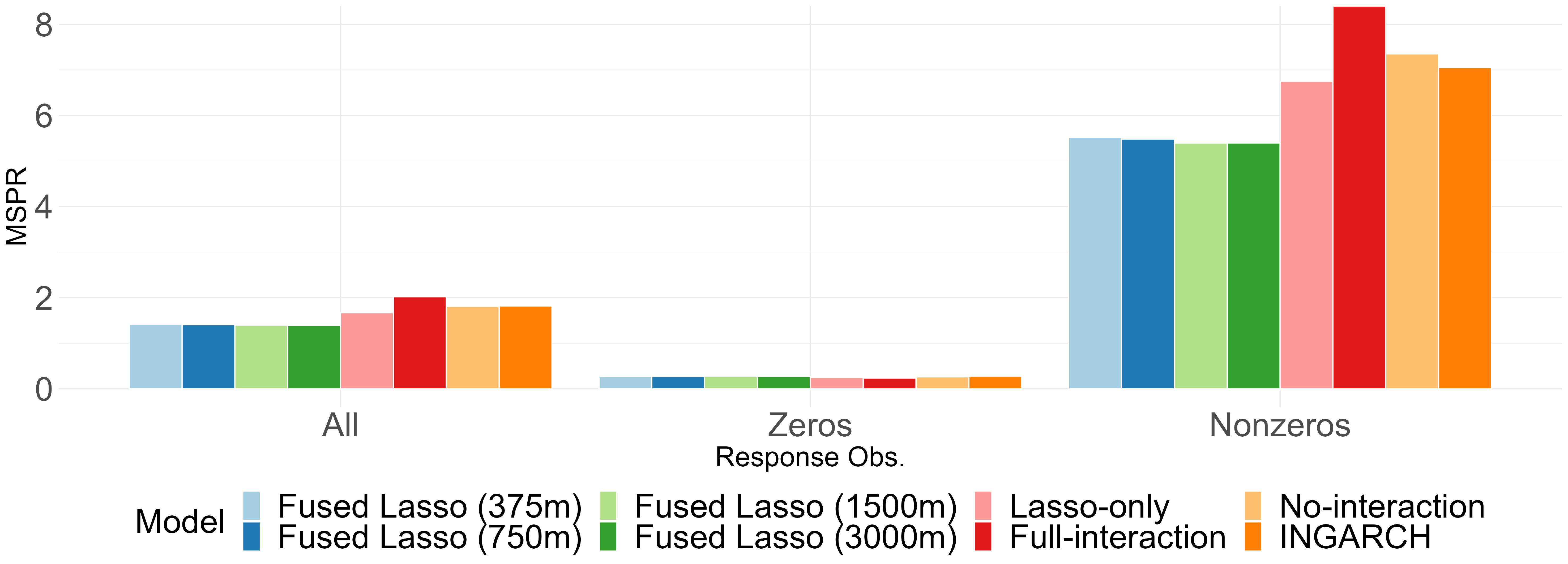}
\caption{San Francisco: Modified MSPRs on the test data (PE). The left panel displays the PE computed from the entire test set, and the middle and right panels display the PE computed with the subsets of the test set where $Y_t(s) = 0$ and $Y_t(s) > 0$, respectively.}
\label{fig:mspr_per_cases_sf}
\end{figure}

\begin{figure}[h!]
\centering
\subfloat{\includegraphics[width = .45\textwidth]{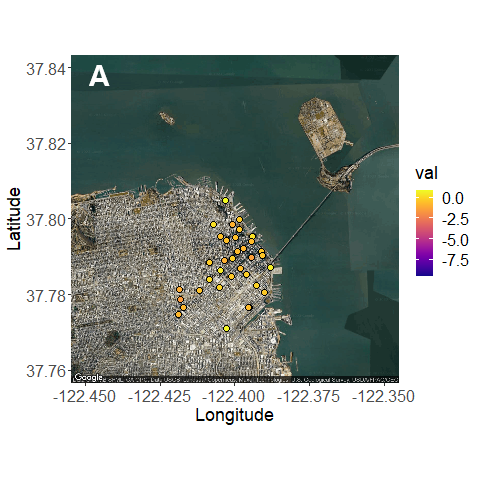}} 
\subfloat{\includegraphics[width = .45\textwidth]{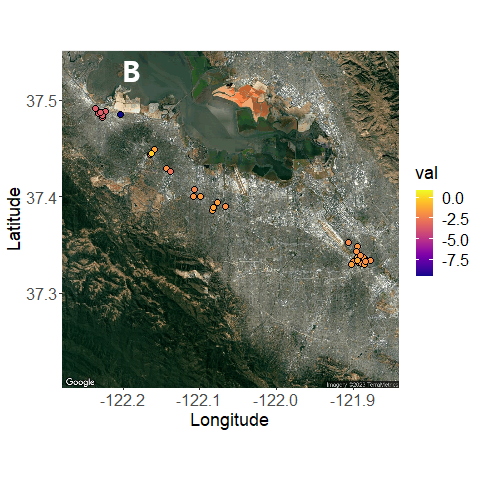}} 
\caption{San Francisco: Estimated station-specific bike demands in log-scale from the model fitted with $r=1500$ at 8pm on Sundays from region  A (left) and region B (right) displayed in Figure~\ref{fig:sf_area}. } 
\label{fig:sf_area_AB}
\end{figure}

Figures~\ref{fig:mspr_sf} and~\ref{fig:mspr_per_cases_sf} show the results closely resembling those reported in the main text, where fused Lasso (regardless of the choice of $r$) outperforms the competitors in predicting the bike usage demand.
This demonstrates the general utility of the proposed multilayer network-based fused Lasso in pooling information across the bike-sharing system.
Additionally, Figure~\ref{fig:sf_area_AB} exhibits a heatmap of the estimated station-specific bike demands.

\subsection{Additional results from prediction exercise}
\label{sec:add:pred}

We report the results complementing the prediction exercise described in Section~\ref{sec:forecasting}, see Tables~\ref{tab:forecast_seoul}--\ref{tab:forecast_SF}.
In addition to the modified MSPR in~\eqref{eq:mspr:pe}, we consider the mean squared prediction error (MSPE) and mean absolute prediction error (MAPE) measured as
\begin{align}
\label{eq:pred:err}
\text{MSPE} = \frac{1}{S \vert \mc T_{\text{test}} \vert} \sum_{s = 1}^S \sum_{t \in \mc T_{\text{test}}} (Y_t(s) - \wh{\mu}_t(s))^2 \text{ \ and \ } 
\text{MAPE} = \frac{1}{S \vert \mc T_{\text{test}} \vert} \sum_{s = 1}^S \sum_{t \in \mc T_{\text{test}}} \vert Y_t(s) - \wh{\mu}_t(s) \vert.
\end{align}

\begin{table}[htbp]
    \caption{Seoul: Prediction errors computed from the entire test set $\mc T_{\text{test}}$ (`All'), as well as those from the subsets of $\mc T_{\text{test}}$ where $Y_t(s) = 0$ (`Zeros') and $Y_t(s) > 0$ (`Non-zeros').}
    \label{tab:forecast_seoul}
      \belowrulesep = -.5pt
      \aboverulesep = 0pt
      \hspace{-1.5cm}
      \begin{center}
      \resizebox{\textwidth}{!}{
        \begin{tabular}{c  ccc ccc ccc}
        \toprule
        & \multicolumn{3}{c}{All} & \multicolumn{3}{c}{Zeros}  & \multicolumn{3}{c}{Non-zeros}  \\
    Method & MSPR & MSPE & MAPE & MSPR & MSPE & MAPE & MSPR & MSPE & MAPE \\
    \cmidrule(lr){1-1} \cmidrule(lr){2-4} \cmidrule(lr){5-7} \cmidrule(lr){8-10} 
         Fused (375m)  & 1.514  &  5.141  &  1.386  &  1.000  &  2.260  &  1.000  &  1.862  &  7.084  &  1.646 \\  
         Fused (750m)  & 1.501  &  5.144  &  1.389  &  1.012  &  2.271  &  1.012  &  1.832  &  7.083  &  1.643 \\
         Fused (1500m)  & 1.497  &  5.148  &  1.391  &  1.022  &  2.278  &  1.022  &  1.818  &  7.084  &  1.641  \\
         Fused (3000m)  & 1.499  &  5.167  &  1.407  &  1.062  &  2.324  &  1.062  &  1.794  &  7.086  &  1.640 \\
         \cmidrule(lr){1-1} \cmidrule(lr){2-4} \cmidrule(lr){5-7} \cmidrule(lr){8-10}
                  Lasso-only & 1.535  &  5.154  &  1.383  &  0.987  &  2.257  &  0.987  &  1.904  &  7.109  &  1.651 \\
                  Full-interaction & 1.723  &  5.214  &  1.375  &  0.939  &  2.241  &  0.939  &  2.251  &  7.219  &  1.669  \\
                  No-interaction & 1.841  &  6.384  &  1.529  &  1.113  &  2.707  &  1.113  &  2.331  &  8.865  &  1.809 \\
                  \cmidrule(lr){1-1} \cmidrule(lr){2-4} \cmidrule(lr){5-7} \cmidrule(lr){8-10}
                  INGARCH & 1.758  &  8.854  &  1.496  &  0.952  &  2.442  &  0.952  &  2.302  &  13.180  &  1.864  \\
                  Random Forest & 3.100  &  5.340  &  1.338  &  0.743  &  1.175  &  0.743  &  4.690  &  8.150  &  1.739 \\
                  XGBoost & 2.442  &  5.176  &  1.316  &  0.752  &  1.248  &  0.752  &  3.582  &  7.825  &  1.696   \\
            \bottomrule
        \end{tabular}}
        \end{center}
\end{table}

\begin{table}[htbp]
    \caption{New York: Prediction errors computed from the entire test set $\mc T_{\text{test}}$ (`All'), as well as those from the subsets of $\mc T_{\text{test}}$ where $Y_t(s) = 0$ (`Zeros') and $Y_t(s) > 0$ (`Non-zeros').}
    \label{tab:forecast_NYC}
      \belowrulesep = -.5pt
      \aboverulesep = 0pt
      \hspace{-1.5cm}
      \begin{center}
      \resizebox{\textwidth}{!}{
        \begin{tabular}{c  ccc ccc ccc}
        \toprule
        & \multicolumn{3}{c}{All} & \multicolumn{3}{c}{Zeros}  & \multicolumn{3}{c}{Non-zeros}  \\
    Method & MSPR & MSPE & MAPE & MSPR & MSPE & MAPE & MSPR & MSPE & MAPE \\
    \cmidrule(lr){1-1} \cmidrule(lr){2-4} \cmidrule(lr){5-7} \cmidrule(lr){8-10} 
    Fused (375m)  & 2.822  &  13.443  &  1.921  &  0.940  &  3.666  &  0.940  &  4.082  &  19.993  &  2.578  \\  
    Fused (750m)  & 2.779  &  13.461  &  1.929  &  0.961  &  3.671  &  0.961  &  3.997  &  20.019  &  2.577   \\
    Fused (1500m)  & 2.726  &  13.490  &  1.934  &  0.973  &  3.657  &  0.973  &  3.901  &  20.077  &  2.577   \\
    Fused (3000m)  & 2.785  &  13.835  &  1.973  &  1.022  &  3.684  &  1.022  &  3.966  &  20.636  &  2.610  \\
        \cmidrule(lr){1-1} \cmidrule(lr){2-4} \cmidrule(lr){5-7} \cmidrule(lr){8-10} 
        Lasso-only & 2.601  &  19.800  &  2.347  &  1.202  &  6.759  &  1.202  &  3.538  &  28.536  &  3.114 \\
        Full-interaction & 2.947  &  20.078  &  2.350  &  1.173  &  6.850  &  1.174  &  4.135  &  28.940  &  3.139   \\
        No-interaction & 3.074  &  21.105  &  2.528  &  1.310  &  7.098  &  1.310  &  4.256  &  30.488  &  3.344 \\
            \cmidrule(lr){1-1} \cmidrule(lr){2-4} \cmidrule(lr){5-7} \cmidrule(lr){8-10} 
        INGARCH & 2.235  &  14.977  &  2.040  &  0.945  &  4.156  &  0.920  &  3.105  &  22.270  &  2.794    \\
        Random Forest & 5.333  &  13.974  &  1.879  &  0.837  &  3.682  &  0.838  &  8.345  &  20.87  &  2.576 \\
        XGBoost & 4.093  &  13.406  &  1.888  &  0.910  &  3.532  &  0.910  &  6.225  &  20.022  &  2.543  \\
            \bottomrule
        \end{tabular}}
        \end{center}
\end{table}

\begin{table}[htbp]
    \caption{San Francisco: Prediction errors computed from the entire test set $\mc T_{\text{test}}$ (`All'), as well as those from the subsets of $\mc T_{\text{test}}$ where $Y_t(s) = 0$ (`Zeros') and $Y_t(s) > 0$ (`Non-zeros').}
    \label{tab:forecast_SF}
      \belowrulesep = -.5pt
      \aboverulesep = 0pt
      \hspace{-1.5cm}
      \begin{center}
      \resizebox{\textwidth}{!}{
        \begin{tabular}{c  ccc ccc ccc}
        \toprule
        & \multicolumn{3}{c}{All} & \multicolumn{3}{c}{Zeros}  & \multicolumn{3}{c}{Non-zeros}  \\
    Method & MSPR & MSPE & MAPE & MSPR & MSPE & MAPE & MSPR & MSPE & MAPE \\
    \cmidrule(lr){1-1} \cmidrule(lr){2-4} \cmidrule(lr){5-7} \cmidrule(lr){8-10} 
    Fused (375m)    & 1.362  &  1.031  &  0.491  &  0.285  &  0.435  &  0.285  &  5.020  &  3.056  &  1.191  \\  
    Fused (750m)  & 1.363  &  1.030  &  0.491  &  0.285  &  0.433  &  0.285  &  5.026  &  3.059  &  1.191  \\
    Fused (1500m)  & 1.352  &  1.036  &  0.494  &  0.288  &  0.427  &  0.288  &  4.966  &  3.102  &  1.195   \\
    Fused (3000m)  & 1.348  &  1.034  &  0.492  &  0.286  &  0.431  &  0.286  &  4.953  &  3.080  &  1.192  \\
     \cmidrule(lr){1-1} \cmidrule(lr){2-4} \cmidrule(lr){5-7} \cmidrule(lr){8-10} 
     Lasso-only & 1.605  &  1.010  &  0.463  &  0.258  &  0.457  &  0.259  &  6.178  &  2.888  &  1.158 \\
     Full-interaction & 1.950  &  1.016  &  0.449  &  0.244  &  0.488  &  0.244  &  7.745  &  2.808  &  1.147   \\
     No-interaction & 1.731  &  1.307  &  0.515  &  0.277  &  0.387  &  0.277  &  6.667  &  4.430  &  1.319  \\
     \cmidrule(lr){1-1} \cmidrule(lr){2-4} \cmidrule(lr){5-7} \cmidrule(lr){8-10} 
     INGARCH & 1.819  &  1.116  &  0.476  &  0.278  &  0.604  &  0.278  &  7.050  &  2.857  &  1.151   \\
     Random Forest & 2.586  &  1.027  &  0.450  &  0.234  &  0.477  &  0.234  &  10.571  &  2.895  &  1.185   \\
     XGBoost & 1.945  &  1.026  &  0.465  &  0.259  &  0.478  &  0.259  &  7.669  &  2.887  &  1.164   \\
            \bottomrule
        \end{tabular}}
        \end{center}
\end{table}

\end{document}